\documentclass[%
reprint,
amsmath,
amssymb,
aps,
prc,
floatfix,
]{revtex4-2}
\usepackage{graphicx}% Include figure files
\usepackage{epstopdf}
\usepackage{dcolumn}% Align table columns on decimal point
\usepackage{bm}% bold math
\usepackage[colorlinks=true, urlcolor=blue, linkcolor=blue,citecolor=blue]{hyperref}% add hypertext capabilities
\usepackage[mathlines]{lineno}% Enable numbering of text and display math
\usepackage{longtable}
\usepackage{amsmath}
\usepackage{natbib}
\usepackage{float}
\usepackage{booktabs}
\usepackage{etoolbox}
\usepackage{blindtext}
\UseRawInputEncoding
\makeatletter
\newif\ifLT@nocaption
\preto\longtable{\LT@nocaptiontrue}
\appto\endlongtable{%
\ifLT@nocaption
\addtocounter{table}{\m@ne}%
\fi}
\preto\LT@caption{%
\noalign{\global\LT@nocaptionfalse}}
\makeatother
\begin{document}
\preprint{APS/123-QED}

\title{Significance of the compound nucleus surface energy coefficients in the synthesis of the superheavy nuclei with {$\textbf{Z=112-120}$}}% Force line breaks with \\

\author{R. Zargini}
\email{reza.zargini@gmail.com,r.zargini@student.pnu.ac.ir}
\affiliation{%
 Department of Physics, Payame Noor University (PNU), P.O. Box 19395-4697, Tehran, Iran}%
 %\altaffiliation[Also at ]{Physics Department, Payame Noor University.}%Lines break automatically or can be forced with \\
\author{S. A. Seyyedi}%
\email{a.seyyedi@pnu.ac.ir
}
\affiliation{%
Department of Physics, Payame Noor University (PNU), P.O. Box 19395-4697, Tehran, Iran}%

%\collaboration{}%\noaffiliation

%\affiliation{
%\affiliation{
%Third institution, the second for Charlie Author
%\affiliation{%


\date{\today}% It is always \today, today,
\begin{abstract}
This paper investigates the impacts of the different surface energy coefficients on the compound nucleus decay modes during heavy ion fusion reactions, with focus given to the superheavy nuclei (SHN) in the range of $Z=112-118$. The evaporation-residue (ER) cross sections were calculated for different surface asymmetric constants, $k_{s}$ and surface energy constants, $\gamma_{0}$. In these calculations, the di-nuclear system model and proximity potential, along with considering deformed nuclei, were employed. Comparing the experimental data and this theoretical approach, the best values of $k_{s}$ and $\gamma_{0}$ are $0.7546$ and $0.9180~\mathrm{MeV~fm^{-2}}$, respectively. Furthermore, this new model was used to investigate the probability of synthesis of experimentally unknown heavier systems with $Z=119$ and 120. There exist five promising combinations to synthesize SHN with $Z=119$: a) ${^{249}}\mathrm{Cf}({^{45}}\mathrm{Sc},3n){^{291}}119$ with the ER cross section, $\sigma_{3n}=417.1~\mathrm{fb}$ at the incident energy $E_{\mathrm{c.m.}}=219~\mathrm{MeV} (E^*=39.84~\mathrm{MeV})$, b) ${^{249}}\mathrm{Cf}({^{45}}\mathrm{Sc},4n){^{290}}119$ with the ER cross section, $\sigma_{4n}=138.5~\mathrm{fb}$ at the incident energy $E_{\mathrm{c.m.}}=221~\mathrm{MeV} (E^*=41.84~\mathrm{MeV})$, c) ${^{247}}\mathrm {Bk}({^{50}}\mathrm{Ti},3n){^{294}}119$ with the ER cross section, $\sigma_{3n}=11.2~\mathrm{fb}$ at the incident energy $E_{\mathrm{c.m.}}=226~\mathrm{MeV} (E^*=35.15~\mathrm{MeV})$, d) ${^{254}}\mathrm{Es}({^{48}}\mathrm{Ca},3n){^{299}}119$ with the ER cross section, $\sigma_{3n}=9115.15~\mathrm{fb}$ at the incident energies $E_{\mathrm{c.m.}}=208~\mathrm{MeV} (E^*=32.14~\mathrm{MeV})$, and e) ${^{254}}\mathrm{Es}({^{48}}\mathrm{Ca},4n){^{298}}119$ with the ER cross section, $\sigma_{4n}=735.46~\mathrm{fb}$ at the incident energies $E_{\mathrm{c.m.}}=210~\mathrm{MeV} (E^*=34.14~\mathrm{MeV})$. In addition, it is found that the best combinations to synthesize SHN with $Z=120$ are ${^{249}}\mathrm{Cf}({^{50}}\mathrm{Ti},3n){^{296}}120$ with the ER cross section, $\sigma_{3n}=51.19~\mathrm{fb}$ at the incident energy $E_{\mathrm{c.m.}}=228~\mathrm{MeV} (E^*=33.19~\mathrm{MeV})$, and ${^{251}}\mathrm{Cf}({^{50}}\mathrm{Ti},3n){^{298}}120$ with the ER cross section,  $\sigma_{3n}=43.17~\mathrm{fb}$ at the incident energy $E_{\mathrm{c.m.}}=227~\mathrm{MeV} (E^*=33.82~\mathrm{MeV})$.
\end{abstract}
\maketitle
\section{Introduction}
Superheavy nuclei (SHN) investigations, have attracted a great deal of attention among researchers in the field of nuclear physics. Fusion reactions are among the important tools in the synthesis of SHN. Nuclei with $Z<112$ are produced via cold fusion reactions using double magic target ${^{208}}\mathrm{Pb}$ whereas nuclei with $Z=112-118$ are generated via hot fusion reactions and employing double magic projectile ${^{48}}\mathrm{Ca}$ and actinide targets \cite{RN309,RN311,RN312,RN313,RN314,RN315,RN316,RN317,RN318,RN319,RN320,RN323,RN324,RN325,RN326,RN327}. Theoretical studies could provide extremely valuable insights into the optimization of expensive experimental efforts towards the synthesis of SHN \cite{RN328,RN329,RN330,RN331,RN332,RN333,RN334,RN335,RN336,RN338,RN339,RN341,RN342,RN343,RN344,RN345,RN347,RN348,RN349,RN350,RN352,RN354,RN355,RN357,RN359,RN361,RN362}. 
Different models, such as the two step model \cite{RN363}, the di-nuclear system (DNS) model \cite{RN328,RN329,RN330,RN331}, the fusion by diffusion model \cite{RN341,RN364,RN365}, and the nuclear collectivization model \cite{RN352} have been previously employed to calculate the evaporation-residue (ER) cross sections. In the DNS model, the calculation of total potential is of great importance for which different models such as double folding, Woods-Saxon, Skyrme energy density formalism, and proximity have been developed \cite{RN362,RN366}. Proximity potential is employed as an analytical description for colliding nuclei (an approach based on the phenomenology model). All of these potentials' foundations are based on the proximity theorem which the nuclear part of the interaction potential will produce various factors. These factors are independent of the colliding nuclei masses and are dependent on the inverted harmonic oscillator and universal function. This concept provides an important simplification in the study of nuclear potential \cite{RN362}.
One of the important parameters in the potential proximity is the surface energy coefficient, $\gamma$. 13 coefficients for surface asymmetric constant, $k_{s}$ and surface energy constant, $\gamma_{0}$ have been introduced \cite{RN368,RN369,RN370,RN371,RN372,RN373,RN374,RN375}, which contribute to factors such as the height of the potential barrier, the position of the potential barrier, and inverted harmonic oscillator potential, which consequently impact the calculated ER cross sections.\\
To study SHN synthesis with the DNS model, one needs to know the total potential of colliding nuclei. For the calculation of the ER cross section, there are three fundamental stages: a) capturing the projectile by the target and overcoming potential barriers, b) combination of the projectile and the target nuclei and formation of excited compound nucleus, and c) evaporation of one or more neutrons from the excited compound nucleus reaching a state of equilibrium and survive against fission \cite{RN328}. In the past four-decades, different theoretical models predicted that there is island of stability on the top of the nuclear chart with a relatively high half-life \cite{RN376,RN377}. These models have predicted that beyond closed proton shell number with $Z=82$, proton numbers with $Z=114$, 120, and 124 are also closed-proton shells. Similarly, beyond closed neutron shell number with $N=126$, neutron number with $N=184$ can be introduced as a new closed neutron shell in the nuclear chart \cite{RN378}. At the moment, the californium (Cf) nucleus is the heaviest available target that can be used for SHN synthesis, and the production of targets heavier than Cf is very difficult. For example, einsteinium(${^{254}}\mathrm{Es}$) can be produced in microgram quantities, approximately three orders of magnitude less than typically required for SHN synthesis \cite{RN379}. Therefore, to produce nuclei with $Z=119$ and 120, the projectile heavier than ${^{48}}\mathrm{Ca}$ such as ${^{50}}\mathrm{Ti}$, ${^{54}}\mathrm{Cr}$, ${^{58}}\mathrm{Fe}$ and ${^{64}}\mathrm{Ni}$ should be used \cite{RN342}. In 2008, one experiment was set up to synthesize nuclei with $Z=120$, by the combination of  ${^{238}}\mathrm{U}({^{64}}\mathrm{Ni},xn){^{302-x}}120$, where the ER cross section, $\sigma=90~\mathrm{fb}$ at the excited compound nucleus (CN) energy $E^\ast=36.4~\mathrm{MeV}$, was reported \cite{RN345}. In 2009, another attempt was performed in order to synthesize nuclei with $Z=119$, by the combination of ${^{249}}\mathrm{Bk}({^{50}}\mathrm{Ti},xn){^{299-x}}119$, where the maximal measured ER cross section, 50 fb was reported \cite{RN381}. In the same year, for the synthesis of $Z=120$ SHN, the combination of ${^{244}}\mathrm{Pu}({^{58}}\mathrm{Fe},xn){^{302-x}}120$, the ER cross section was measured, $\sigma=400~\mathrm{fb}$ \cite{RN380}. In 2015 Hofmann \textit{et al}, tried to use the combination of ${^{248}}\mathrm{Cm}({^{54}}\mathrm{Cr},xn){^{302-x}}120$, to synthesize $Z=120$, for which they reported the maximum ER cross section equal to ${5.3}_{-2.1}^{+3.6}~\mathrm{pb}$ \cite{RN395}. In 2020, another group tried to produce nuclei with $Z=119$, 120 by combinations of ${^{249}}\mathrm{Bk}({^{50}}\mathrm{Ti},xn){^{299-x}}119$ and ${^{249}}\mathrm{Cf}({^{50}}\mathrm{Ti},xn){^{299-x}}120$. In those attempts, the ER cross sections were reported equal to $65~\mathrm{fb}$ and $200~\mathrm{fb}$ at the excited CN energies $E^\ast=43.2~\mathrm{MeV}$ and $E^\ast=37.6~\mathrm{MeV}$, respectively \cite{RN396}. No evidence confirming the formation of SHN with $Z=119$ and 120 was observed. Some more theoretical studies have been performed on the synthesis of SHN with $Z=119$ and 120 \cite{RN330,RN333,RN338,RN339,RN342,RN344,RN345,RN348,RN354,RN355,RN357,RN359,RN364,RN382,RN383,RN384,RN385,RN389}. In these researches, different combinations have been suggested to synthesis $Z=119$, 120; the most suggested combination for $Z=119$ is ${^{249}}\mathrm{Bk}({^{50}}\mathrm{Ti},xn){^{299-x}}119$ with ER cross section in $x=3$, 4 neutron channels and ranges from $12~\mathrm{fb}$ to $480~\mathrm{fb}$ and the most suggested combination for $Z=120$ is ${^{249}}\mathrm{Cf}({^{50}}\mathrm{Ti},xn){^{299-x}}120$ with ER cross section in $x=3$, 4 neutron channels and ranges from 1.5 fb  to 760 fb.\\
This current paper presents the model and employed theory in Sec.~\ref{Section2}. The results of our study on the probability of the synthesis of SHN in the range of $112\le Z\le120$ are described in Sec.~\ref{Section3-1}. Moreover, using the obtained model, our results of the probability of the synthesis of nuclei with $Z=119$, 120 will be discussed in Sec.~\ref{Section3-2}. 
\section{\label{Section2}Model and theory}
\subsection{Total interaction potential}
\begin{figure}[htbp]
\includegraphics[width=70mm]{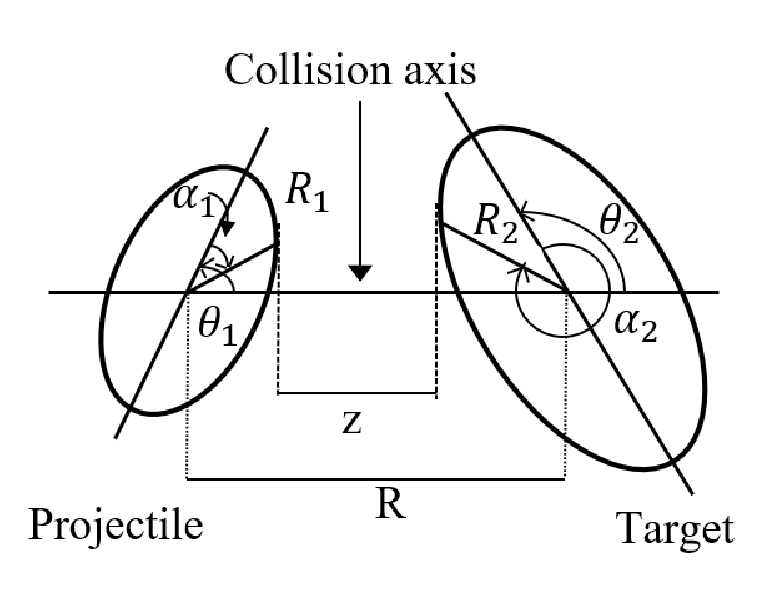}
\caption{\label{fig:fig1} Interaction angles of two nuclei during fusion.}
\end{figure}

The potential barrier characteristics, such as barrier height, position, curvature, and shape are important parameters that should be calculated. Total potential is the function of two nuclei radius distance, colliding angle, and deformation parameters of two nuclei. Total potential is equal to the summation of the coulomb long range repulsive potential, nuclear short range attractive potential, and centrifugal potential \cite{RN362}:
\begin{eqnarray}
V_{Total}=&&{V_C\left(R.Z_i,\beta_{\lambda i},\theta_i\right)+V}_N\left(R,A_i,\beta_{\lambda i},\theta_i\right)\nonumber\\+&&V_l(R,A_i,\beta_{\lambda i},\theta_i),
\end{eqnarray}
where $ V_C\left(R.Z_i,\beta_{\lambda i},\theta_i\right)$ is the Coulomb potential, $V_N\left(R,A_i,\beta_{\lambda i},\theta_i\right)$ shows the nuclear potential and $V_l(R,A_i,\beta_{\lambda i},\theta_i)$ denotes the centrifugal potential. Considering the multipole deformations, the Coulomb potential is \cite{RN390}
\begin{eqnarray}
&&V_C\left(R.Z_i,\beta_{\lambda i},\theta_i\right)=\frac{Z_1Z_2e^2}{r}+{3Z}_1Z_2e^2\nonumber\\&&\times\sum_{\lambda,i=1,2}{\frac{R_i^\lambda\left(\alpha_i,T\right)}{\left(2\lambda+1\right)R^{\lambda+1}}Y_\lambda^{(0)}(\theta_i)\left[\beta_{\lambda i}+\frac{4}{7}\beta_{\lambda i}^2Y_\lambda^{(0)}(\theta_i)\right]}.\nonumber\\\label{eq:2}
\end{eqnarray} 
In Eq. ~(\ref{eq:2}), $Z_1Z_2$ is the Coulomb factor and is related to projectile and target, $r$ is the center of two nuclei distance, $R_i^\lambda\left(\alpha_i,T\right)$ is the deformed radius of the projectile and target, $\beta_{\lambda i}$ represents multipole deformations of the projectile and target, $\theta_i$ is the angle between the nuclear symmetry axis and the collision $Z$-axis, and  $\alpha_i$ is the angle between the symmetry axis and the radius vector of the colliding nucleus \cite{RN332,RN336}. It should be noted that $\alpha_i$ is measured in the clockwise direction from the symmetry axis while $\theta_i$ is measured in the counterclockwise direction, as shown in Fig.~\ref{fig:fig1}.

For the calculation of the nuclear potential, the proximity potential is employed as follows \cite{RN366}:
\begin{eqnarray}
V_N\left(R,A_i,\beta_{\lambda i},\theta_i\right)=4\pi\gamma\ bR\phi(\frac{z}{b}),\label{eq:3}
\end{eqnarray}
where $b=\left(\frac{\pi}{\sqrt3}\right)0.55\approx0.99~\mathrm{fm}$, $R=\frac{C_1{.C}_2}{C_1+C_2}$ , $C_i={R_i\left(1-\left(\frac{b}{R_i}\right)^2\right)}_{i=1,2}$ are Susmann central radii for projectile and target and $\phi(\frac{z}{b})$ is the universal function where in $\xi=\frac{z}{b}=\frac{r-C_1-C_2}{b}$. $z$ is the minimum distance of projectile and target during colliding. In Eq.~(\ref{eq:3}), the CN surface energy coefficient is obtained from \cite{RN366} 
\begin{eqnarray}
\gamma=\gamma_0[1-k_s\frac{\left(N-Z\right)^2}{A^2}].\label{eq:4}
\end{eqnarray}
In Eq.~(\ref{eq:4}), $\gamma_0$ is the surface energy constant which is obtained from  $\gamma_0=\frac{a_s}{4\pi r_0^2}$ where $a_s$ is the usual liquid drop model surface energy coefficient and $r_0$ is the nuclear radius constant, and $k_s$ is the surface asymmetric constant \cite{RN367}. $N$, $Z$, and $A$ are the neutron number, proton number, and mass number of CN, respectively. The universal function is obtained as \cite{RN366}
\begin{eqnarray}
\phi\left(\xi\right)=&&-1.7817+0.9270\xi+0.143\xi^2-0.09\xi^3,\nonumber\\ &&\mathrm{for}\ \xi\le0,\nonumber
\end{eqnarray}
\begin{eqnarray}
\phi\left(\xi\right)=&&-1.7817+0.9270\xi+0.01696\xi^2-0.05148\xi^3,\\ &&\mathrm{for}\ 0\le\xi\le1.9475,\nonumber
\end{eqnarray}
\begin{eqnarray}
\phi\left(\xi\right)=-4.41\mathrm{exp}\left(\frac{-\xi}{0.7176}\right), \mathrm{for}\ \xi\geq1.9475,\nonumber
\end{eqnarray}
Considering the nuclear deformations, the projectile and target radii are given as \cite{RN390}:
\begin{eqnarray}
R_i\left(\alpha_i,T\right)=R_{0i}(T)\left[1+\sum_{\lambda}{\beta_{\lambda i}Y_\lambda^{(0)}(\alpha_i)}\right].\label{eq:6}
\end{eqnarray}
In Eq.~(\ref{eq:6}), $R_{0i}(T)\\=\left(1.28{A_i}^\frac{1}{3}-0.76+0.8{A_i}^{-\frac{1}{3}}\right)_{i=1,2}$ is the nuclear radii of the colliding participant, where the nuclear deformations are not taken into the account. Centrifugal potential is given as:
\begin{eqnarray}
V_l=\frac{\hbar^2l(l+1)}{2I_{NS}}.
\end{eqnarray}
Here, $I_{NS}=\ \mu\ r^2$ is the nonsticking moment of inertia, and $l$ is the angular momentum. As $l$ increases, the depth of the saddle point will decrease in the graph of the scattering potential versus distance. In other words, the quasifission barrier will continue to decrease, until the saddle point in the potential graph vanishes. This results in reduced values for the survival probability of CN and consequently decay of the compound nucleus to fission fragments.\\
The barrier penetration model developed by Wong has been widely used to describe the fusion reactions at the energies close to, or greater than that of, the barrier \cite{RN390}. The capture cross section is expressed as the sum of the cross sections for each partial wave, $l$ \cite{RN328},
\begin{eqnarray}
\sigma_{cap}=\frac{\pi\hbar^2}{2\mu E_{\mathrm{c.m.}}}\sum_{l=0}^{l_{max}}{(2l+1)}T_l\left(E_{\mathrm{c.m.}},l\right).\label{eq:8}
\end{eqnarray}
In Eq.~(\ref{eq:8}), $\mu$ is the reduced mass of the interacting nuclei, $E_{\mathrm{c.m.}}$ is the center of mass of the colliding systems, $l$ is the angular momentum and $T_l\left(E_{\mathrm{c.m.}},l\right)$ denotes the penetration probability for the $l^{th}$ partial wave and calculated according to the Hill-Wheeler Equation \cite{RN391}:
\begin{eqnarray}
T_l\left(E_{\mathrm{c.m.}}\right)={1+\mathrm{exp}{\left[2\pi\left(E_l-E_{\mathrm{c.m.}}\right)/\hbar\omega_l\right]}}^{-1}.
\end{eqnarray}
For calculating the capture cross section, the Wong formula is used \cite{RN335,RN390}:
\begin{eqnarray}
\sigma_{cap}(E)=\frac{10R_0^2\hbar\omega_0}{2E_{\mathrm{c.m.}}}\ln(1+\exp{\left[\frac{2\pi\left(E_{\mathrm{c.m.}}-E_0\right)}{\hbar\omega_0}\right]}),\nonumber\\
\end{eqnarray}
where $E_0$ is the height of the total potential and $\hbar\omega_0$ shows the inverted harmonic oscillator potential and is given as: 
\begin{eqnarray}
\hbar\omega_0=\frac{\hbar}{\sqrt\mu}\sqrt{\left|\frac{d^2V(r)}{dr^2}\right|_{R_l}}.
\end{eqnarray}
Fusion cross section is one of the main parameters, which is described as:
\begin{eqnarray}
\sigma_{fus}=\frac{\pi\hbar^2}{2\mu E_{\mathrm{c.m.}}}\sum_{l=0}^{l_{max}}{(2l+1)}T_l\left(E_{\mathrm{c.m.}},l\right)P_{CN}\left(E_{\mathrm{c.m.}},l\right).\nonumber\\\label{eq:12}
\end{eqnarray}
In Eq.~(\ref{eq:12}), $P_{CN}\left(E_{\mathrm{c.m.}},l\right)$ is the fusion probability for which various models have been used \cite{RN329,RN333,RN357}. In this work to calculate the fusion probability, the developed model by Zagrebaev and Greiner \cite{RN357} was employed:
\begin{eqnarray}
P_{CN}\left(E^*,l\right)=\frac{P_{CN}^0}{1+\mathrm{exp}\left[\frac{E_B^\ast-E^\ast}{\Delta}\right]},\label{eq:13}
\end{eqnarray}
\begin{eqnarray}
P_{CN}^0=\frac{1}{1+\mathrm{exp}\left[\frac{Z_1Z_2-\zeta}{\tau}\right]},\label{eq:14}
\end{eqnarray}
In Eq.~(\ref{eq:13}) and~(\ref{eq:14}), $E^\ast$ is the CN excitation energy, $E_B^\ast$ is the energy of the CN at the center-of-mass beam energy equal to the Bass barrier, $\Delta$ is an adjustable parameter and is often set around  $4~\mathrm{MeV}$, $\zeta\approx1760$ and $\tau\approx45$ are the fitted parameters \cite{RN357}. The compound nucleus excitation energy which also defines the damping of the shell correction to the fission barrier of CN, is obtained from
\begin{eqnarray}
E_{CN}^\ast=E_{\mathrm{c.m.}}+Q_{val}-E_{rot},
\end{eqnarray}
where $E_{rot}$ is the rotational energy of CN and is obtained from $E_{rot}=\frac{\hbar^2l(l+1)}{2\ \mu r^2}$ and $Q_{val}$ is obtained from \cite{RN349}
\begin{eqnarray}
Q_{val}=\mathrm{\Delta M}\left(A,Z\right)-\mathrm{\Delta M}\left(A_1,Z_1\right)-\mathrm{\Delta M}\left(A_2,Z_2\right).\nonumber\\\label{eq:16}
\end{eqnarray}
In Eq.~(\ref{eq:16}), $\mathrm{\Delta M}\left(A,Z\right)$, $\mathrm{\Delta M}\left(A_1,Z_1\right)$, and $\mathrm{\Delta M}\left(A_2,Z_2\right)$ are mass excess values of CN, projectile, and target, respectively.
In this work, to calculate $Q_{val}$, the mass excess values, were obtained from Möller\textit{et al}\cite{RN392} tables. 
\subsection{The evaporation residue cross section}
Once the excited CN is formed the equilibrium condition is achieved via the evaporation of one or more neutrons, before decay. Therefore, one should consider the competition between neutron emission and decay to fission fragments in CN. In cold fusion reactions, the formed CN has an excitation energy of $E_B^\ast<15~\mathrm{MeV}$ where the equilibrium condition is reached by the evaporation of one or two neutrons, where as in hot fusion reactions with the excitation energy of $E_B^\ast>15~\mathrm{MeV}$, the evaporation of three or four neutrons is involved \cite{RN350}. The evaporation residue cross section is given by
\begin{eqnarray}
\sigma_{ER}^{xn}=&&\frac{\pi\hbar^2}{2\mu E_{\mathrm{c.m.}}}\sum_{l=0}^{l_{max}}{\left(2l+1\right)T_l\left(E_{\mathrm{c.m.}},l\right)}\nonumber\\&&\times P_{CN}(E_{\mathrm{c.m.}},l)W_{sur}^{xn}(E^\ast,l),
\end{eqnarray}
where $W_{sur}^{xn}(E^\ast,l)$ denotes the survival probability of CN. Because of the high Coulomb barrier for the emission of charged particles from the excited heavy nucleus, the widths for the emission of a proton or an $\alpha$ particle are much smaller than the neutron emission width $\mathrm{\Gamma}_n$. Under these circumstances, ${\mathrm{\Gamma}_t\approx\mathrm{\Gamma}}_n+\mathrm{\Gamma}_f$ was set and the survival probability $W_{sur}^{xn}(E^\ast,l)$ reflects the competition between neutron evaporation and fission of the excited CN. The survival probability is given by \cite{RN361}  
\begin{eqnarray}
W_{sur}^{xn}(E^\ast,l)=P_{xn}(E_{CN}^\ast)\prod_{i=1}^{i_{max}=x}{(\frac{\mathrm{\Gamma}_n}{\mathrm{\Gamma}_n+\mathrm{\Gamma}_f})}_{i,E^\ast},
\end{eqnarray}
where $\mathrm{\Gamma}_n$ and $\mathrm{\Gamma}_f$ are partial neutron emission width and partial fission width, respectively. At $E_{CN}^\ast>10~\mathrm{MeV}$, the $\mathrm{\Gamma}_n$ is much smaller than the $\mathrm{\Gamma}_f$. Therefore, the survival probability $W_{sur}^{xn}(E^\ast,l)$ is mainly determined by the realization probability $P_{xn}(E_{CN}^\ast)$ and the ratio $\frac{\mathrm{\Gamma}_n}{\mathrm{\Gamma}_f}$ \cite{RN331,RN393}:
\begin{eqnarray}
&&\frac{\mathrm{\Gamma}_n}{\mathrm{\Gamma}_f}=\frac{4A^\frac{2}{3}a_f\left(E_{CN}^*-B_n\right)}{K_0a_n[2a_f^{\frac{1}{2}}\left[\left(E_{CN}^\ast-B_f\right)\right]^\frac{1}{2}-1)]}\nonumber\\&&\times \mathrm{exp}\left[2a_n^{1/2}\left(E_{CN}^*-B_n\right)^\frac{1}{2}-2a_f^{1/2}\left(E_{CN}^*-B_f\right)^\frac{1}{2}\right],\nonumber\\\label{eq:19}
\end{eqnarray}
In Eq.~(\ref{eq:19}), $B_n$ denotes the neutron binding energy, $B_f$ is the fission barrier, $K_0$ is set to a constant value of $10~\mathrm{MeV}$, $a_n=A/10$ and $a_f=1.1a_n$ are the level density parameters of the fissioning nucleus at the ground state and saddle configurations, respectively \cite{RN393}. The fission barrier is given by \cite{RN329}
\begin{eqnarray}
B_f\left(E^\ast\right)=B_f^{LD}+S\exp{\left(\frac{-E^\ast}{E_D}\right)}.\label{eq:20}
\end{eqnarray}
In Eq.~(\ref{eq:20}), $B_f^{LD}$ is the liquid drop (LD) fission barrier (macroscopic) and $S$ is the shell correction term \cite{RN348}. The liquid drop fission barrier is very low or equal to zero for heavy elements with $Z\geq109$ \cite{RN330,RN361}. The shell damping energy is given by: 
\begin{eqnarray}
E_D=\frac{5.48A^\frac{1}{3}}{1+1.3A^{-\frac{1}{3}}}.
\end{eqnarray}
To calculate $P_{xn}(E_{CN}^\ast)$, an equation, which was developed by Jackson \cite{RN394}, was employed. $P_{xn}(E_{CN}^\ast)$ represent the probability of emitting an exact number of neutrons, $x$, from the CN and is given by
\begin{eqnarray}
P_{xn}\left(E_{CN}^\ast\right)=I\left(\mathrm{\Delta}_x,2x-3\right)-\ I(\mathrm{\Delta}_{x+1},2x-1),\label{eq:22}
\end{eqnarray} 
where $I\left(z,n\right)$ is the Pearson’s incomplete $\gamma$ function, and is obtained by $I\left(z,n\right)=\left(\frac{1}{n!}\right)\int_{0}^{z}m^ne^{-m}dm$, in Eq.~ (\ref{eq:22}), $\mathrm{\Delta}_x=\frac{\left(E_{CN}^\ast-\sum_{1}^{x}{B_n(i)}\right)}{T}$, $B_n(i)$ is the binding energy of the $i{th}$ evaporated neutron, and $T$ is the compound nucleus temperature that is given by $E_{CN}^\ast=E_{\mathrm{c.m.}}+Q_{val}=\frac{1}{a}{AT}^2-T$. It should be noted that Eq.~(\ref{eq:22}) is valid for calculation of the evaporation probability of two neutrons and more.
 
\begin{figure}[b]
\includegraphics[width=85mm]{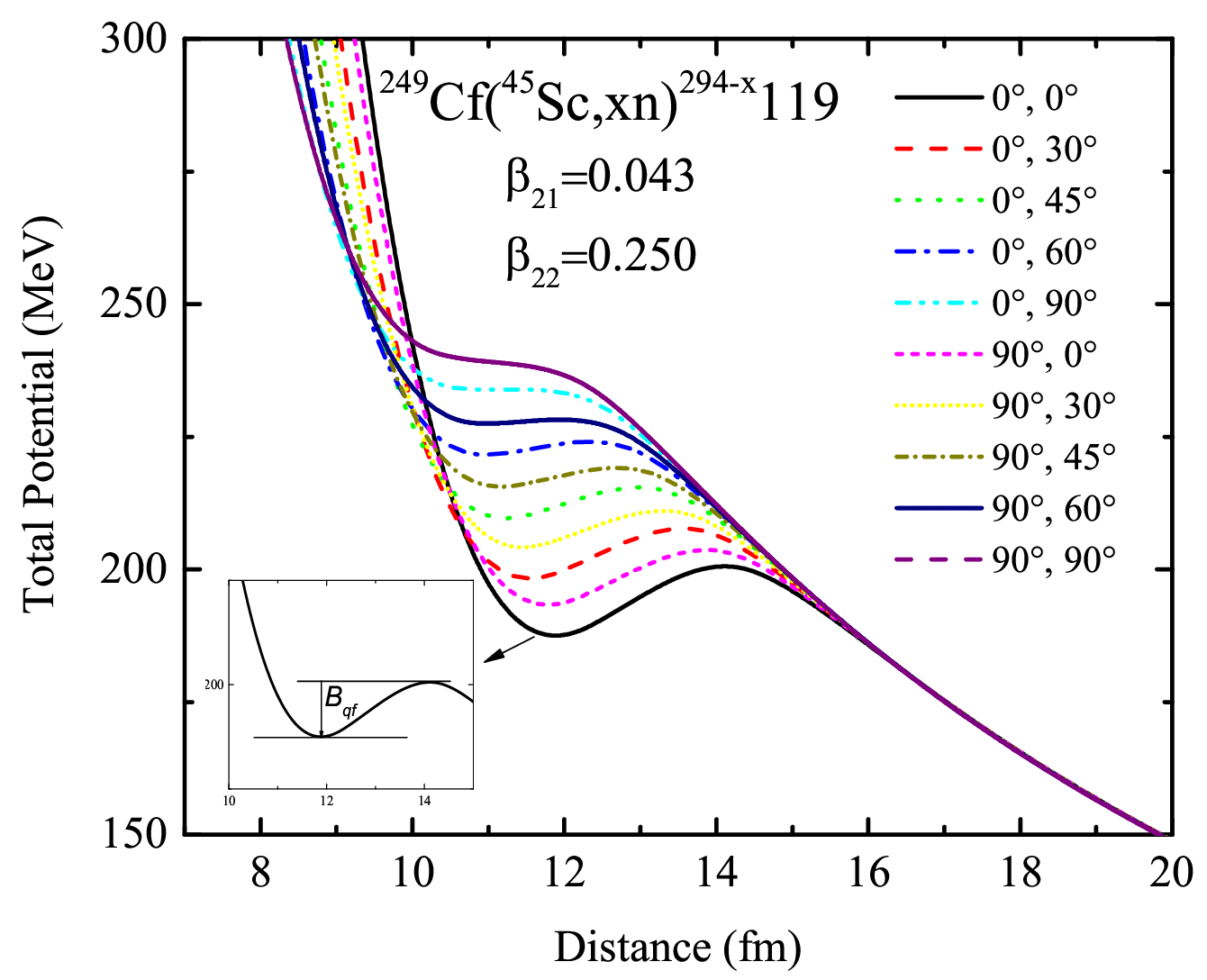}
\caption{\label{fig:fig2}(Color online) the distribution of total potential versus distance for ${^{249}}\mathrm{Cf}({^{45}}\mathrm{Sc},xn){^{294-x}}119$ combination, along with deformations, $\beta_{2i}$ and rotations, $\alpha_i$, of projectile and target.}
\end{figure}
\begin{figure}[htbp]
\includegraphics[width=85mm]{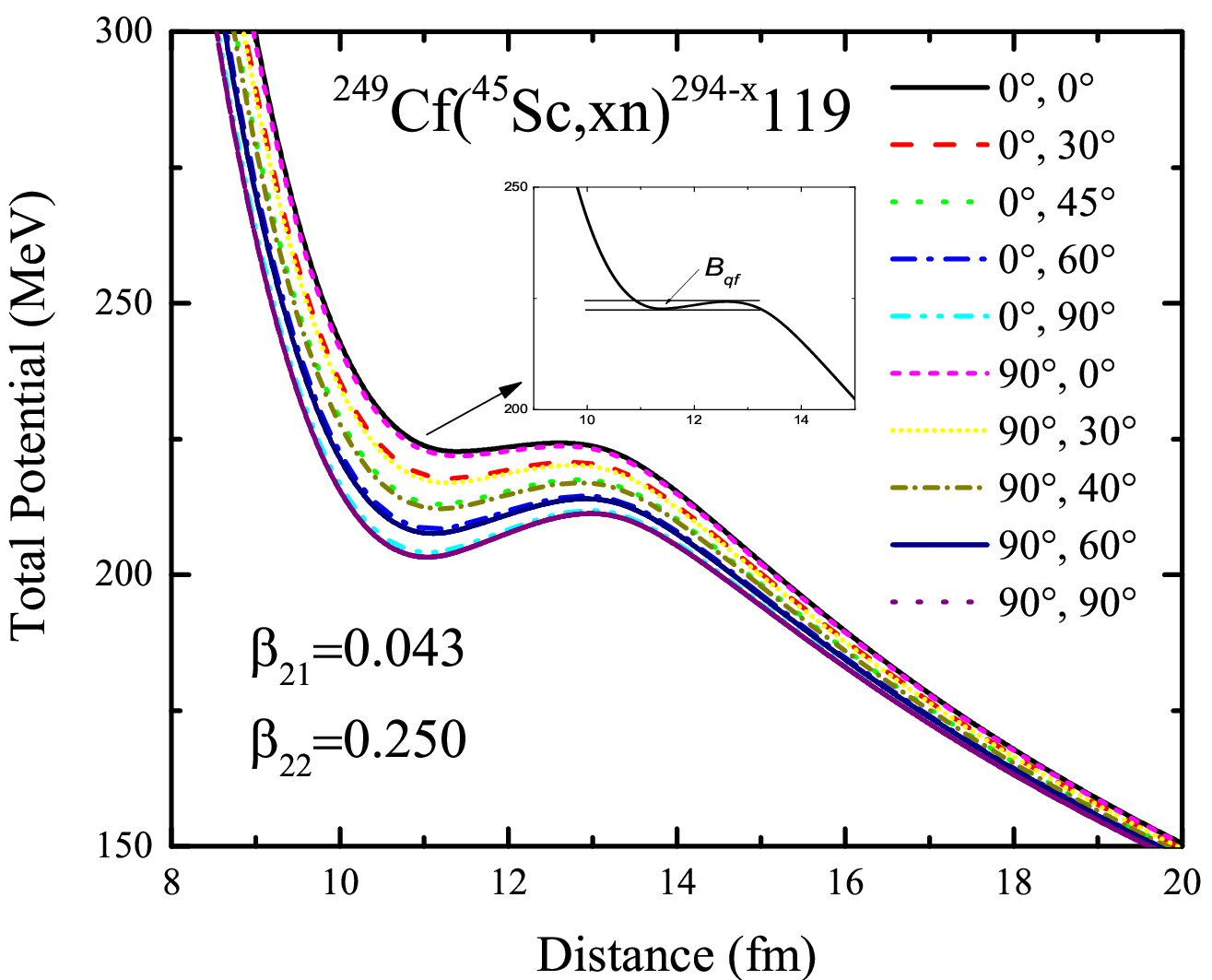}
\caption{\label{fig:fig3}(Color online) the distribution of total potential versus distance for ${^{249}}\mathrm{Cf}({^{45}}\mathrm{Sc},xn){^{294-x}}119$ combination, along with deformations, $\beta_{2i}$ and directions, $\theta_i$, of projectile and target.}
\end{figure}
\begin{table}[b]%The best place to locate the table environment is directly after its first reference in text
\caption{\label{tab:table1}
The different surface energy coefficients available from the literature.
}
\renewcommand{\arraystretch}{1.2}
\begin{ruledtabular}
\begin{tabular}{lcccc}
\textrm{$r_0\left(\mathrm{fm}\right)$}&
\textrm{$a_s\left(\mathrm{MeV}\right)$}&
\multicolumn{1}{c}{\textrm{$k_s$}}&
\textrm{$\gamma_0\left(\mathrm{MeVfm^{-2}}\right)$}&
\textrm{Ref.}\\
\colrule
1.2049 & 18.56 & 1.79 & 1.01734 & \cite{RN368}\\
1.2249 & 17.9439 & 1.7826 & 0.9517 & \cite{RN369}\\
1.16 & 24.7 & 4.0 & 1.460734 & \cite{RN370}\\
1.18 & 21.7 & 3.0 & 1.2402 & \cite{RN371}\\
1.18 & 20.57 & 2.2 & 1.1756 & \cite{RN372}\\
1.16 & 21.53 & 2.5 & 1.27326 & \cite{RN372}\\
1.16 & 21.14 & 2.4 & 1.2502 & \cite{RN372}\\
1.16 & 21.13 & 2.3 & 1.2496 & \cite{RN372}\\
1.2249 & 17.9439 & 2.6 & 0.9517 & \cite{RN373}\\
1.16 & 21.18466 & 2.345 & 1.25284 & \cite{RN374}\\
1.18995 & 19.3859 & 1.9830 & 1.08948 & \cite{RN375}\\
1.21610 & 17.0603 & 0.7546 & 0.9180 & \cite{RN375}\\
1.21725 & 16.9707 & 2.2938 & 0.911445 & \cite{RN375}\\
\end{tabular}
\end{ruledtabular}
\end{table}

\begin{figure*}[htbp]
\includegraphics[width=59mm]{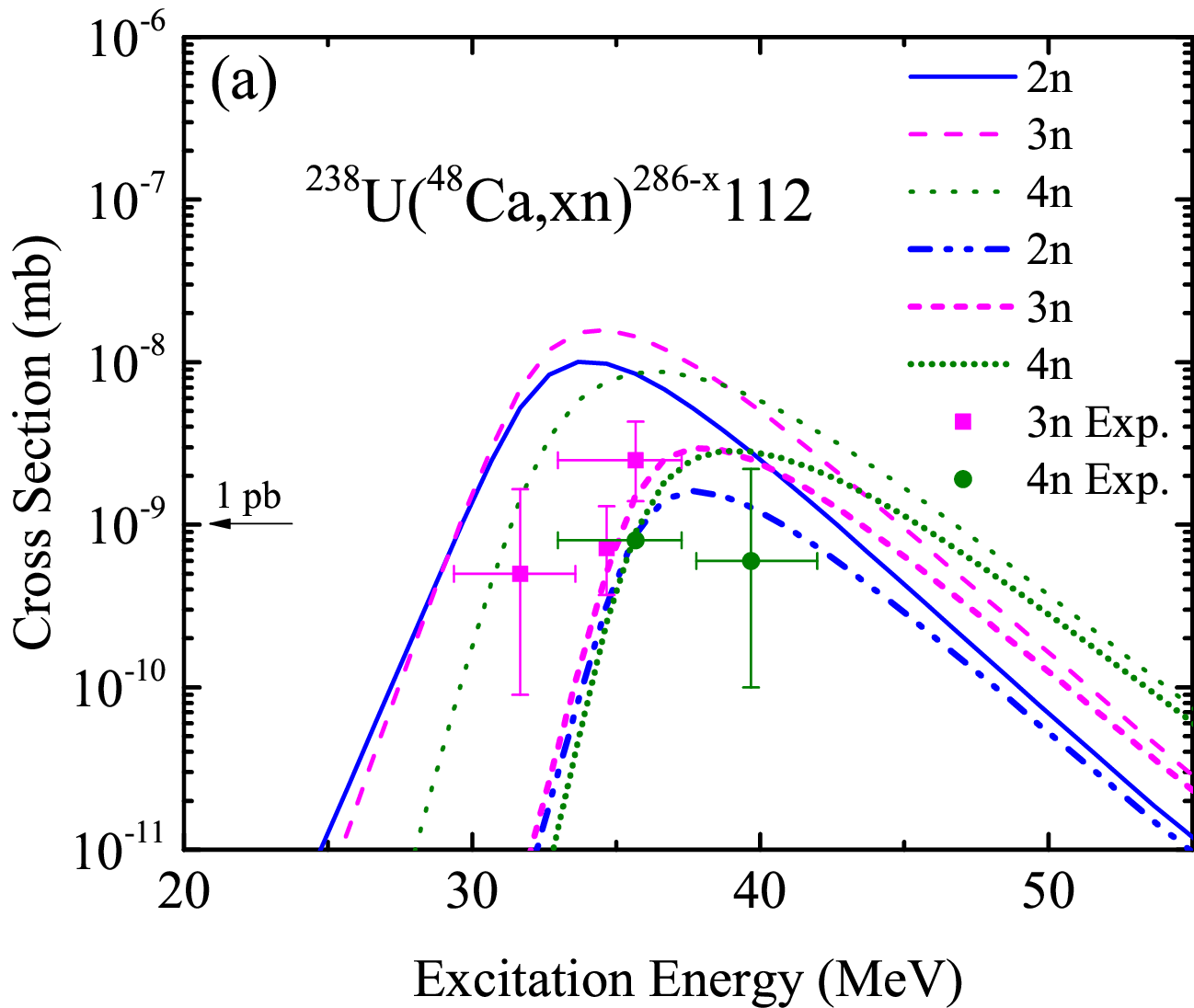}
\includegraphics[width=59mm]{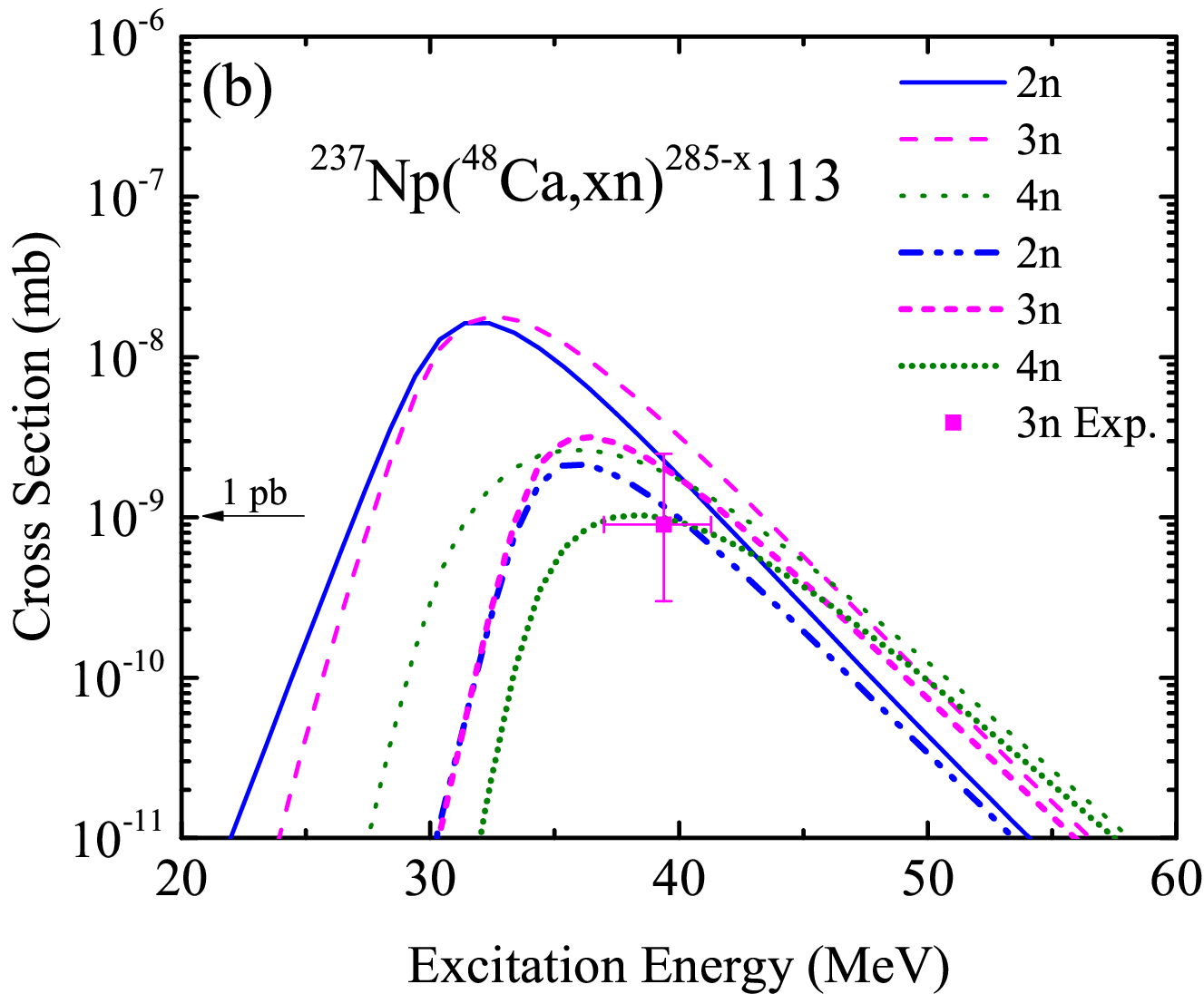}
\includegraphics[width=59mm]{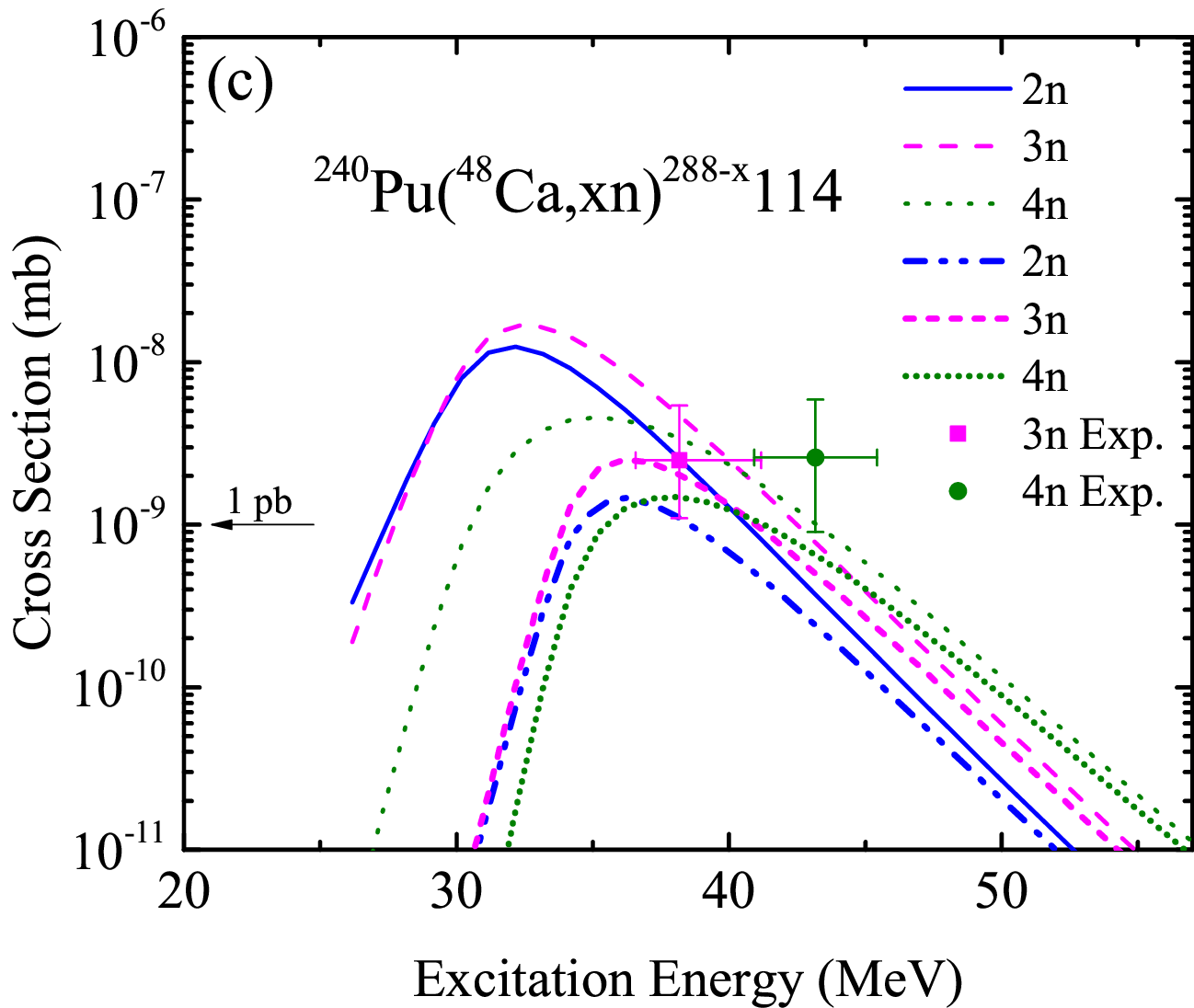}
\includegraphics[width=59mm]{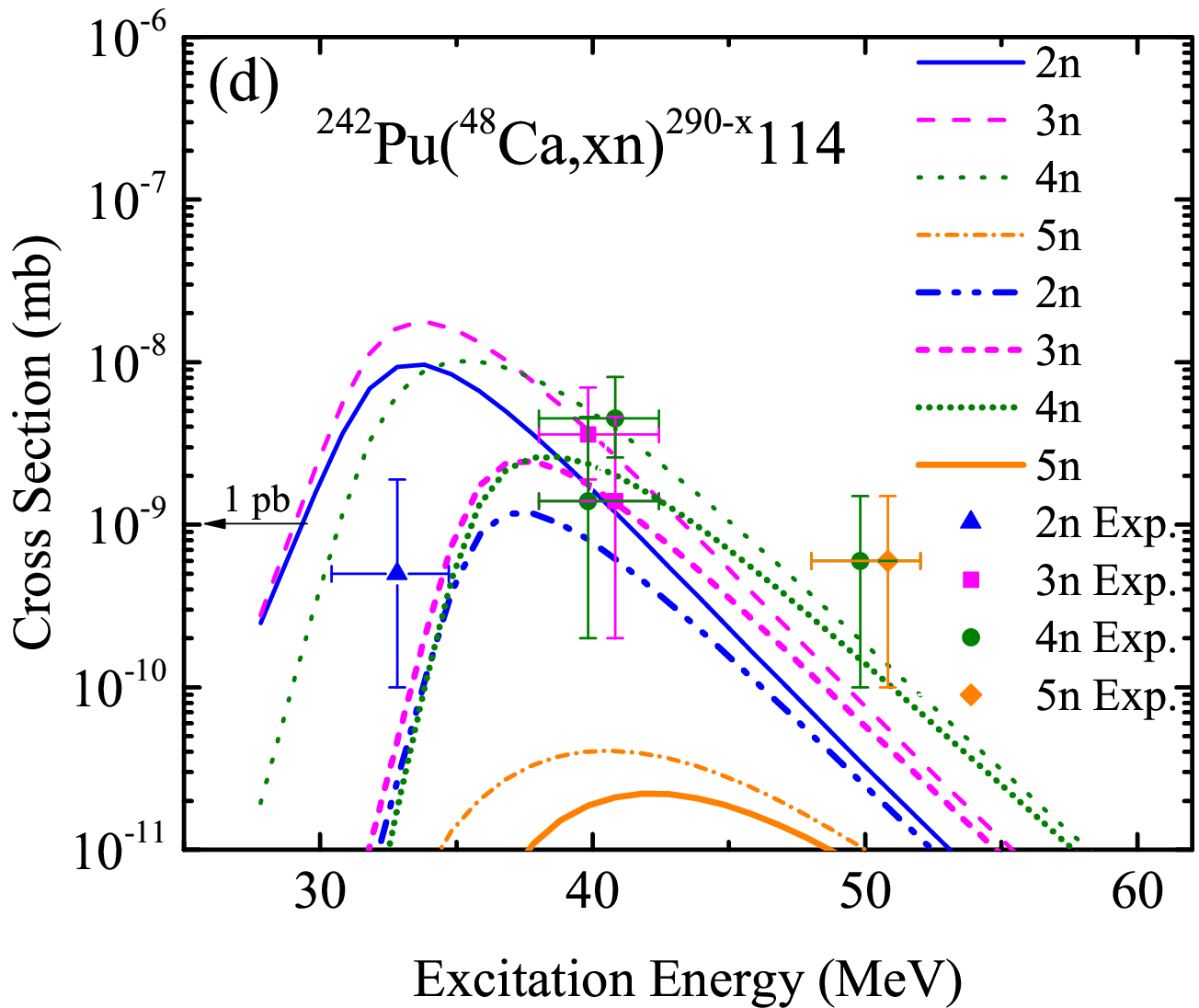}
\includegraphics[width=59mm]{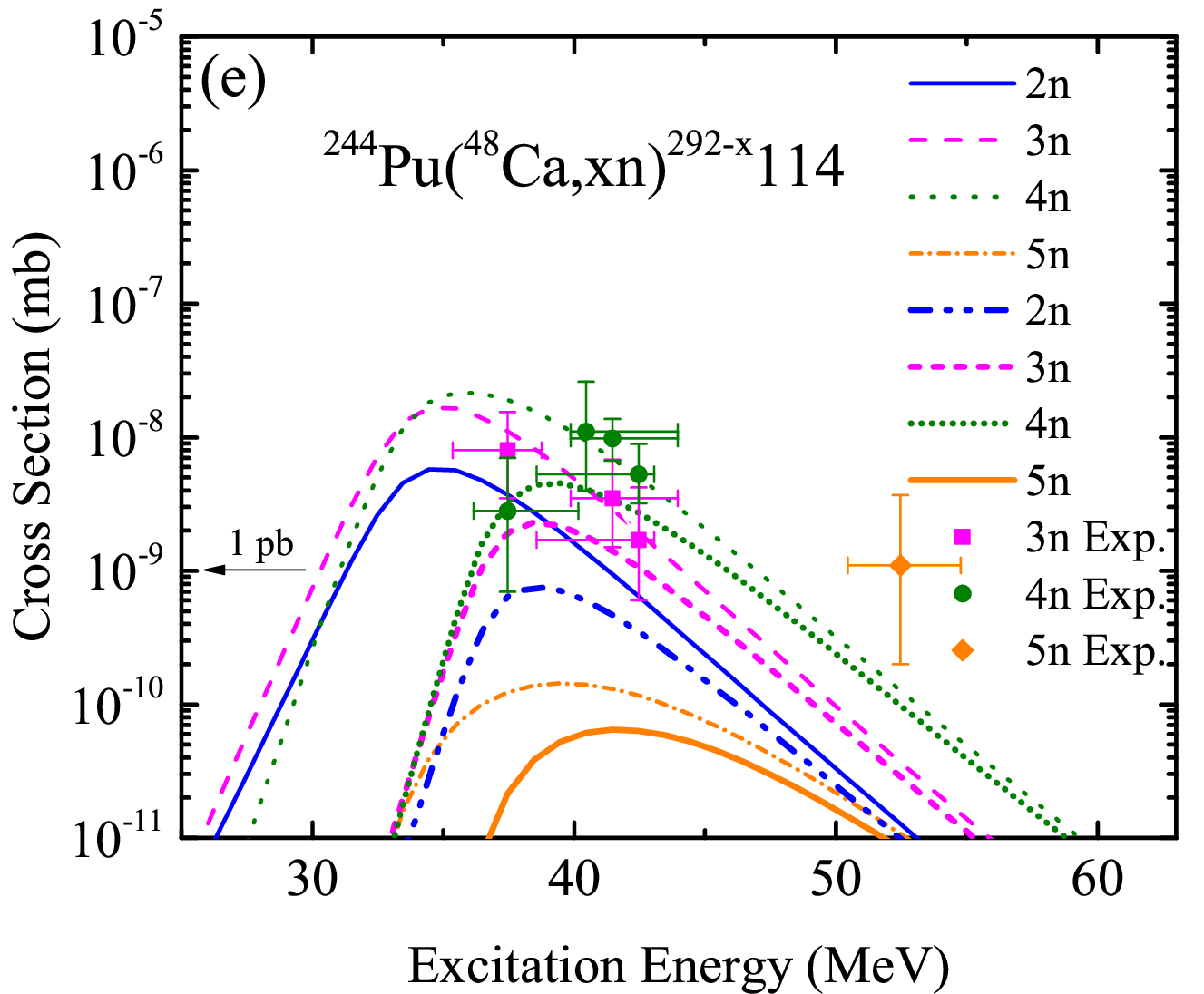}
\includegraphics[width=59mm]{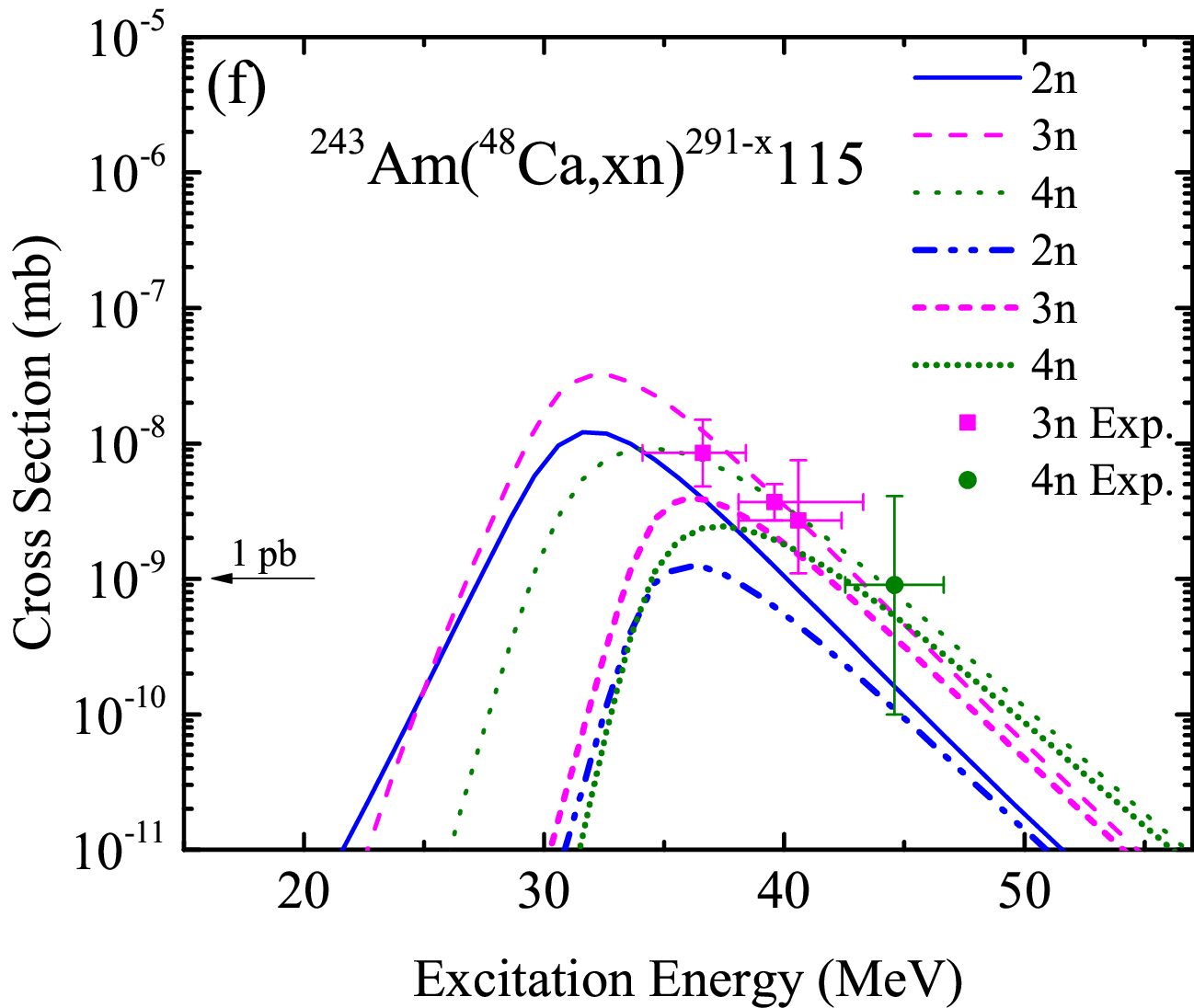}
\includegraphics[width=59mm]{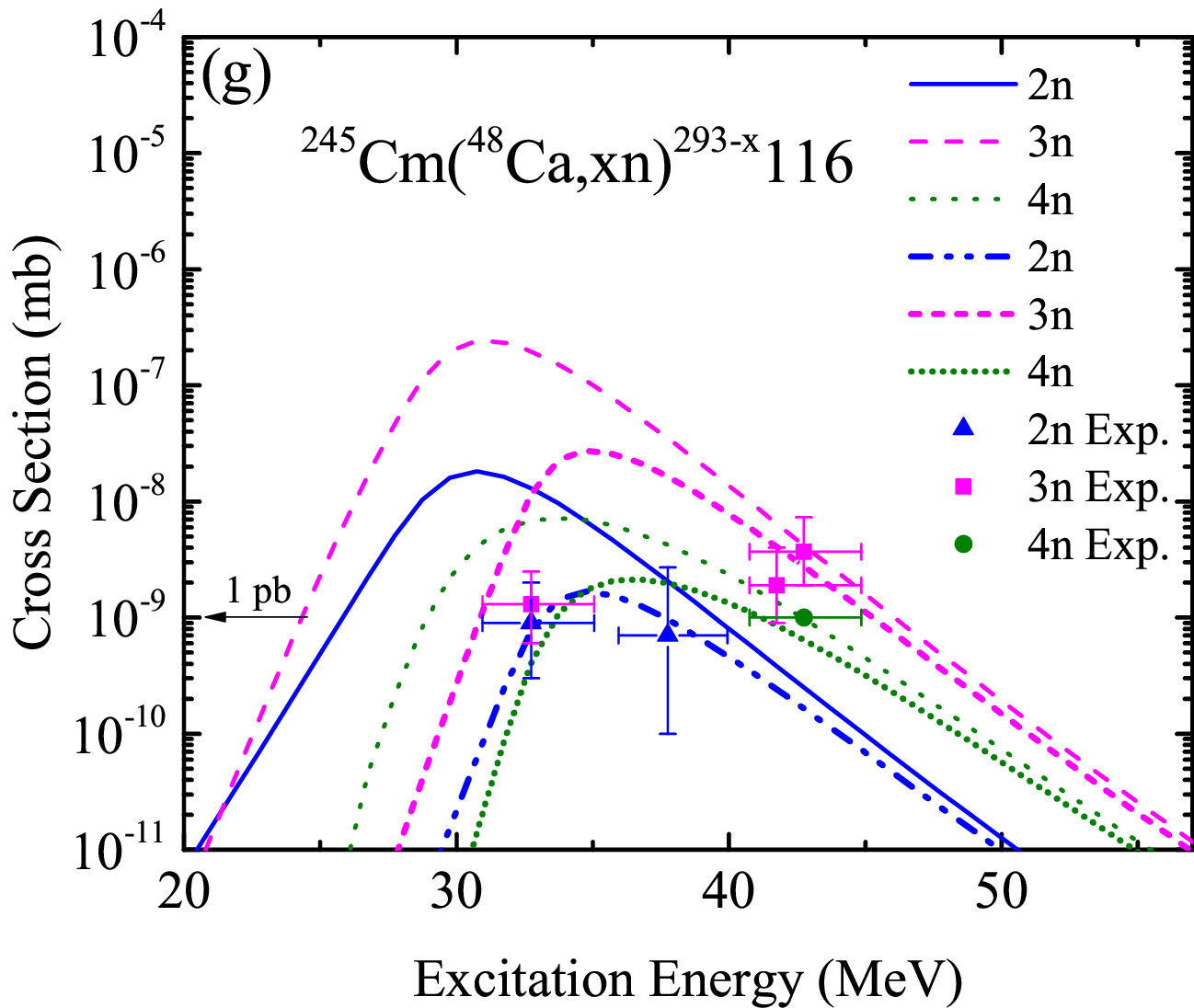}
\includegraphics[width=59mm]{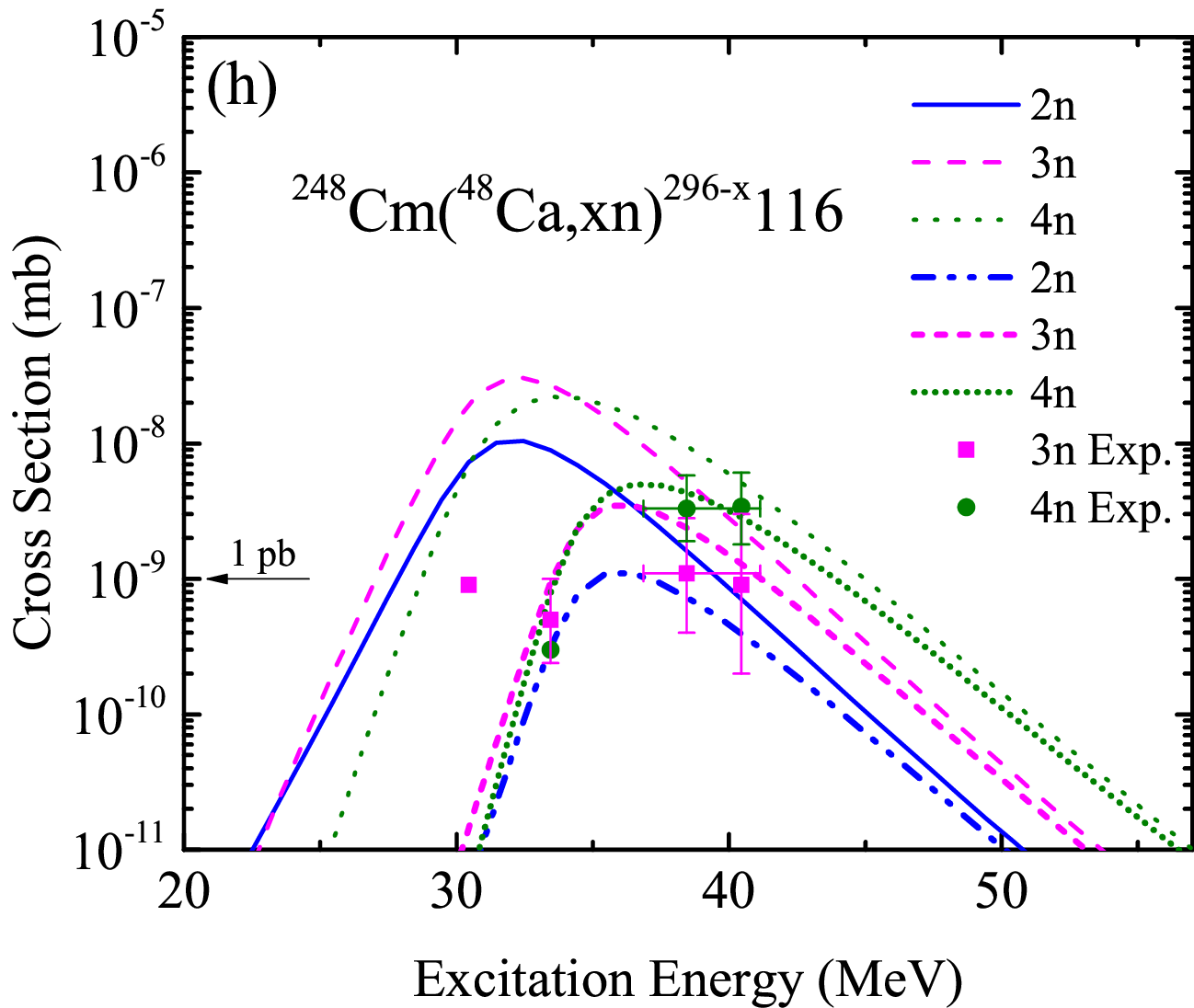}
\includegraphics[width=59mm]{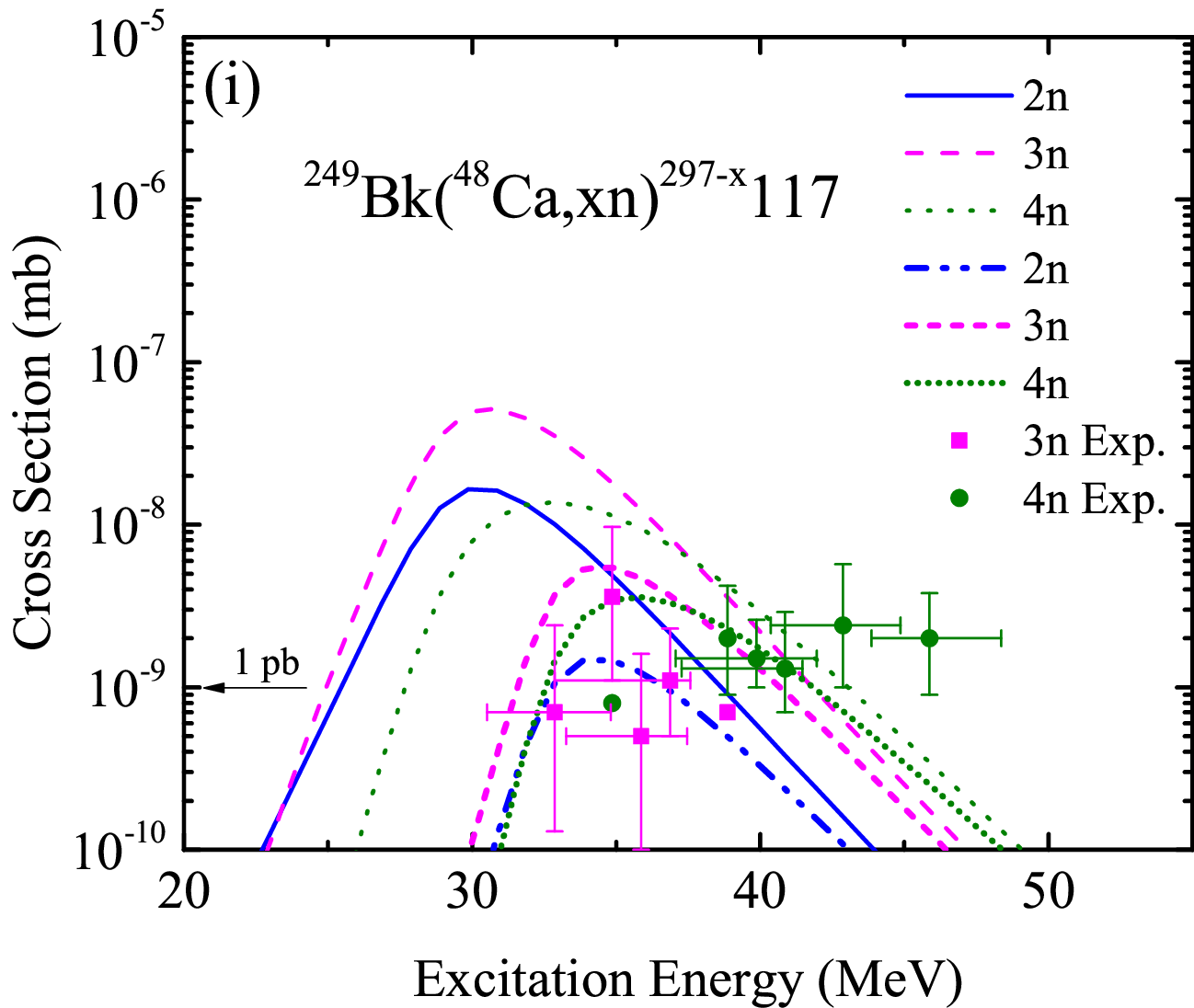}
\includegraphics[width=59mm]{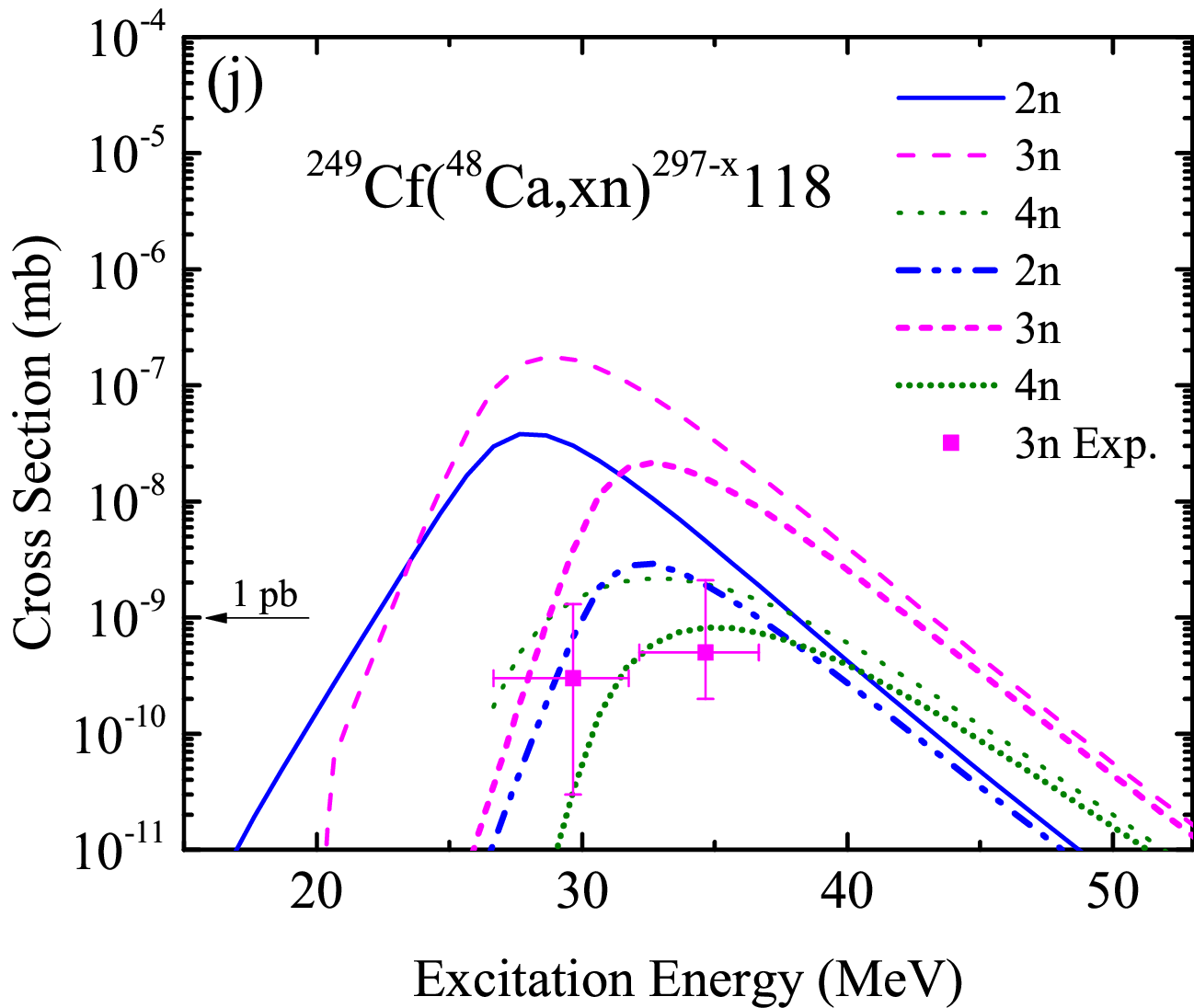}
\caption{\label{fig:fig4}(Color online) the ER cross section vs. the CN excitation energy for combinations of a) ${^{238}}\mathrm{U}({^{48}}\mathrm{Ca},xn){^{286-x}}112$, b) ${^{237}}\mathrm{Np}({^{48}}\mathrm{Ca},xn){^{285-x}}113$, c) ${^{240}}\mathrm{Pu}({^{48}}\mathrm{Ca},xn){^{288-x}}114$, d) ${^{242}}\mathrm{Pu}({^{48}}\mathrm{Ca},xn){^{290-x}}114$, e) ${^{244}}\mathrm{Pu}({^{48}}\mathrm{Ca},xn){^{292-x}}114$, f) ${^{243}}\mathrm{Am}({^{48}}\mathrm{Ca},xn){^{291-x}}115$, g) ${^{245}}\mathrm{Cm}({^{48}}\mathrm{Ca},xn){^{293-x}}116$, h) ${^{248}}\mathrm{Cm}({^{48}}\mathrm{Ca},xn){^{296-x}}116$, i) ${^{249}}\mathrm{Bk}({^{48}}\mathrm{Ca},xn){^{297-x}}117$ and j) ${^{249}}\mathrm{Cf}({^{48}}\mathrm{Ca},xn){^{297-x}}118$. The thin lines included 2n solid(blue), 3n dash(magenta), 4n dot(olive), and 5n short dash dot(orange), denote the calculated cross sections for values of $k_s=4.0$ and $\gamma_0=1.460734~\mathrm{MeV{fm}}^{-2}$ and the thick lines included 2n dash dot dot(blue), 3n short dash(magenta), 4n short dot(olive), and 5n solid(orange) show the calculated cross sections for values of $k_s=0.7546$ and $\gamma_0=0.9180~\mathrm{MeV{fm}^{-2}}$ in different neutrons channel.The experimental data are given by Ref.~\cite{RN309,RN311,RN312,RN313,RN314,RN315,RN316,RN317,RN318,RN319,RN320,RN323,RN324,RN325,RN326,RN327}.}
\end{figure*}

\begin{table*}[htbp]
\caption{\label{tab:table2}Selected combinations along with the calculated values and experimental data for the surface energy coefficient $\gamma$, the CN excitation energy $E_{CN}^\ast$, and the ER cross section, $\sigma_{ER}$ in different neutron channels.}
\setlength{\tabcolsep}{2pt}
\renewcommand{\arraystretch}{1.5}
\begin{ruledtabular}	
\begin{tabular}{ccccccccc}
&&&\multicolumn{3}{c}{Calculated values}&\multicolumn{2}{c}{Experimental data}&\\
\cmidrule(rl){4-6} \cmidrule(rl){7-8}
Combination&$k_s$&$\gamma_0(\mathrm{MeV{fm}^{-2}})$&$\gamma(\mathrm{MeV{fm}^{-2}})$
&$E_{CN}^\ast(\mathrm{MeV})$&$\sigma_{ER}(\mathrm{pb})$&$E_{CN}^\ast(\mathrm{MeV})$&$\sigma_{ER}(\mathrm{pb})$&Ref.\\ \hline
${^{48}}\mathrm{Ca}+{^{238}}\mathrm{U}$&4.0&1.460734&1.8614&31.6&$\sigma_{3n}=6.83$&29.3-33.5&${\sigma_{3n}=0.5}_{-0.41}^{+1.15}$&\cite{RN309} \\
&&&&34.6&$\sigma_{3n}=15.77$&34.6&${\sigma_{3n}=0.7}_{-0.35}^{+0.58}$&\cite{RN323}\\
&&&&35.6&$\sigma_{3n}=14.34$&32.9-37.2&${\sigma_{3n}=2.5}_{-1.1}^{+1.8}$&\cite{RN309}\\
&&&&35.6&$\sigma_{4n}=8.61$&32.9-37.2&$\sigma_{4n}=0.8$&\cite{RN309}\\
&&&&39.6&$\sigma_{4n}=6.16$&37.7-41.9&${\sigma_{4n}=0.6}_{-0.5}^{+1.6}$&\cite{RN309}\\
&0.7546&0.9180&0.8854&31.6&$\sigma_{3n}=0.005$&29.3-33.5&	${\sigma_{3n}=0.5}_{-0.41}^{+1.15}$&\cite{RN309}\\
&&&&34.6&$\sigma_{3n}=0.51$&34.6&${\sigma_{3n}=0.7}_{-0.35}^{+0.58}$&\cite{RN323}\\
&&&&35.6&$\sigma_{3n}=1.50$&32.9-37.2&${\sigma_{3n}=2.5}_{-1.1}^{+1.8}$&\cite{RN309}\\
&&&&35.6&$\sigma_{4n}=0.90$&32.9-37.2&$\sigma_{4n}=0.8$&\cite{RN309}\\
&&&&39.6&$\sigma_{4n}=2.81$&37.7-41.9&${\sigma_{4n}=0.6}_{-0.5}^{+1.6}$&\cite{RN309}\\
${^{48}}\mathrm{Ca}+{^{237}}\mathrm{Np}$&4.0&1.460734&1.2103&39.3&$\sigma_{3n}=3.93$&36.9-41.2&${\sigma_{3n}=0.9}_{-0.6}^{+1.6}$	&\cite{RN311}\\
&0.7546&0.9180&0.8883&39.3&$\sigma_{3n}=2.03$&36.9-41.2&${\sigma_{3n}=0.9}_{-0.6}^{+1.6}$&\cite{RN311}\\
${^{48}}\mathrm{Ca}+{^{240}}\mathrm{Pu}$&4.0&1.460734&1.2071&38.1&$\sigma_{3n}=4.70$&36.5-41.1&$\sigma_{3n}{=2.5}_{-1.4}^{+2.9}$&\cite{RN324}\\
&&&&43.1&$\sigma_{4n}=1.01$&40.9-45.4&${\sigma_{4n}=2.6}_{-1.7}^{+3.3}$&\cite{RN324}\\
&0.7546&0.9180&0.8879&38.1&$\sigma_{3n}=2.04$&36.5-41.1&$\sigma_{3n}{=2.5}_{-1.4}^{+2.9}$&\cite{RN324}\\
&&&&43.1&$\sigma_{4n}=0.65$&40.9-45.4&${\sigma_{4n}=2.6}_{-1.7}^{+3.3}$&\cite{RN324}\\
${^{48}}\mathrm{Ca}+{^{242}}\mathrm{Pu}$&4.0&1.460734&1.1936&32.8&$\sigma_{2n}=9.38$&30.4-34.7&${\sigma_{2n}=0.5}_{-0.4}^{+1.4}$&\cite{RN309}\\
&&&&39.8&$\sigma_{3n}=3.81$&38-42.4&$\sigma_{3n}{=1.4}_{-1.2}^{+3.2}$&\cite{RN319}\\
&&&&40.8&$\sigma_{4n}=3.90$&38-42.4&$\sigma_{4n}{=1.4}_{-1.2}^{+3.2}$&\cite{RN319}\\
&&&&39.8&$\sigma_{3n}=3.81$&38-42.4&$\sigma_{3n}{=3.6}_{-1.7}^{+3.4}$&\cite{RN309}\\
&&&&40.8&$\sigma_{4n}=3.90$&38-42.4&${\sigma_{4n}=4.5}_{-1.9}^{+3.6}$&\cite{RN309}\\
&&&&49.8&$\sigma_{4n}=0.19$&48-52&${\sigma_{4n}=0.6}_{-0.5}^{+0.9}$&\cite{RN318}\\
&&&&50.8&$\sigma_{5n}=0.008$&48-52&${\sigma_{5n}=0.6}_{-0.5}^{+0.9}$&\cite{RN318}\\
&0.7546&0.9180&0.8863&32.8&$\sigma_{2n}=0.02$&30.4-34.7&${\sigma_{2n}=0.5}_{-0.4}^{+1.4}$&\cite{RN309}\\
&&&&39.8&$\sigma_{3n}=1.77$&38-42.4&$\sigma_{3n}{=1.4}_{-1.2}^{+3.2}$&\cite{RN319}\\
&&&&40.8&$\sigma_{4n}=2.02$&38-42.4&$\sigma_{4n}{=1.4}_{-1.2}^{+3.2}$&\cite{RN319}\\
&&&&39.8&$\sigma_{3n}=1.77$&38-42.4&$\sigma_{3n}{=3.6}_{-1.7}^{+3.4}$&\cite{RN309}\\
&&&&40.8&	$\sigma_{4n}=2.02$	&38-42.4	&${\sigma_{4n}=4.5}_{-1.9}^{+3.6}	$&\cite{RN309}\\
&&&&49.8&	$\sigma_{4n}=0.14$	&48-52&	${\sigma_{4n}=0.6}_{-0.5}^{+0.9}$	&\cite{RN318}\\
&&&&50.8&	$\sigma_{5n}=0.006$	&48-52&	${\sigma_{5n}=0.6}_{-0.5}^{+0.9}$	&\cite{RN318}\\
${^{48}}\mathrm{Ca}+{^{244}}\mathrm{Pu}$	&4.0	&1.460734	&1.1800	&37.4&	$\sigma_{3n}=11.15$&	36.1-39.5&	$\sigma_{3n}{=8.0}_{-4.5}^{+7.4}$	&\cite{RN320}\\
&&&&41.4&	$\sigma_{3n}=2.86$	&39.8-43.9&	$\sigma_{3n}{=3.5}_{-2.0}^{+3.3}$	&\cite{RN316}\\
&&&&42.4&	$\sigma_{3n}=1.95$	&39-43&	$\sigma_{3n}{=1.7}_{-1.1}^{+2.5}$	&\cite{RN326}\\
&&&&37.4&	$\sigma_{4n}=19.09$	&36.1-40.1&	${\sigma_{4n}=2.8}_{-2.1}^{+4.2}$	&\cite{RN316}\\
&&&&40.4&	$\sigma_{4n}=9.42$	&39.8-43.9&	${\sigma_{4n}=11}_{-7}^{+15}$	&\cite{RN316}\\
&&&&41.4&	$\sigma_{4n}=6.94$	&39.8-43.9&	${\sigma_{4n}=9.8}_{-3.1}^{+3.9}$	&\cite{RN320}\\
&&&&42.4&	$\sigma_{4n}=5.01$	&39-43&	${\sigma_{4n}=5.3}_{-2.1}^{+3.6}$	&\cite{RN326}\\
&&&&52.4&	$\sigma_{5n}=0.01$	&50.4-54.7&	${\sigma_{5n}=1.1}_{-0.9}^{+2.6}$	&\cite{RN326}\\
\end{tabular}
\end{ruledtabular}
\end{table*}
\begin{table*}
Continued table~\ref{tab:table2}\\\hspace{-2.3cm}
\setlength{\tabcolsep}{2pt}
\renewcommand{\arraystretch}{1.5}
\begin{ruledtabular}
\begin{tabular}{ccccccccc}
&&&\multicolumn{3}{c}{Calculated values}&\multicolumn{2}{c}{Experimental data}&\\
\cmidrule(rl){4-6} \cmidrule(rl){7-8}
Combination&$k_s$&$\gamma_0(\mathrm{MeV{fm}^{-2}})$&$\gamma(\mathrm{MeV{fm}^{-2}})$
&$E_{CN}^\ast(\mathrm{MeV})$&$\sigma_{ER}(\mathrm{pb})$&$E_{CN}^\ast(\mathrm{MeV})$&$\sigma_{ER}(\mathrm{pb})$&Ref.\\ \hline
&0.7546	&0.9180	&0.8847	&37.4&	$\sigma_{3n}=1.93$	&36.1-39.5&	$\sigma_{3n}{=8.0}_{-4.5}^{+7.4}$	&\cite{RN320}\\
&&&&41.4&	$\sigma_{3n}=1.41$	&39.8-43.9&	$\sigma_{3n}{=3.5}_{-2.0}^{+3.3}$	&\cite{RN316}\\
&&&&42.4&	$\sigma_{3n}=1.06$	&39-43&	$\sigma_{3n}{=1.7}_{-1.1}^{+2.5}$	&\cite{RN326}\\
&&&&37.4&	$\sigma_{4n}=3.32$	&36.1-40.1&	${\sigma_{4n}=2.8}_{-2.1}^{+4.2}$	&\cite{RN316}\\
&&&&40.4&	$\sigma_{4n}=4.10$	&39.8-43.9&	${\sigma_{4n}=11}_{-7}^{+15}$	&\cite{RN316}\\
&&&&41.4&	$\sigma_{4n}=3.46$	&39.8-43.9&	${\sigma_{4n}=9.8}_{-3.1}^{+3.9}$	&\cite{RN320}\\
&&&&42.4&	$\sigma_{4n}=2.73$	&39-43&	${\sigma_{4n}=5.3}_{-2.1}^{+3.6}$	&\cite{RN326}\\
&&&&52.4&	$\sigma_{5n}=0.008$	&50.4-54.7&	${\sigma_{5n}=1.1}_{-0.9}^{+2.6}$	&\cite{RN326}\\
${^{48}}\mathrm{Ca}+{^{243}}\mathrm{Am}$	&4.0	&1.460734	&1.2039	&36.5&	$\sigma_{3n}=12.35$	&34-38.3&	${\sigma_{3n}=8.5}_{-3.7}^{+6.4}$	&\cite{RN327}\\
&&&&39.5&	$\sigma_{3n}=4.07$	&38-43.2&	${\sigma_{3n}=3.7}_{-1.0}^{+1.3}$	&\cite{RN312}\\
&&&&40.5&	$\sigma_{3n}=2.74$	&38-43.2&	${\sigma_{3n}=2.7}_{-1.6}^{+4.8}$	&\cite{RN312}\\
&&&&44.5&	$\sigma_{4n}=0.78$	&42.4-46.5&	${\sigma_{4n}=0.9}_{-0.8}^{+3.2}$	&\cite{RN312}\\
&0.7546&	0.9180&	0.8875&	36.5&	$\sigma_{3n}=3.89$	&34-38.3	&${\sigma_{3n}=8.5}_{-3.7}^{+6.4}$	&\cite{RN327}\\
&&&&39.5&	$\sigma_{3n}=2.06$	&38-43.2&	${\sigma_{3n}=3.7}_{-1.0}^{+1.3}$	&\cite{RN312}\\
&&&&40.5&	$\sigma_{3n}=1.51$	&38-43.2&	${\sigma_{3n}=2.7}_{-1.6}^{+4.8}$	&\cite{RN312}\\
&&&&44.5&	$\sigma_{4n}=0.53$	&42.4-46.5&	${\sigma_{4n}=0.9}_{-0.8}^{+3.2}$	&\cite{RN312}\\
${^{48}}\mathrm{Ca}+{^{245}}\mathrm{Cm}$	&4.0&	1.460734&	1.2074&	32.7&	$\sigma_{2n}=12.97$	&30.9-35&	${\sigma_{2n}=0.9}_{-0.6}^{+1.1}$	&\cite{RN326}\\
&&&&37.7&	$\sigma_{2n}=2.06$	&35.9-39.9&	${\sigma_{2n}=0.7}_{-0.6}^{+2.0}$	&\cite{RN313}\\
&&&&32.7&	$\sigma_{3n}=197.3$	&30.9-35&	${\sigma_{3n}=1.3}_{-0.7}^{+1.2}$	&\cite{RN326}\\
&&&&41.7&	$\sigma_{3n}=6.45$	&40.7-44.8&	${\sigma_{3n}=1.9}_{-1.0}^{+2.1}$	&\cite{RN313}\\
&&&&42.7&	$\sigma_{3n}=4.18$	&40.7-44.8&	${\sigma_{3n}=3.7}_{-1.8}^{+3.6}$	&\cite{RN313}\\
&&&&42.7&	$\sigma_{4n}=0.97$	&40.7-44.8&	${\sigma_{4n}=1.0}_{-0.0}^{+0.0}$	&\cite{RN313}\\
&0.7546&	0.9180&	0.8879&	32.7&	$\sigma_{2n}=0.76$	&30.9-35&	${\sigma_{2n}=0.9}_{-0.6}^{+1.1}$	&\cite{RN326}\\
&&&&37.7&	$\sigma_{2n}=0.96$	&35.9-39.9&	${\sigma_{2n}=0.7}_{-0.6}^{+2.0}$	&\cite{RN313}\\
&&&&32.7&	$\sigma_{3n}=11.63$	&30.9-35&	${\sigma_{3n}=1.3}_{-0.7}^{+1.2}$	&\cite{RN326}\\
&&&&41.7&	$\sigma_{3n}=4.06$	&40.7-44.8&	${\sigma_{3n}=1.9}_{-1.0}^{+2.1}$	&\cite{RN313}\\
&&&&42.7&	$\sigma_{3n}=2.75$	&40.7-44.8&	${\sigma_{3n}=3.7}_{-1.8}^{+3.6}$	&\cite{RN313}\\
&&&&42.7&	$\sigma_{4n}=0.64$	&40.7-44.8&	${\sigma_{4n}=1.0}_{-0.0}^{+0.0}$	&\cite{RN313}\\
${^{48}}\mathrm{Ca}+{^{248}}\mathrm{Cm}$	&4.0	&1.460734	&1.1875	&30.4	&$\sigma_{3n}=19.51$	&30.5&	$\sigma_{3n}=0.9$	&\cite{RN309}\\
&&&&33.4&	$\sigma_{3n}=26.96$	&33&	${\sigma_{3n}=0.5}_{-0.26}^{+0.5}$	&\cite{RN309}\\
&&&&38.4&	$\sigma_{3n}=5.25$	&36.8-41.1&	${\sigma_{3n}=1.1}_{-0.7}^{+1.7}$	&\cite{RN325}\\
&&&&40.4&	$\sigma_{3n}=2.32$	&40.9&	${\sigma_{3n}=0.9}_{-0.7}^{+2.1}$	&\cite{RN325}\\
&&&&33.4&	$\sigma_{4n}=22.13$	&33&	$\sigma_{4n}=0.3$	&\cite{RN309}\\
&&&&38.4&	$\sigma_{4n}=9.60$	&36.8-41.1&	${\sigma_{4n}=3.3}_{-1.4}^{+2.5}$	&\cite{RN325}\\
&&&&40.4&	$\sigma_{4n}=5.12$	&40.9&	${\sigma_{4n}=3.4}_{-1.6}^{+2.7}$	&\cite{RN325}\\
&0.7546&	0.9180&	0.8856&	30.4&	$\sigma_{3n}=0.014$	&30.5&	$\sigma_{3n}=0.9$	&\cite{RN309}\\
&&&&33.4&	$\sigma_{3n}=0.9$	&33&	${\sigma_{3n}=0.5}_{-0.26}^{+0.5}$	&\cite{RN309}\\
&&&&38.4&	$\sigma_{3n}=2.38$	&36.8-41.1&	${\sigma_{3n}=1.1}_{-0.7}^{+1.7}$	&\cite{RN325}\\
\end{tabular}
\end{ruledtabular}
\end{table*}
\begin{table*}
Continued table~\ref{tab:table2}\\\hspace{-2.3cm}
\setlength{\tabcolsep}{2pt}
\renewcommand{\arraystretch}{1.5}
\begin{ruledtabular}
\begin{tabular}{ccccccccc}
&&&\multicolumn{3}{c}{Calculated values}&\multicolumn{2}{c}{Experimental data}&\\
\cmidrule(rl){4-6} \cmidrule(rl){7-8}
Combination&$k_s$&$\gamma_0(\mathrm{MeV{fm}^{-2}})$&$\gamma(\mathrm{MeV{fm}^{-2}})$
&$E_{CN}^\ast(\mathrm{MeV})$&$\sigma_{ER}(\mathrm{pb})$&$E_{CN}^\ast(\mathrm{MeV})$&$\sigma_{ER}(\mathrm{pb})$&Ref.\\ \hline
&&&&40.4&	$\sigma_{3n}=1.29$	&40.9&	${\sigma_{3n}=0.9}_{-0.7}^{+2.1}$	&\cite{RN325}\\
&&&&33.4&	$\sigma_{4n}=0.7$	&33&	$\sigma_{4n}=0.3$	&\cite{RN309}\\
&&&&38.4&	$\sigma_{4n}=4.36$	&36.8-41.1&	${\sigma_{4n}=3.3}_{-1.4}^{+2.5}$	&\cite{RN325}\\
&&&&40.4&	$\sigma_{4n}=2.84$	&40.9&	${\sigma_{4n}=3.4}_{-1.6}^{+2.7}$	&\cite{RN325}\\
${^{48}}\mathrm{Ca}+{^{249}}\mathrm{Bk}$	&4.0	&1.460734	&1.1978&	32.8&	$\sigma_{3n}=35.24$	&30.4-34.7&	$\sigma_{3n}{=0.7}_{-0.57}^{+1.7}$	&\cite{RN315}\\
&&&&34.8&	$\sigma_{3n}=18.11$	&33.2-37.4&	$\sigma_{3n}{=0.5}_{-0.4}^{+1.1}$	&\cite{RN314}\\
&&&&34.8&	$\sigma_{3n}=18.11$	&35&	$\sigma_{3n}={3.6}_{-2.5}^{+6.1}$	&\cite{RN317}\\
&&&&34.8&	$\sigma_{3n}=18.11$	&32.8-37.5&	${\sigma_{3n}=1.1}_{-0.6}^{+1.2}$	&\cite{RN315}\\
&&&&38.8&	$\sigma_{3n}=3.57$	&39&	$\sigma_{3n}=0.7$	&\cite{RN314}\\
&&&&38.8&	$\sigma_{3n}=3.57$	&39&	$\sigma_{3n}=0.32$	&\cite{RN315}\\
&&&&34.8&	$\sigma_{4n}=11.41$	&35&	$\sigma_{4n}=0.8$	&\cite{RN314}\\
&&&&34.8&	$\sigma_{4n}=11.41$	&35&	$\sigma_{4n}=0.59$	&\cite{RN315}\\
&&&&38.8&	$\sigma_{4n}=4.20$	&37.2-41.4&	$\sigma_{4n}={1.3}_{-0.6}^{+1.5}$	&\cite{RN314}\\
&&&&38.8&	$\sigma_{4n}=4.20$	&39&	${\sigma_{4n}=2.0}_{-1.1}^{+2.2}$	&\cite{RN317}\\
&&&&38.8&	$\sigma_{4n}=4.20$	&37-41.9&	$\sigma_{4n}={1.5}_{-0.5}^{+1.1}$	&\cite{RN315}\\
&&&&42.8&	$\sigma_{4n}=1.06$	&40.3-44.8&	$\sigma_{4n}={2.4}_{-1.4}^{+3.3}$	&\cite{RN315}\\
&&&&45.8&	$\sigma_{4n}=0.34$	&43.8-48.3&	${\sigma_{4n}=2.0}_{-1.1}^{+1.8}$	&\cite{RN315}\\
&0.7546&	0.9180&	0.8868&	32.8&	$\sigma_{3n}=3.70$	&30.4-34.7&	$\sigma_{3n}{=0.7}_{-0.57}^{+1.7}$	&\cite{RN315}\\
&&&&34.8&	$\sigma_{3n}=5.45$	&33.2-37.4&	$\sigma_{3n}{=0.5}_{-0.4}^{+1.1}$	&\cite{RN314}\\
&&&&34.8&	$\sigma_{3n}=5.45$	&35&	$\sigma_{3n}={3.6}_{-2.5}^{+6.1}$	&\cite{RN317}\\
&&&&34.8&	$\sigma_{3n}=5.45$	&32.8-37.5&	${\sigma_{3n}=1.1}_{-0.6}^{+1.2}$	&\cite{RN315}\\
&&&&38.8&	$\sigma_{3n}=1.94$	&39&	$\sigma_{3n}=0.7$	&\cite{RN314}\\
&&&&38.8&	$\sigma_{3n}=1.94$	&39&	$\sigma_{3n}=0.32$	&\cite{RN315}\\
&&&&34.8&	$\sigma_{4n}=3.43$	&35&	$\sigma_{4n}=0.8$	&\cite{RN314}\\
&&&&34.8&	$\sigma_{4n}=3.43$	&35&	$\sigma_{4n}=0.59$	&\cite{RN315}\\
&&&&38.8&	$\sigma_{4n}=2.29$	&37.2-41.4&	$\sigma_{4n}={1.3}_{-0.6}^{+1.5}$	&\cite{RN314}\\
&&&&38.8&	$\sigma_{4n}=2.29$	&39&	${\sigma_{4n}=2.0}_{-1.1}^{+2.2}$	&\cite{RN317}\\
&&&&38.8&	$\sigma_{4n}=2.29$	&37-41.9&	$\sigma_{4n}={1.5}_{-0.5}^{+1.1}$	&\cite{RN315}\\
&&&&42.8&	$\sigma_{4n}=0.71$	&40.3-44.8&	$\sigma_{4n}={2.4}_{-1.4}^{+3.3}$	&\cite{RN315}\\
&&&&45.8&	$\sigma_{4n}=0.25$	&43.8-48.3&	${\sigma_{4n}=2.0}_{-1.1}^{+1.8}$	&\cite{RN315}\\
${^{48}}\mathrm{Ca}+{^{249}}\mathrm{Cf}$	&4.0&	1.460734&	1.2142&	29.6&	$\sigma_{3n}=168.2$	&26.6-31.7&	${\sigma_{3n}=0.3}_{-0.27}^{+1.0}$	&\cite{RN313}\\
&&&&34.6&	$\sigma_{3n}=38.06$	&32.1-36.6&	${\sigma_{3n}=0.5}_{-0.3}^{+1.6}$	&\cite{RN313}\\
&0.7546&	0.9180&	0.8887&	29.6&	$\sigma_{3n}=3.77$	&26.6-31.7&	${\sigma_{3n}=0.3}_{-0.27}^{+1.0}$	&\cite{RN313}\\
&&&&34.6&	$\sigma_{3n}=15.93$	&32.1-36.6&	${\sigma_{3n}=0.5}_{-0.3}^{+1.6}$	&\cite{RN313}\\
\end{tabular}
\end{ruledtabular}
\end{table*}

\begin{table}[htbp]
\caption{\label{tab:table3}Combinations, excitation energies of the CN, and the maximum ER cross sections in three and four neutron channels to synthesis of nuclei with $Z=119$}
\renewcommand{\arraystretch}{1.2}
\begin{ruledtabular}
\begin{tabular}{lcr}
Combination&$E_{CN}^\ast(\mathrm{MeV})$&$\sigma_{ER}(\mathrm{fb})$\\ \hline
${^{249}}\mathrm{Cf}({^{45}}\mathrm{Sc},3n){^{291}}119$&39.84&$\sigma_{3n}=417.1$\\
${^{249}}\mathrm{Cf}({^{45}}\mathrm{Sc},4n){^{290}}119$&41.84&$\sigma_{4n}=138.5$\\
${^{249}}\mathrm{Bk}({^{50}}\mathrm{Ti},3n){^{296}}119$&35.60&$\sigma_{3n}=7.96$\\
${^{249}}\mathrm{Bk}({^{50}}\mathrm{Ti},4n){^{295}}119$&37.60&$\sigma_{4n}=3.31$\\
${^{247}}\mathrm{Bk}({^{50}}\mathrm{Ti},3n){^{294}}119$&35.15&$\sigma_{3n}=11.2$\\
${^{247}}\mathrm{Bk}({^{50}}\mathrm{Ti},4n){^{293}}119$&37.15&$\sigma_{4n}=2.46$\\
${^{248}}\mathrm{Cm}({^{51}}\mathrm{V},3n){^{296}}119$&40.37&$\sigma_{3n}=0.17$\\
${^{248}}\mathrm{Cm}({^{51}}\mathrm{V},4n){^{295}}119$&41.37&$\sigma_{4n}=0.13$\\
${^{243}}\mathrm{Am}({^{54}}\mathrm{Cr},3n){^{294}}119$&33.34&$\sigma_{3n}=0.8$\\
${^{243}}\mathrm{Am}({^{54}}\mathrm{Cr},4n){^{293}}119$&36.34&$\sigma_{4n}=0.12$\\
${^{242}}\mathrm{Pu}({^{55}}\mathrm{Mn},3n){^{294}}119$&37.1&$\sigma_{3n}=0.02$\\
${^{254}}\mathrm{Es}({^{48}}\mathrm{Ca},3n){^{299}}119$&32.14&$\sigma_{3n}=9115.15$\\
${^{254}}\mathrm{Es}({^{48}}\mathrm{Ca},4n){^{298}}119$&34.14&$\sigma_{4n}=735.46$\\
\end{tabular}
\end{ruledtabular}
\end{table}

\begin{figure*}[htbp]
\includegraphics[width=59mm]{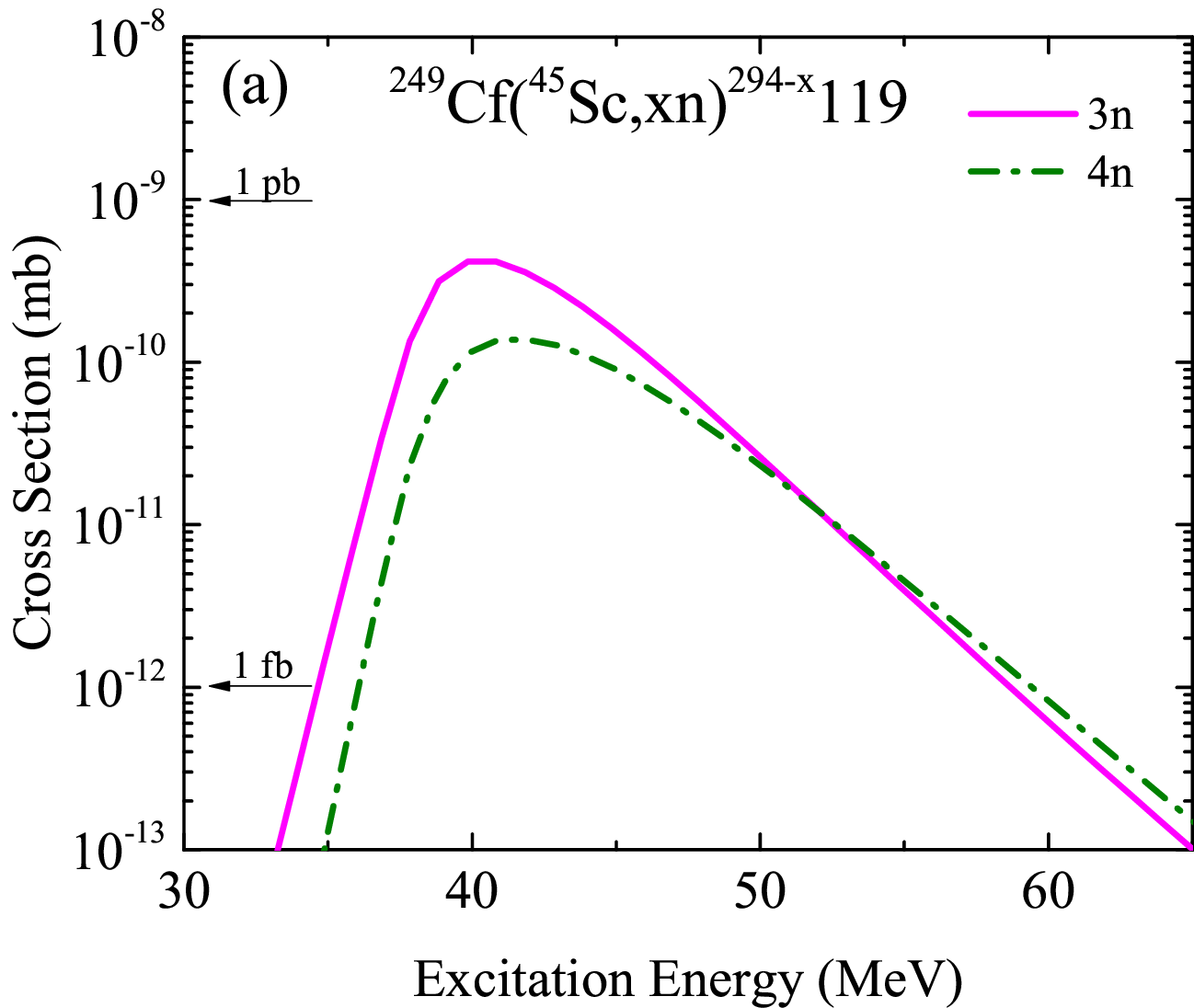}
\includegraphics[width=59mm]{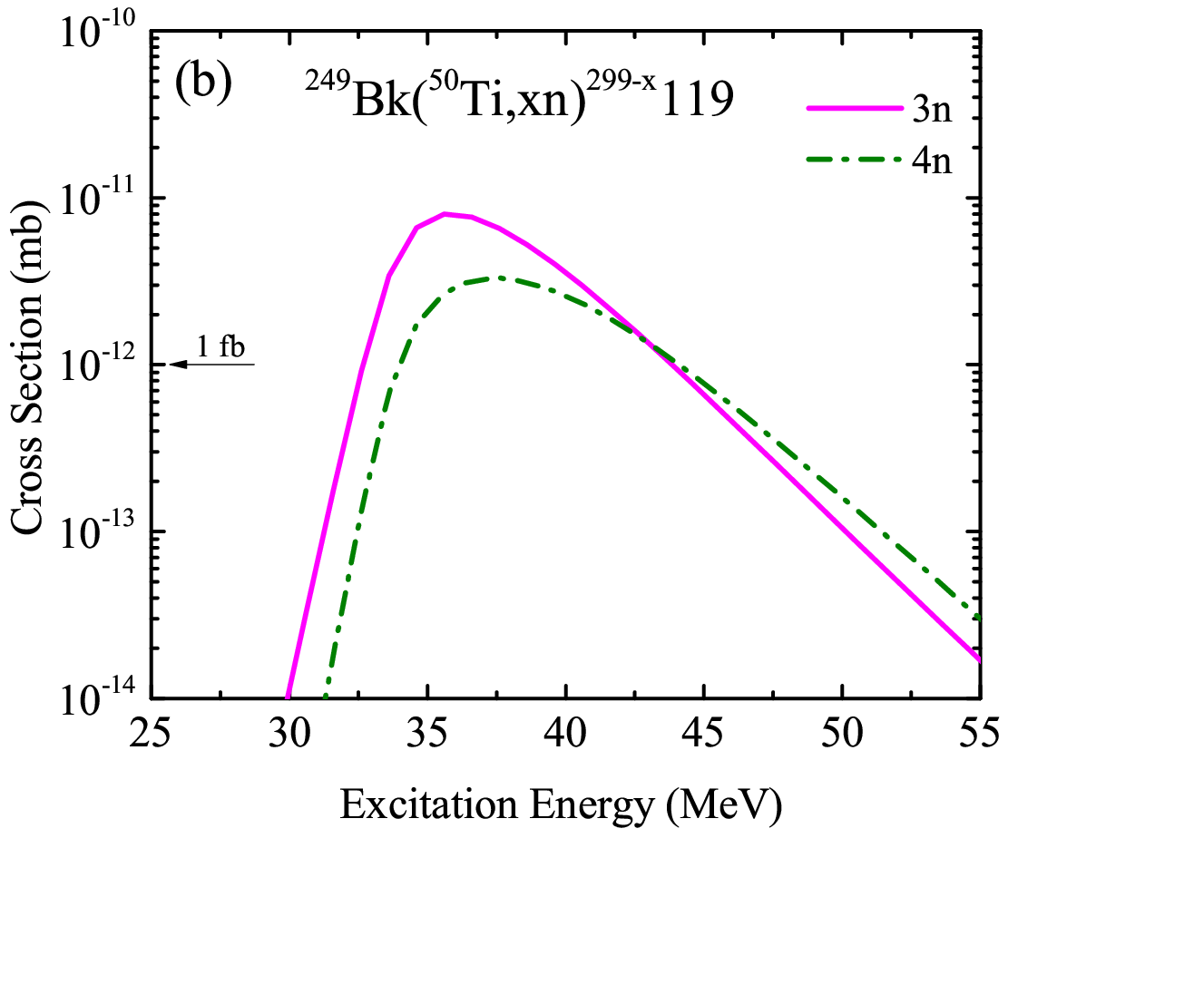}
\includegraphics[width=59mm]{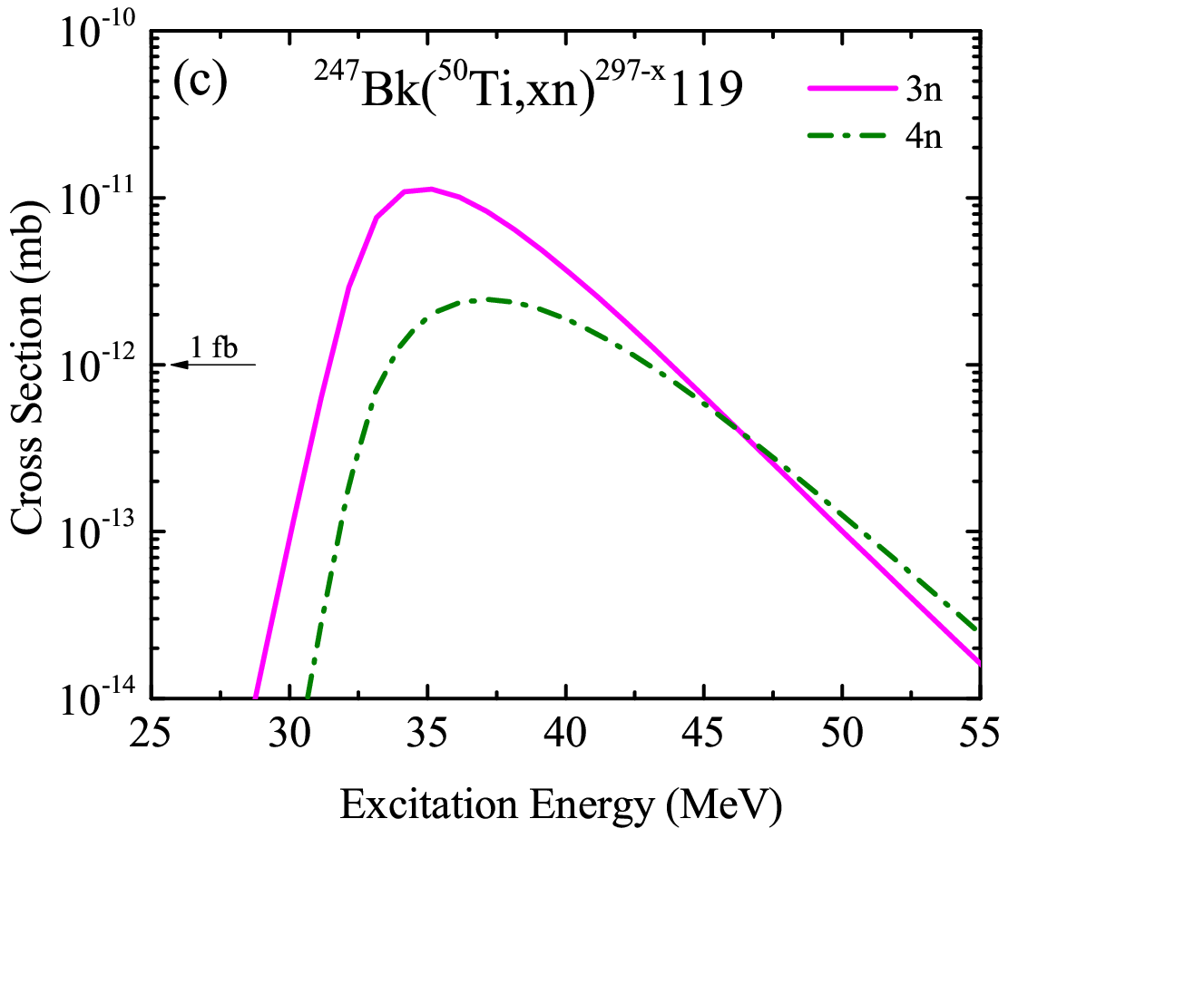}
\includegraphics[width=59mm]{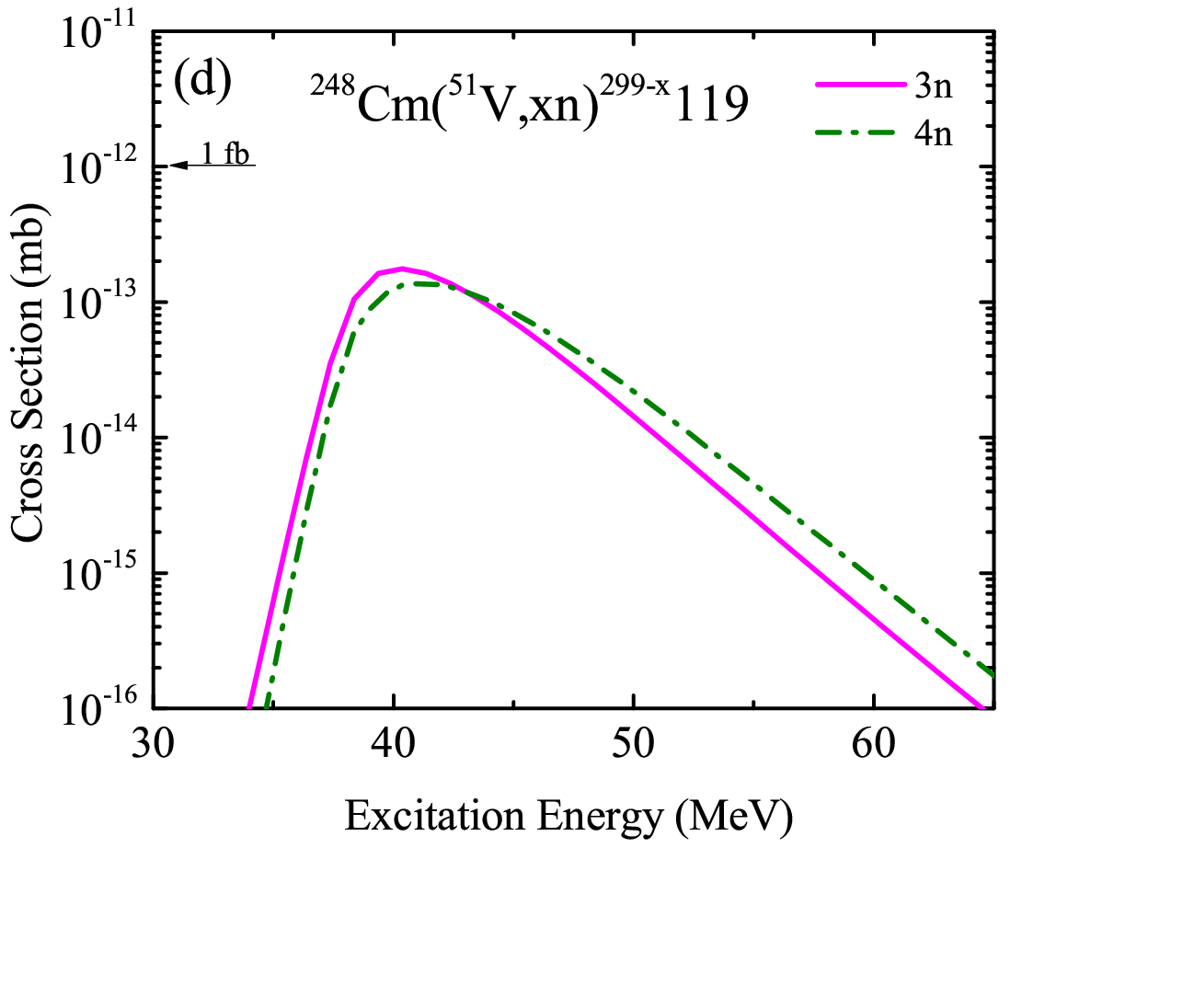}
\includegraphics[width=59mm]{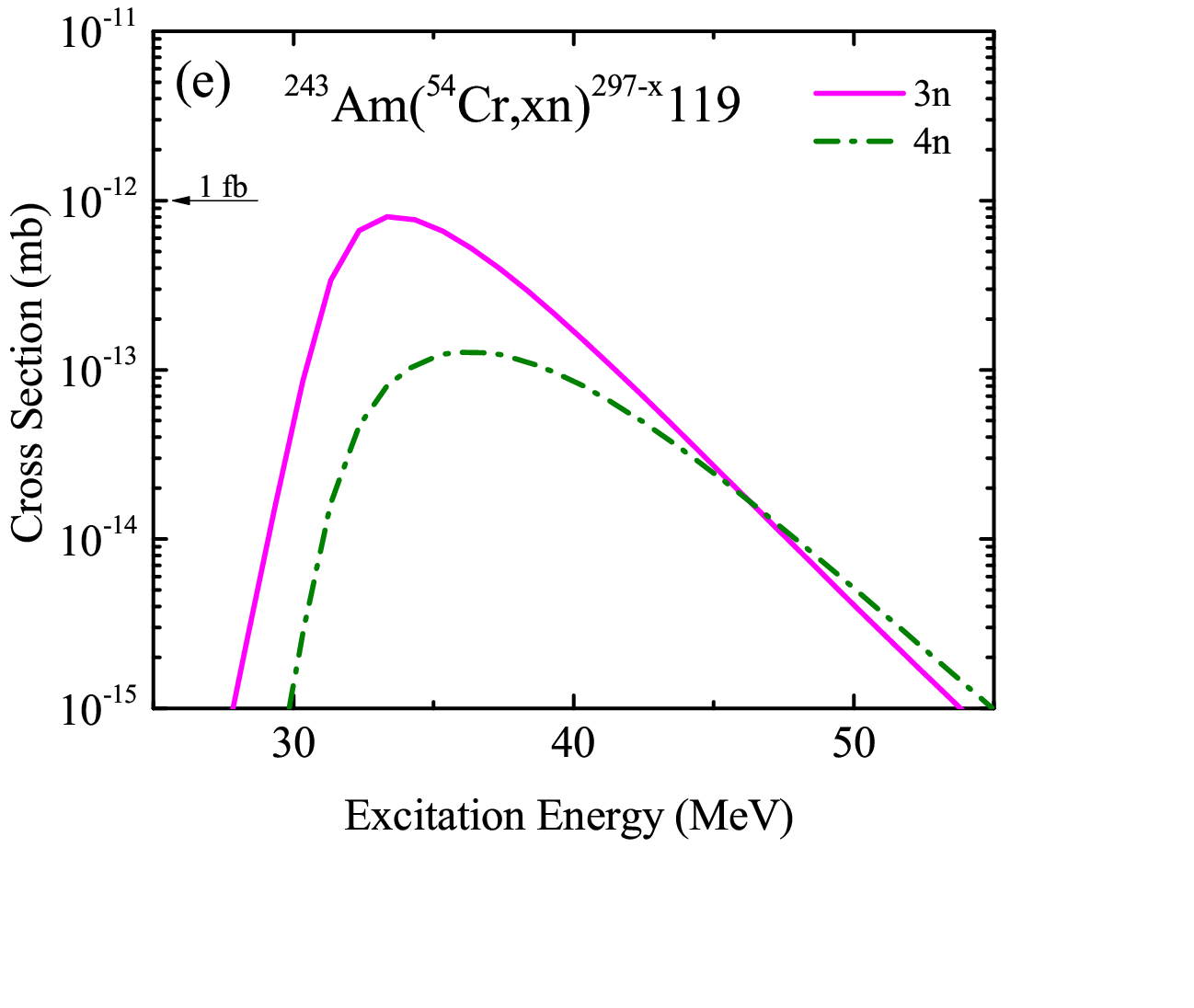}
\includegraphics[width=59mm]{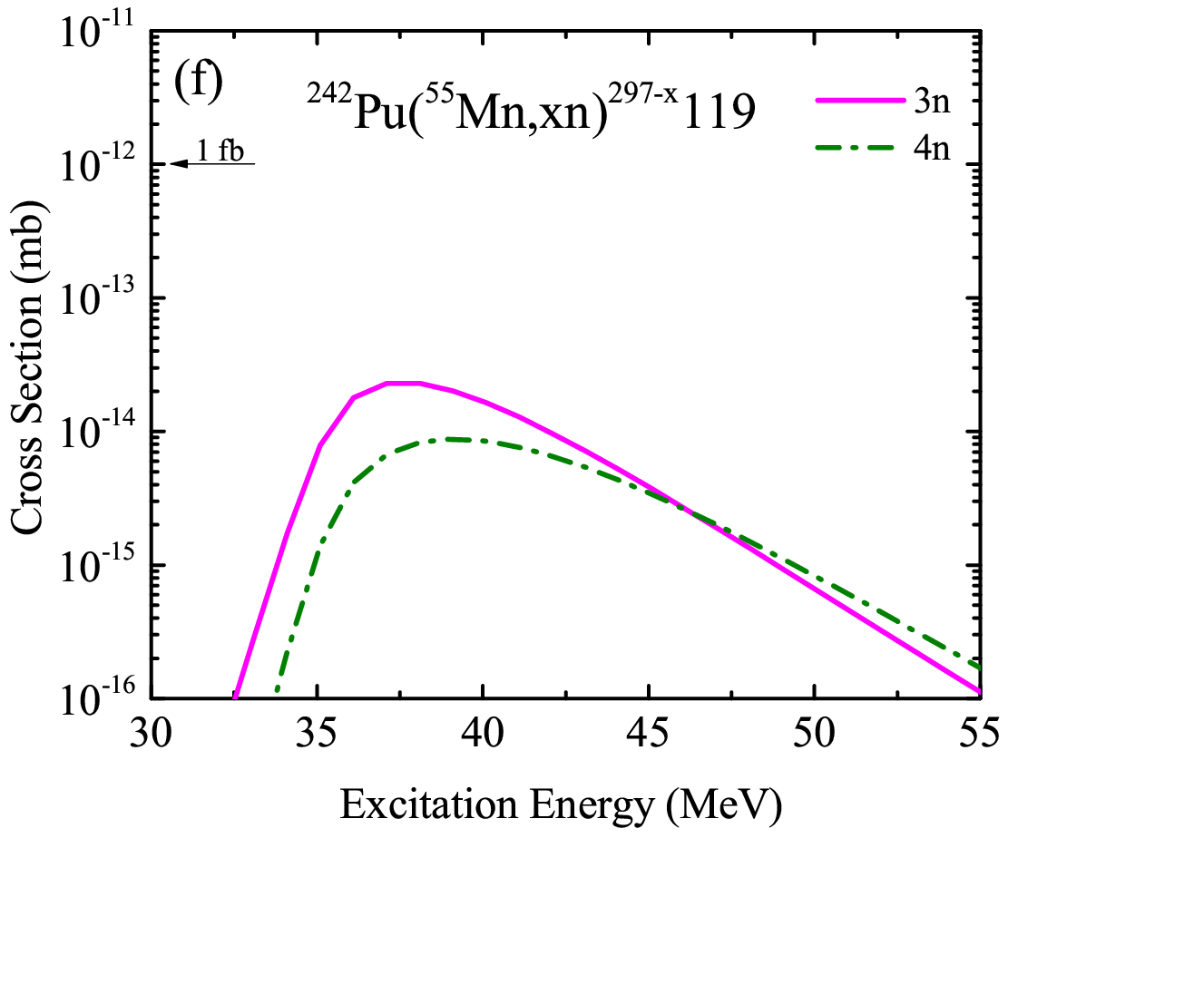}
\includegraphics[width=59mm]{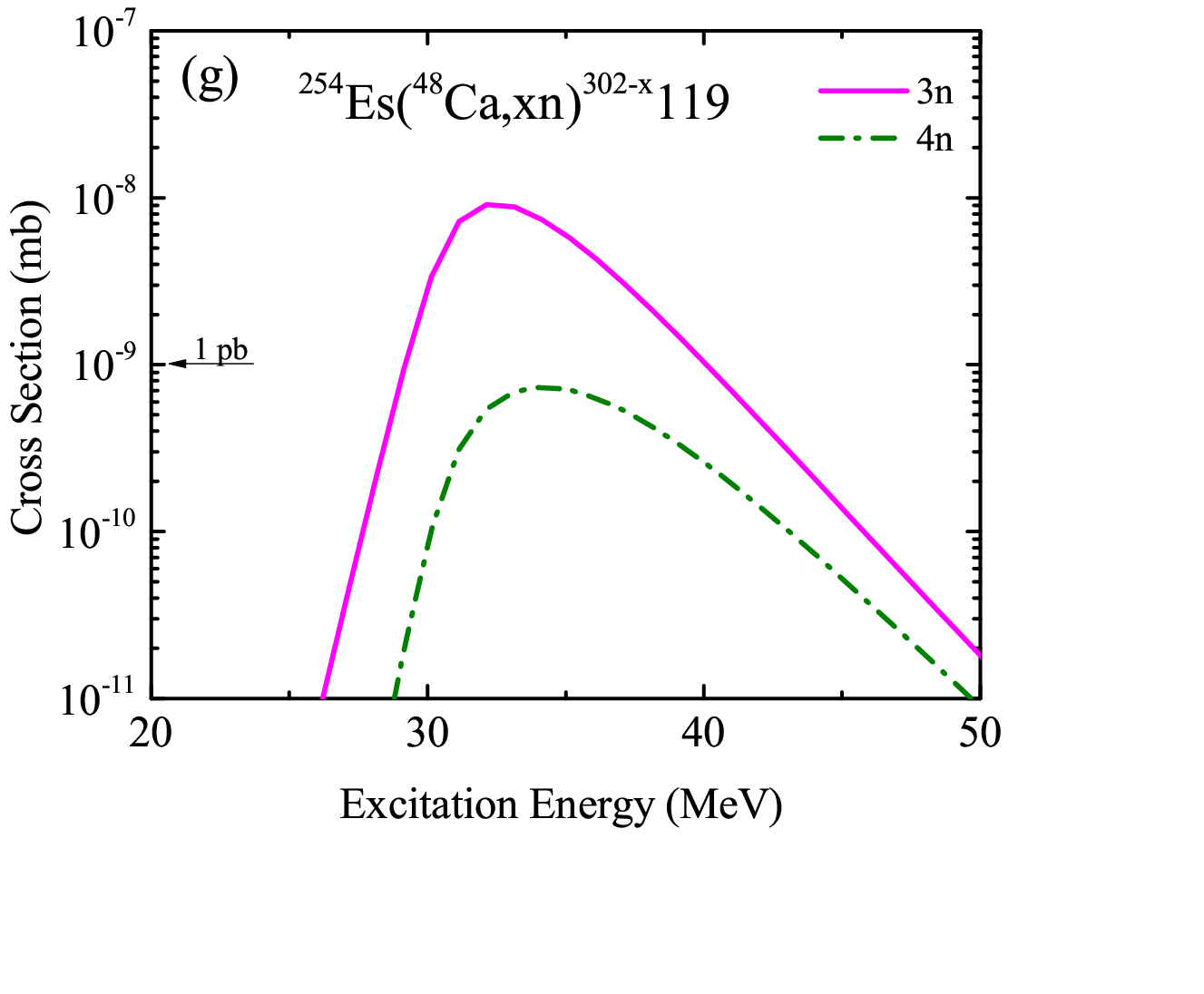}
\caption{\label{fig:fig5}(Color online) the ER cross sections vs. the excitation energies of the CN for combinations of a) ${^{249}}\mathrm{Cf}({^{45}}\mathrm{Sc},xn){^{294-x}}119$, b) ${^{249}}\mathrm{Bk}({^{50}}\mathrm{Ti},xn){^{299-x}}119$, c) ${^{247}}\mathrm{Bk}({^{50}}\mathrm{Ti},xn){^{297-x}}119$, d) ${^{248}}\mathrm{Cm}({^{51}}\mathrm{V},xn){^{299-x}}119$, e)  ${^{243}}\mathrm{Am}({^{54}}\mathrm{Cr},xn){^{297-x}}119$, f) ${^{242}}\mathrm{Pu}({^{55}}\mathrm{Mn},xn){^{297-x}}119$, g) ${^{254}}\mathrm{Es}({^{48}}\mathrm{Ca},xn){^{302-x}}119$.}
\end{figure*}

\begin{table}[htbp]
\caption{\label{tab:table4}Combinations, excitation energies of the CN, and the maximum ER cross sections in three and four neutron channels to synthesize nuclei with $Z=120$}
\renewcommand{\arraystretch}{1.2}	\begin{ruledtabular}
\begin{tabular}{lcr}
Combination&$E_{CN}^\ast(\mathrm{MeV})$&$\sigma_{ER}(\mathrm{fb})$\\ \hline
${^{254}}\mathrm{Es}({^{45}}\mathrm{Sc},4n){^{295}}120$&49.81&$\sigma_{4n}=0.05$\\
${^{249}}\mathrm{Cf}({^{50}}\mathrm{Ti},3n){^{296}}120$&33.19&$\sigma_{3n}=51.19$\\
${^{249}}\mathrm{Cf}({^{50}}\mathrm{Ti},4n){^{295}}120$&36.19&$\sigma_{4n}=1.05$\\
${^{251}}\mathrm{Cf}({^{50}}\mathrm{Ti},3n){^{298}}120$&33.82&$\sigma_{3n}=43.17$\\
${^{251}}\mathrm{Cf}({^{50}}\mathrm{Ti},4n){^{297}}120$&36.82&$\sigma_{4n}=1.14$\\
${^{249}}\mathrm{Bk}({^{51}}\mathrm{V},3n){^{297}}120$&38.86&$\sigma_{3n}=0.06$\\
${^{249}}\mathrm{Bk}({^{51}}\mathrm{V},4n){^{296}}120$&40.86&$\sigma_{4n}=0.03$\\
${^{248}}\mathrm{Cm}({^{54}}\mathrm{Cr},3n){^{299}}120$&33.58&$\sigma_{3n}=0.19$\\
${^{243}}\mathrm{Am}({^{55}}\mathrm{Mn},3n){^{295}}120$&36.26&$\sigma_{4n}=0.007$\\
\end{tabular}
\end{ruledtabular}
\end{table}

\begin{figure*}[htbp]
\includegraphics[width=59mm]{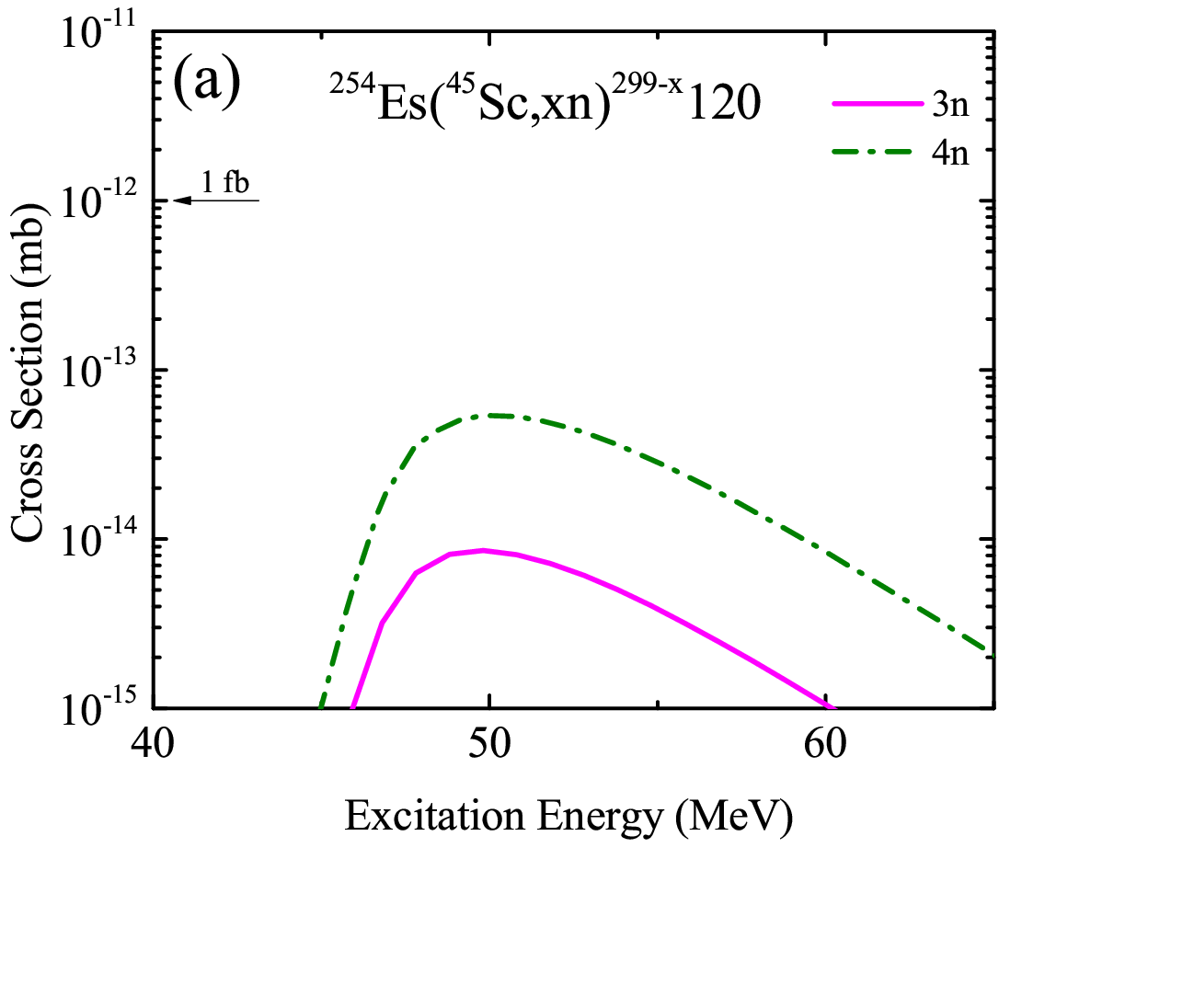}
\includegraphics[width=59mm]{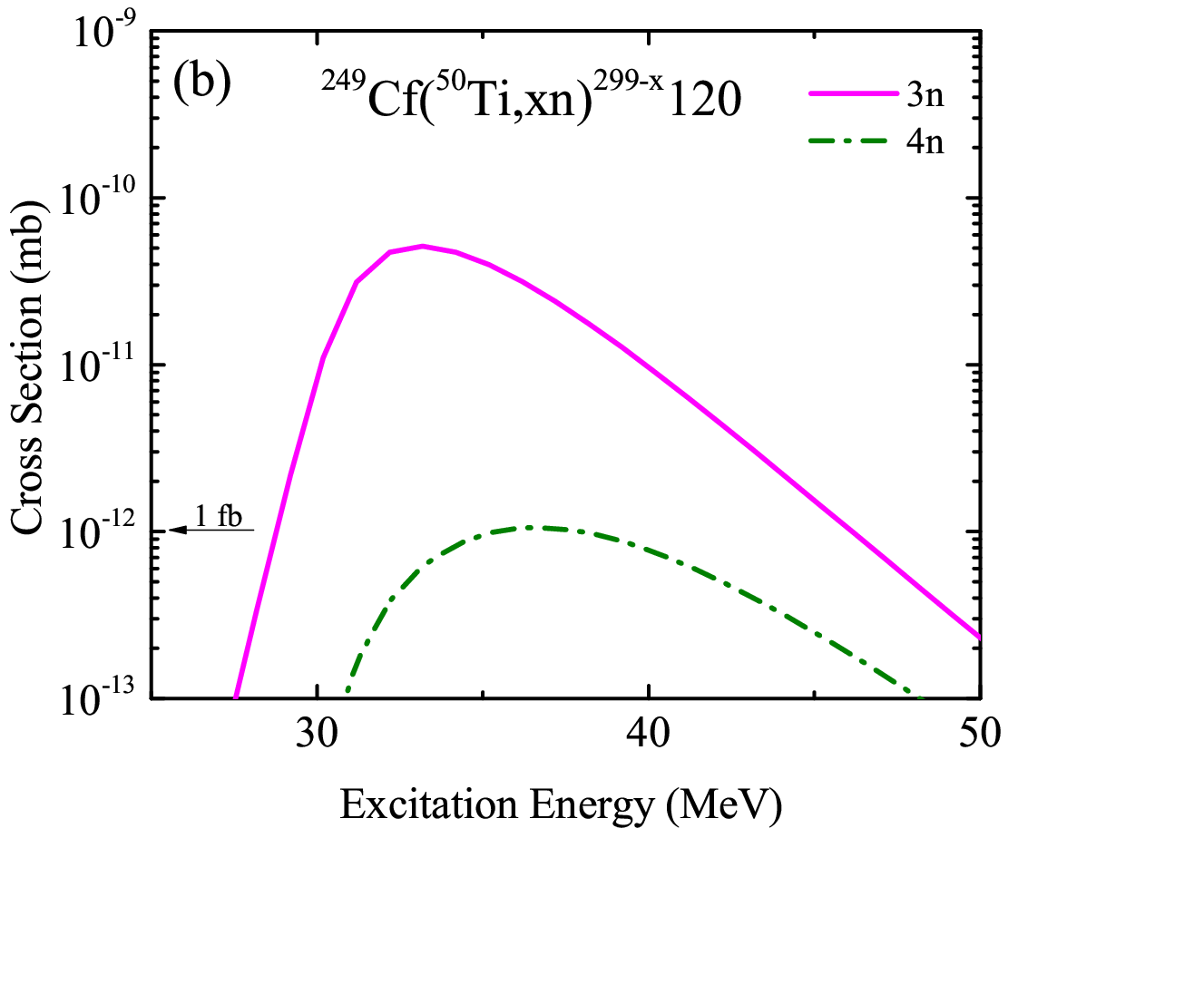}
\includegraphics[width=59mm]{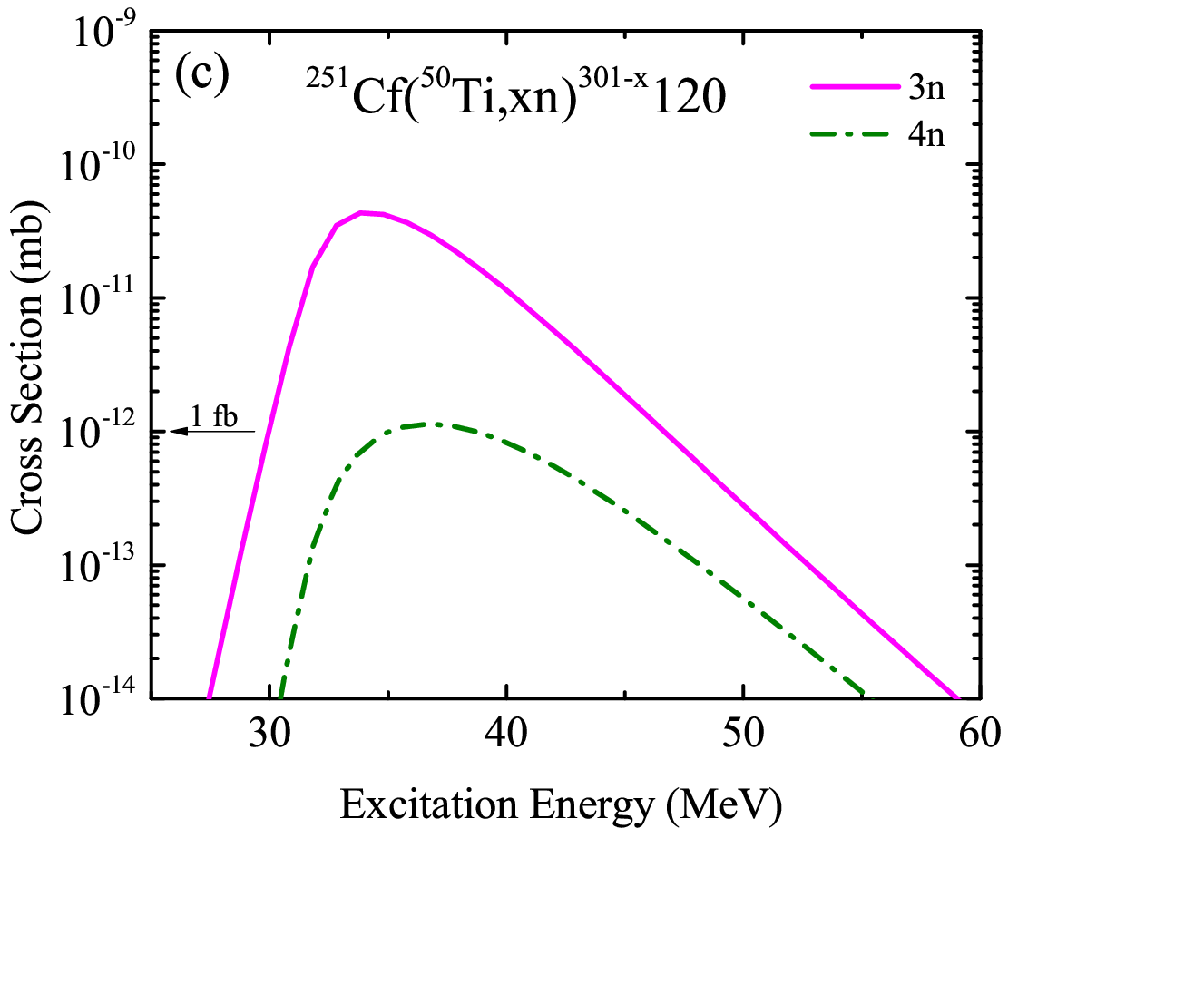}
\includegraphics[width=59mm]{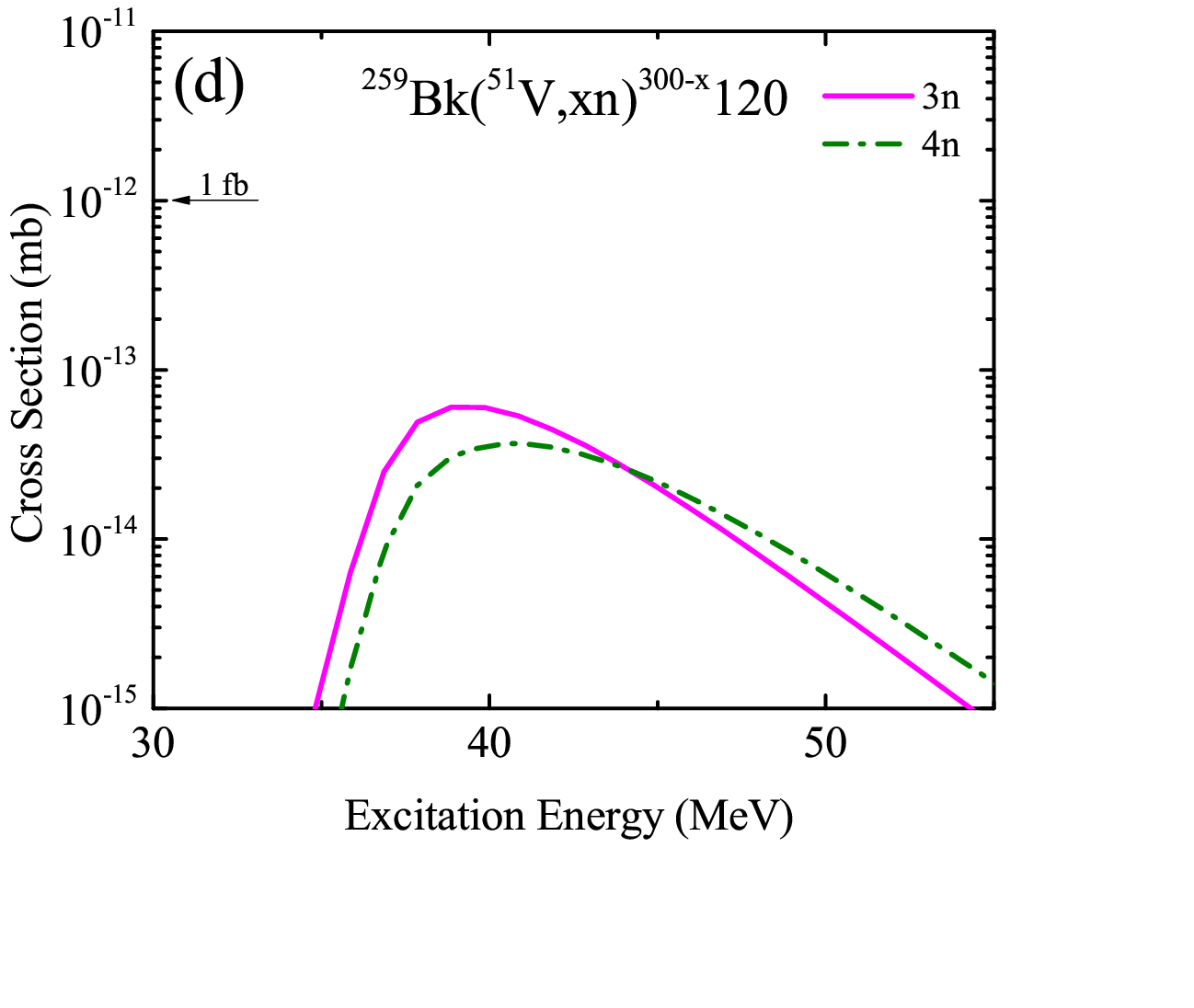}
\includegraphics[width=59mm]{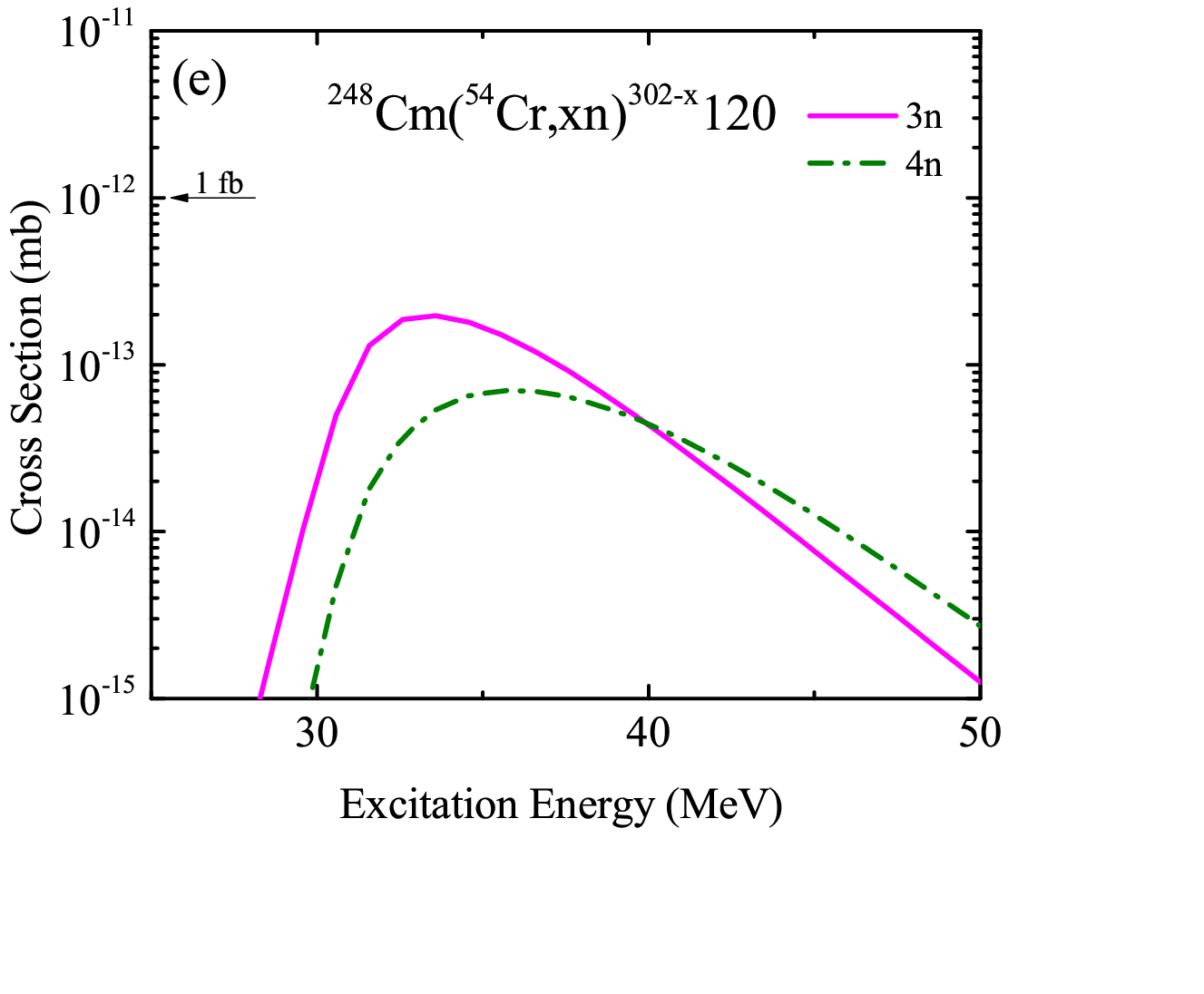}
\includegraphics[width=59mm]{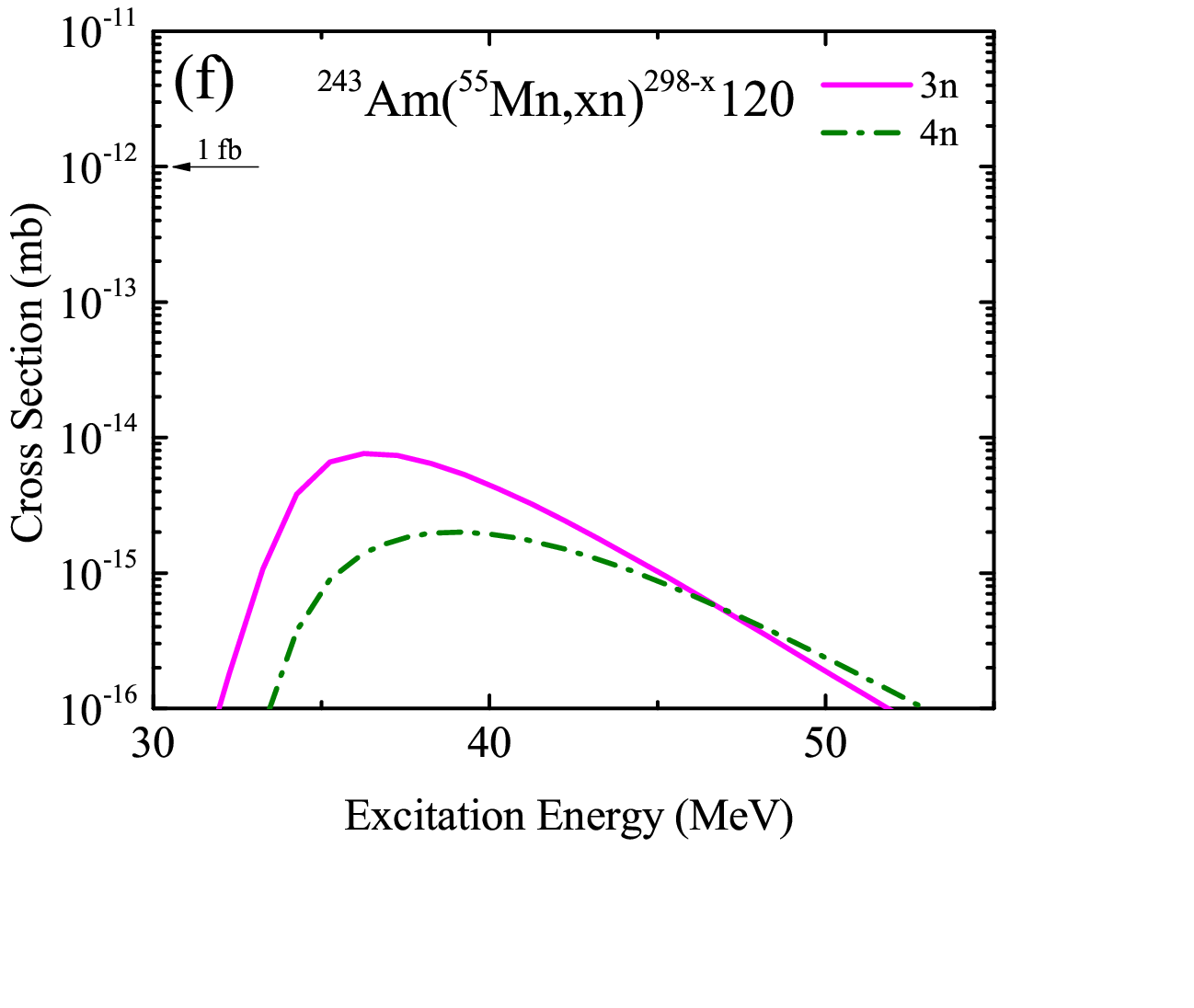}
\caption{\label{fig:fig6}(Color online) the ER cross section vs. the excitations energies of the CN for combinations of (a) ${^{254}}\mathrm{Es}({^{45}}\mathrm{Sc},xn){^{299-x}}120$, (b) ${^{249}}\mathrm{Cf}({^{50}}\mathrm{Ti},xn){^{299-x}}120$, (c) ${^{251}}\mathrm{Cf}({^{50}}\mathrm{Ti},xn){^{301-x}}120$, (d) ${^{249}}\mathrm{Bk}({^{51}}\mathrm{V},xn){^{300-x}}120$, (e) ${^{248}}\mathrm{Cm}({^{54}}\mathrm{Cr},xn){^{302-x}}120$, (f) ${^{243}}\mathrm{Am}({^{55}}\mathrm{Mn},xn){^{298-x}}120$.}
\end{figure*}

\begin{table*}[htbp]
\caption{\label{tab:table5}The comparison between the results from current study with previously reported results employing different models.}
\renewcommand{\arraystretch}{1.2}
\begin{ruledtabular}
\begin{tabular}{cccccc}
${^{45}}\mathrm{Sc}+{^{249}}\mathrm{Cf}$&${^{50}}\mathrm{Ti}+{^{247}}\mathrm{Bk}$&${^{50}}\mathrm{Ti}+{^{249}}\mathrm{Bk}$&${^{50}}\mathrm{Ti}+{^{249}}\mathrm{Cf}$&${^{50}}\mathrm{Ti}+{^{251}}\mathrm{Cf}$&Ref.\\ \hline
417.1($3n$), 138.5($4n$)&11.2($3n$), 2.46($4n$)&7.96($3n$), 3.31($4n$)&51.2($3n$), 1($4n$)&43.2($3n$), 1.1($4n$)&This Work\\
5.47($3n$), 0.72($4n$)&&48.2($3n$), 5.67($4n$)&7.7($3n$)&17.2($3n$)&\cite{RN354}\\
&43($3n$)&&&&	\cite{RN342}\\
&&12($3n$), 64($4n$)&&&\cite{RN338}\\
&&40($3n$), 55($4n$)&40($3n$), 46($4n$)&&\cite{RN357}\\
&&35($3n$), 35($4n$)&20($3n$), 4.5($4n$)&&\cite{RN383}\\
&&60($3n$), 110($4n$)&50($3n$), 3.5($4n$)&&\cite{RN384}\\
&&480($3n$), 310($4n$)&150($3n$), 50($4n$)&&\cite{RN385}\\
&&33($3n$), 15($4n$)&&&\cite{RN342}\\
&&40($3n$), 150($4n$)&&&\cite{RN339}\\
&&340($3n$), 160($4n$)&34($3n$), 2($4n$)&120($3n$), 4($4n$)&\cite{RN338}\\
&&30($3n$), 36($4n$)&5.5($3n$), 6.1($4n$)&&\cite{RN355}\\
&&105.1($3n$), 40.9($4n$)&&&\cite{RN348}\\
&&&100($3n$), 2.5($4n$)&& \cite{RN344}\\
&&&760($3n$), 28($4n$)&&\cite{RN365}\\
&&&20($3n$), 21($2n$)&&\cite{RN359}\\
&&&102.6($3n$), 2.3($4n$)&&\cite{RN345}\\
\end{tabular}
\end{ruledtabular}
\end{table*}
\section{\label{Section3}Discussion and conclusion}
\subsection{\label{Section3-1}Significance of the compound nucleus surface energy coefficients on the ER cross sections}
Four parameters are involved in the calculation of the potential: a) The nuclear deformations b) the surface energy coefficient c) the universal function, and d) the nucleus radius. Nuclear deformations, along with collision angles, play important roles to the obtained ER cross sections. Figures~\ref{fig:fig2} and \ref{fig:fig3} show the distribution of total potential versus distance, along with collision angles for ${^{249}}\mathrm{Cf}({^{45}}\mathrm{Sc},xn){^{294-x}}119$ combination. Figures~\ref{fig:fig2} shows the potential distribution for the rotation of colliding nuclei, $\alpha_i$ (see Fig.~\ref{fig:fig1}). One can consider for collisions of $\alpha_1={90}^\circ$  and $\alpha_2={90}^\circ$, the maximum potential, and for collisions of  $\alpha_1=0^\circ$ and $\alpha_2=0^\circ$, the minimum potential occurs. However, for collisions of $\alpha_1={90}^\circ$ and $\alpha_2={90}^\circ$ the quasi-fission barrier ($B_{qf}$) is very low, therefore the survival probability of CN is very small. It means that in competition between the neutron evaporation and changing to the fission fragments, the fission fragments would occur, and the ER cross section will be low. Figures~\ref{fig:fig3} shows the potential distribution for the direction of colliding nuclei, $\theta_i$. It is clear for collisions of $\theta_1=0^\circ$ and $\theta_2=0^\circ$, the maximum potential, and for collisions of $\theta_1={90}^\circ$ and $\theta_2={90}^\circ$, the minimum potential occurs. It is considered for collisions of $\theta_1=0^\circ$ and $\theta_2=0^\circ$ the $B_{qf}$ is smaller than collisions of $\theta_1={90}^\circ$ and $\theta_2={90}^\circ$. It means that for a collision of $\theta_1=0^\circ$ and $\theta_2=0^\circ$ the probability of changing to fission fragments is more than the neutron evaporation, and of course, the ER cross section also will be low. In actual conditions, nuclei can collide with each other, in different rotation and direction,therefore the average of the capture cross section will be considered.\\
The surface energy coefficient is given by Eq.~(\ref{eq:4}). Based on previous studies, 13 different values have been provided for $k_s$ and $\gamma_0$ \cite{RN368,RN369,RN370,RN371,RN372,RN373,RN374,RN375}. Those coefficients are listed in Table~\ref{tab:table1}.
The combinations which lead to the synthesis of SHN with $Z=112-118$ were chosen. These combinations included
${^{238}}\mathrm{U}({^{48}}\mathrm{Ca},xn){^{286-x}}112$, ${^{237}}\mathrm{Np}({^{48}}\mathrm{Ca},xn){^{285-x}}113$, ${^{240}}\mathrm{Pu}({^{48}}\mathrm{Ca},xn){^{288-x}}114$, ${^{242}}\mathrm{Pu}({^{48}}\mathrm{Ca},xn){^{290-x}}114$, ${^{244}}\mathrm{Pu}({^{48}}\mathrm{Ca},xn){^{292-x}}114$, ${^{243}}\mathrm{Am}({^{48}}\mathrm{Ca},xn){^{291-x}}115$, ${^{245}}\mathrm{Cm}({^{48}}\mathrm{Ca},xn){^{293-x}}116$, ${^{248}}\mathrm{Cm}({^{48}}\mathrm{Ca},xn){^{296-x}}116$, ${^{249}}\mathrm{Bk}({^{48}}\mathrm{Ca},xn){^{297-x}}117$, ${^{249}}\mathrm{Cf}({^{48}}\mathrm{Ca},xn){^{297-x}}118$. Using the values listed for $k_s$ and $\gamma_0$ in Table~\ref{tab:table1}, the values for the surface energy coefficient, the CN excitation energy, and the ER cross section were calculated. Some selected calculated results along with the experimental data are listed in Table~\ref{tab:table2}. The selected results represent the values of $k_s$ and $\gamma_0$ by which the highest agreement ($k_s=0.7546$ and $\gamma_0=0.9180~\mathrm{MeV{fm}^{-2}}$) and lowest agreement ($k_s=4.0$ and $\gamma_0=1.460734~\mathrm{MeV{fm}^{-2}}$) between the experimental data and calculated values were achieved.
\\
From Table~\ref{tab:table2}, one can conclude that the calculated ER cross-sections with the defined values of $k_s=0.7546$ and $\gamma_0=0.9180~\mathrm{MeV{fm}^{-2}}$ is in good agreement with the experimental data. Figures~\ref{fig:fig4} represent the ER cross sections versus the excitation energies for combinations of
${^{238}}\mathrm{U}({^{48}}\mathrm{Ca},xn){^{286-x}}112$, ${^{237}}\mathrm{Np}({^{48}}\mathrm{Ca},xn){^{285-x}}113$, ${^{240}}\mathrm{Pu}({^{48}}\mathrm{Ca},xn){^{288-x}}114$, ${^{242}}\mathrm{Pu}({^{48}}\mathrm{Ca},xn){^{290-x}}114$, ${^{244}}\mathrm{Pu}({^{48}}\mathrm{Ca},xn){^{292-x}}114$, ${^{243}}\mathrm{Am}({^{48}}\mathrm{Ca},xn){^{291-x}}115$, ${^{245}}\mathrm{Cm}({^{48}}\mathrm{Ca},xn){^{293-x}}116$, ${^{248}}\mathrm{Cm}({^{48}}\mathrm{Ca},xn){^{296-x}}116$, ${^{249}}\mathrm{Bk}({^{48}}\mathrm{Ca},xn){^{297-x}}117$, ${^{249}}\mathrm{Cf}({^{48}}\mathrm{Ca},xn){^{297-x}}118$.
\subsection{\label{Section3-2}Probability of synthesis of SHN with $\textbf{Z=119,120}$}
Many attempts are in progress to synthesize SHN with $Z=119$, 120 \cite{RN396}. In this part, using the obtained model in Sec.~\ref{Section3-1}, the synthesis probability of SHN with $Z=119$, 120 was investigated by calculating the ER cross sections for several combinations. To synthesize nuclei beyond $Z=118$, targets heavier than Californium (Cf) should be used. To produce the nuclei with Z=119 the seven combinations, namely 
${^{249}}\mathrm{Cf}({^{45}}\mathrm{Sc},xn){^{294-x}}119$, ${^{249}}\mathrm{Bk}({^{50}}\mathrm{Ti},xn){^{299-x}}119$, ${^{247}}\mathrm{Bk}({^{50}}\mathrm{Ti},xn){^{297-x}}119$, ${^{248}}\mathrm{Cm}({^{51}}\mathrm{V},xn){^{299-x}}119$, ${^{243}}\mathrm{Am}({^{54}}\mathrm{Cr},xn){^{297-x}}119$, ${^{242}}\mathrm{Pu}({^{55}}\mathrm{Mn},xn){^{297-x}}119$ and ${^{254}}\mathrm{Es}({^{48}}\mathrm{Ca},xn){^{302-x}}119$
were initially selected. In these combinations, the projectiles heavier than ${^{48}}\mathrm{Ca}$, with stable nuclei (${^{45}}\mathrm{Sc}$, ${^{50}}\mathrm{Ti}$, ${^{51}}\mathrm{V}$, ${^{54}}\mathrm{Cr}$ and ${^{55}}\mathrm{Mn}$), were chosen. The actinide targets which had already been employed to the synthesis of SHN with $Z=114-118$, were used. The ${^{254}}\mathrm{Es}({^{48}}\mathrm{Ca},xn){^{302-x}}119$ combination was chosen due to the potential interest for ${^{254}}\mathrm{Es}$ as a new target in the near future \cite{RN379}. The results are presented in Fig.~\ref{fig:fig5} and the optimum conditions are summarized in Table~\ref{tab:table3}.
From Table~\ref{tab:table3} one can consider that the most promising combinations for synthesis of SHN with $Z=119$ are ${^{249}}\mathrm{Cf}({^{45}}\mathrm{Sc},3n){^{291}}119$, ${^{249}}\mathrm{Cf}({^{45}}\mathrm{Sc},4n){^{290}}119$, ${^{247}}\mathrm{Bk}({^{50}}\mathrm{Ti},3n){^{294}}119$, ${^{254}}\mathrm{Es}({^{48}}\mathrm{Ca},3n){^{299}}119$, and ${^{254}}\mathrm{Es}({^{48}}\mathrm{Ca},4n){^{298}}119$.\\   
For SHN with $Z=120$, six combinations included ${^{254}}\mathrm{Es}({^{45}}\mathrm{Sc},xn){^{299-x}}120$, ${^{249}}\mathrm{Cf}({^{50}}\mathrm{Ti},xn){^{299-x}}120$, ${^{251}}\mathrm{Cf}({^{50}}\mathrm{Ti},xn){^{301-x}}120$, ${^{249}}\mathrm{Bk}({^{51}}\mathrm{V},xn){^{300-x}}120$, ${^{248}}\mathrm{Cm}({^{54}}\mathrm{Cr},xn){^{302-x}}120$ and ${^{243}}\mathrm{Am}({^{55}}\mathrm{Mn},xn){^{298-x}}120$ were investigated, The results are presented in Figure~\ref{fig:fig6} and the ideal conditions are summarized in Table~\ref{tab:table4}.
From Table~\ref{tab:table4} one can consider that the most promising combinations to synthesize nuclei with Z=120, are ${^{249}}\mathrm{Cf}({^{50}}\mathrm{Ti},3n){^{296}}120$ and ${^{251}}\mathrm{Cf}({^{50}}\mathrm{Ti},3n){^{298}}120$. 
\\
The obtained ER cross sections (in femtobarn) from this work in comparison with some common combinations previously reported by other research teams are summarized in Table~\ref{tab:table5}.
\section{\label{Section4}Summary}
In this research the roles of the different surface energy coefficients on the compound nucleus decay modes, were investigated. The fuscous was given to the superheavy nuclei in the range of $Z=112-118$ which were synthesized via heavy ion fusion reactions. With employing di-nuclear system model, proximity potential, and considering deformed nuclei, the evaporation residue cross sections were calculated. Comparing results from performed calculations and experimental data, the best values for surface asymmetric constant, $k_s$ and surface energy constant, $\gamma_0$ were obtained to be 0.7546 and $0.9180~\mathrm{MeV{fm}^{-2}}$, respectively. Furthermore, this new model was used to investigate the probability of synthesis of super heavynuclei with $Z=119$, 120. The most promising combinations to synthesize nuclei with $Z=119$ are ${^{249}}\mathrm{Cf}({^{45}}\mathrm{Sc},3n){^{291}}119$ with the ER cross section, $\sigma_{3n}=417.1~\mathrm{fb}$ at the incident energy $E_{\mathrm{c.m.}}=219~\mathrm{MeV} (E^\ast=39.84~\mathrm{MeV})$, ${^{249}}\mathrm{Cf}({^{45}}\mathrm{Sc},4n){^{290}}119$ with the ER cross section, $\sigma_{4n}=138.5~\mathrm{fb}$ at the incident energy $E_{\mathrm{c.m.}}=221~\mathrm{MeV} (E^*=41.84~\mathrm{MeV})$, ${^{247}}\mathrm {Bk}({^{50}}\mathrm{Ti},3n){^{294}}119$ with the ER cross section, $\sigma_{3n}=11.2~\mathrm{fb}$ at the incident energy $E_{\mathrm{c.m.}}=226~\mathrm{MeV} (E^*=35.15~\mathrm{MeV})$, ${^{254}}\mathrm{Es}({^{48}}\mathrm{Ca},3n){^{299}}119$ with the ER cross section, $\sigma_{3n}=9115.15~\mathrm{fb}$ at the incident energies $E_{\mathrm{c.m.}}=208~\mathrm{MeV} (E^*=32.14~\mathrm{MeV})$, and ${^{254}}\mathrm{Es}({^{48}}\mathrm{Ca},4n){^{298}}119$ with the ER cross section, $\sigma_{4n}=735.46~\mathrm{fb}$ at the incident energies $E_{\mathrm{c.m.}}=210~\mathrm{MeV} (E^*=34.14~\mathrm{MeV})$. In addition, it was found out that the best combinations to synthesis of SHN with $Z=120$ are ${^{249}}\mathrm{Cf}({^{50}}\mathrm{Ti},3n){^{296}}120$ with the ER cross section, $\sigma_{3n}=51.19~\mathrm{fb}$ at the incident energy $E_{\mathrm{c.m.}}=228~\mathrm{MeV} (E^*=33.19~\mathrm{MeV})$, and ${^{251}}\mathrm{Cf}({^{50}}\mathrm{Ti},3n){^{298}}120$  with the ER cross section, $\sigma_{3n}=43.17~\mathrm{fb}$ at the incident energy $E_{\mathrm{c.m.}}=227~\mathrm{MeV} (E^*=33.82~\mathrm{MeV})$.

\nocite{*}
\bibliographystyle{apsrev4-2}
\bibliography{Surface-Energy-Coeff.}% Produces the bibliography via BibTeX.

\begin{thebibliography}{74}%
\makeatletter
\providecommand \@ifxundefined [1]{%
 \@ifx{#1\undefined}
}%
\providecommand \@ifnum [1]{%
 \ifnum #1\expandafter \@firstoftwo
 \else \expandafter \@secondoftwo
 \fi
}%
\providecommand \@ifx [1]{%
 \ifx #1\expandafter \@firstoftwo
 \else \expandafter \@secondoftwo
 \fi
}%
\providecommand \natexlab [1]{#1}%
\providecommand \enquote  [1]{``#1''}%
\providecommand \bibnamefont  [1]{#1}%
\providecommand \bibfnamefont [1]{#1}%
\providecommand \citenamefont [1]{#1}%
\providecommand \href@noop [0]{\@secondoftwo}%
\providecommand \href [0]{\begingroup \@sanitize@url \@href}%
\providecommand \@href[1]{\@@startlink{#1}\@@href}%
\providecommand \@@href[1]{\endgroup#1\@@endlink}%
\providecommand \@sanitize@url [0]{\catcode `\\12\catcode `\$12\catcode
  `\&12\catcode `\#12\catcode `\^12\catcode `\_12\catcode `\%12\relax}%
\providecommand \@@startlink[1]{}%
\providecommand \@@endlink[0]{}%
\providecommand \url  [0]{\begingroup\@sanitize@url \@url }%
\providecommand \@url [1]{\endgroup\@href {#1}{\urlprefix }}%
\providecommand \urlprefix  [0]{URL }%
\providecommand \Eprint [0]{\href }%
\providecommand \doibase [0]{https://doi.org/}%
\providecommand \selectlanguage [0]{\@gobble}%
\providecommand \bibinfo  [0]{\@secondoftwo}%
\providecommand \bibfield  [0]{\@secondoftwo}%
\providecommand \translation [1]{[#1]}%
\providecommand \BibitemOpen [0]{}%
\providecommand \bibitemStop [0]{}%
\providecommand \bibitemNoStop [0]{.\EOS\space}%
\providecommand \EOS [0]{\spacefactor3000\relax}%
\providecommand \BibitemShut  [1]{\csname bibitem#1\endcsname}%
\let\auto@bib@innerbib\@empty
%</preamble>
\bibitem [{\citenamefont {Oganessian}\ \emph
  {et~al.}(2004{\natexlab{a}})\citenamefont {Oganessian}, \citenamefont
  {Utyonkov}, \citenamefont {Lobanov}, \citenamefont {Abdullin}, \citenamefont
  {Polyakov}, \citenamefont {Shirokovsky}, \citenamefont {Tsyganov},
  \citenamefont {Gulbekian}, \citenamefont {Bogomolov}, \citenamefont {Gikal},
  \citenamefont {Mezentsev}, \citenamefont {Iliev}, \citenamefont {Subbotin},
  \citenamefont {Sukhov}, \citenamefont {Voinov}, \citenamefont {Buklanov},
  \citenamefont {Subotic}, \citenamefont {Zagrebaev}, \citenamefont {Itkis},
  \citenamefont {Patin}, \citenamefont {Moody}, \citenamefont {Wild},
  \citenamefont {Stoyer}, \citenamefont {Stoyer}, \citenamefont {Shaughnessy},
  \citenamefont {Kenneally}, \citenamefont {Wilk}, \citenamefont {Lougheed},
  \citenamefont {Il’kaev},\ and\ \citenamefont {Vesnovskii}}]{RN309}%
  \BibitemOpen
  \bibfield  {author} {\bibinfo {author} {\bibfnamefont {Y.~T.}\ \bibnamefont
  {Oganessian}}, \bibinfo {author} {\bibfnamefont {V.~K.}\ \bibnamefont
  {Utyonkov}}, \bibinfo {author} {\bibfnamefont {Y.~V.}\ \bibnamefont
  {Lobanov}}, \bibinfo {author} {\bibfnamefont {F.~S.}\ \bibnamefont
  {Abdullin}}, \bibinfo {author} {\bibfnamefont {A.~N.}\ \bibnamefont
  {Polyakov}}, \bibinfo {author} {\bibfnamefont {I.~V.}\ \bibnamefont
  {Shirokovsky}}, \bibinfo {author} {\bibfnamefont {Y.~S.}\ \bibnamefont
  {Tsyganov}}, \bibinfo {author} {\bibfnamefont {G.~G.}\ \bibnamefont
  {Gulbekian}}, \bibinfo {author} {\bibfnamefont {S.~L.}\ \bibnamefont
  {Bogomolov}}, \bibinfo {author} {\bibfnamefont {B.~N.}\ \bibnamefont
  {Gikal}}, \bibinfo {author} {\bibfnamefont {A.~N.}\ \bibnamefont
  {Mezentsev}}, \bibinfo {author} {\bibfnamefont {S.}~\bibnamefont {Iliev}},
  \bibinfo {author} {\bibfnamefont {V.~G.}\ \bibnamefont {Subbotin}}, \bibinfo
  {author} {\bibfnamefont {A.~M.}\ \bibnamefont {Sukhov}}, \bibinfo {author}
  {\bibfnamefont {A.~A.}\ \bibnamefont {Voinov}}, \bibinfo {author}
  {\bibfnamefont {G.~V.}\ \bibnamefont {Buklanov}}, \bibinfo {author}
  {\bibfnamefont {K.}~\bibnamefont {Subotic}}, \bibinfo {author} {\bibfnamefont
  {V.~I.}\ \bibnamefont {Zagrebaev}}, \bibinfo {author} {\bibfnamefont {M.~G.}\
  \bibnamefont {Itkis}}, \bibinfo {author} {\bibfnamefont {J.~B.}\ \bibnamefont
  {Patin}}, \bibinfo {author} {\bibfnamefont {K.~J.}\ \bibnamefont {Moody}},
  \bibinfo {author} {\bibfnamefont {J.~F.}\ \bibnamefont {Wild}}, \bibinfo
  {author} {\bibfnamefont {M.~A.}\ \bibnamefont {Stoyer}}, \bibinfo {author}
  {\bibfnamefont {N.~J.}\ \bibnamefont {Stoyer}}, \bibinfo {author}
  {\bibfnamefont {D.~A.}\ \bibnamefont {Shaughnessy}}, \bibinfo {author}
  {\bibfnamefont {J.~M.}\ \bibnamefont {Kenneally}}, \bibinfo {author}
  {\bibfnamefont {P.~A.}\ \bibnamefont {Wilk}}, \bibinfo {author}
  {\bibfnamefont {R.~W.}\ \bibnamefont {Lougheed}}, \bibinfo {author}
  {\bibfnamefont {R.~I.}\ \bibnamefont {Il’kaev}},\ and\ \bibinfo {author}
  {\bibfnamefont {S.~P.}\ \bibnamefont {Vesnovskii}},\ }\href
  {https://doi.org/10.1103/PhysRevC.70.064609} {\bibfield  {journal} {\bibinfo
  {journal} {Physical Review C}\ }\textbf {\bibinfo {volume} {70}},\ \bibinfo
  {pages} {064609} (\bibinfo {year} {2004}{\natexlab{a}})}\BibitemShut
  {NoStop}%
\bibitem [{\citenamefont {Oganessian}\ \emph {et~al.}(2007)\citenamefont
  {Oganessian}, \citenamefont {Utyonkov}, \citenamefont {Lobanov},
  \citenamefont {Abdullin}, \citenamefont {Polyakov}, \citenamefont {Sagaidak},
  \citenamefont {Shirokovsky}, \citenamefont {Tsyganov}, \citenamefont
  {Voinov}, \citenamefont {Gulbekian}, \citenamefont {Bogomolov}, \citenamefont
  {Gikal}, \citenamefont {Mezentsev}, \citenamefont {Subbotin}, \citenamefont
  {Sukhov}, \citenamefont {Subotic}, \citenamefont {Zagrebaev}, \citenamefont
  {Vostokin}, \citenamefont {Itkis}, \citenamefont {Henderson}, \citenamefont
  {Kenneally}, \citenamefont {Landrum}, \citenamefont {Moody}, \citenamefont
  {Shaughnessy}, \citenamefont {Stoyer}, \citenamefont {Stoyer},\ and\
  \citenamefont {Wilk}}]{RN311}%
  \BibitemOpen
  \bibfield  {author} {\bibinfo {author} {\bibfnamefont {Y.~T.}\ \bibnamefont
  {Oganessian}}, \bibinfo {author} {\bibfnamefont {V.~K.}\ \bibnamefont
  {Utyonkov}}, \bibinfo {author} {\bibfnamefont {Y.~V.}\ \bibnamefont
  {Lobanov}}, \bibinfo {author} {\bibfnamefont {F.~S.}\ \bibnamefont
  {Abdullin}}, \bibinfo {author} {\bibfnamefont {A.~N.}\ \bibnamefont
  {Polyakov}}, \bibinfo {author} {\bibfnamefont {R.~N.}\ \bibnamefont
  {Sagaidak}}, \bibinfo {author} {\bibfnamefont {I.~V.}\ \bibnamefont
  {Shirokovsky}}, \bibinfo {author} {\bibfnamefont {Y.~S.}\ \bibnamefont
  {Tsyganov}}, \bibinfo {author} {\bibfnamefont {A.~A.}\ \bibnamefont
  {Voinov}}, \bibinfo {author} {\bibfnamefont {G.~G.}\ \bibnamefont
  {Gulbekian}}, \bibinfo {author} {\bibfnamefont {S.~L.}\ \bibnamefont
  {Bogomolov}}, \bibinfo {author} {\bibfnamefont {B.~N.}\ \bibnamefont
  {Gikal}}, \bibinfo {author} {\bibfnamefont {A.~N.}\ \bibnamefont
  {Mezentsev}}, \bibinfo {author} {\bibfnamefont {V.~G.}\ \bibnamefont
  {Subbotin}}, \bibinfo {author} {\bibfnamefont {A.~M.}\ \bibnamefont
  {Sukhov}}, \bibinfo {author} {\bibfnamefont {K.}~\bibnamefont {Subotic}},
  \bibinfo {author} {\bibfnamefont {V.~I.}\ \bibnamefont {Zagrebaev}}, \bibinfo
  {author} {\bibfnamefont {G.~K.}\ \bibnamefont {Vostokin}}, \bibinfo {author}
  {\bibfnamefont {M.~G.}\ \bibnamefont {Itkis}}, \bibinfo {author}
  {\bibfnamefont {R.~A.}\ \bibnamefont {Henderson}}, \bibinfo {author}
  {\bibfnamefont {J.~M.}\ \bibnamefont {Kenneally}}, \bibinfo {author}
  {\bibfnamefont {J.~H.}\ \bibnamefont {Landrum}}, \bibinfo {author}
  {\bibfnamefont {K.~J.}\ \bibnamefont {Moody}}, \bibinfo {author}
  {\bibfnamefont {D.~A.}\ \bibnamefont {Shaughnessy}}, \bibinfo {author}
  {\bibfnamefont {M.~A.}\ \bibnamefont {Stoyer}}, \bibinfo {author}
  {\bibfnamefont {N.~J.}\ \bibnamefont {Stoyer}},\ and\ \bibinfo {author}
  {\bibfnamefont {P.~A.}\ \bibnamefont {Wilk}},\ }\href
  {https://doi.org/10.1103/PhysRevC.76.011601} {\bibfield  {journal} {\bibinfo
  {journal} {Physical Review C}\ }\textbf {\bibinfo {volume} {76}},\ \bibinfo
  {pages} {011601} (\bibinfo {year} {2007})}\BibitemShut {NoStop}%
\bibitem [{\citenamefont {Oganessian}\ \emph {et~al.}(2005)\citenamefont
  {Oganessian}, \citenamefont {Utyonkov}, \citenamefont {Dmitriev},
  \citenamefont {Lobanov}, \citenamefont {Itkis}, \citenamefont {Polyakov},
  \citenamefont {Tsyganov}, \citenamefont {Mezentsev}, \citenamefont {Yeremin},
  \citenamefont {Voinov}, \citenamefont {Sokol}, \citenamefont {Gulbekian},
  \citenamefont {Bogomolov}, \citenamefont {Iliev}, \citenamefont {Subbotin},
  \citenamefont {Sukhov}, \citenamefont {Buklanov}, \citenamefont {Shishkin},
  \citenamefont {Chepygin}, \citenamefont {Vostokin}, \citenamefont {Aksenov},
  \citenamefont {Hussonnois}, \citenamefont {Subotic}, \citenamefont
  {Zagrebaev}, \citenamefont {Moody}, \citenamefont {Patin}, \citenamefont
  {Wild}, \citenamefont {Stoyer}, \citenamefont {Stoyer}, \citenamefont
  {Shaughnessy}, \citenamefont {Kenneally}, \citenamefont {Wilk}, \citenamefont
  {Lougheed}, \citenamefont {Gäggeler}, \citenamefont {Schumann},
  \citenamefont {Bruchertseifer},\ and\ \citenamefont {Eichler}}]{RN312}%
  \BibitemOpen
  \bibfield  {author} {\bibinfo {author} {\bibfnamefont {Y.~T.}\ \bibnamefont
  {Oganessian}}, \bibinfo {author} {\bibfnamefont {V.~K.}\ \bibnamefont
  {Utyonkov}}, \bibinfo {author} {\bibfnamefont {S.~N.}\ \bibnamefont
  {Dmitriev}}, \bibinfo {author} {\bibfnamefont {Y.~V.}\ \bibnamefont
  {Lobanov}}, \bibinfo {author} {\bibfnamefont {M.~G.}\ \bibnamefont {Itkis}},
  \bibinfo {author} {\bibfnamefont {A.~N.}\ \bibnamefont {Polyakov}}, \bibinfo
  {author} {\bibfnamefont {Y.~S.}\ \bibnamefont {Tsyganov}}, \bibinfo {author}
  {\bibfnamefont {A.~N.}\ \bibnamefont {Mezentsev}}, \bibinfo {author}
  {\bibfnamefont {A.~V.}\ \bibnamefont {Yeremin}}, \bibinfo {author}
  {\bibfnamefont {A.~A.}\ \bibnamefont {Voinov}}, \bibinfo {author}
  {\bibfnamefont {E.~A.}\ \bibnamefont {Sokol}}, \bibinfo {author}
  {\bibfnamefont {G.~G.}\ \bibnamefont {Gulbekian}}, \bibinfo {author}
  {\bibfnamefont {S.~L.}\ \bibnamefont {Bogomolov}}, \bibinfo {author}
  {\bibfnamefont {S.}~\bibnamefont {Iliev}}, \bibinfo {author} {\bibfnamefont
  {V.~G.}\ \bibnamefont {Subbotin}}, \bibinfo {author} {\bibfnamefont {A.~M.}\
  \bibnamefont {Sukhov}}, \bibinfo {author} {\bibfnamefont {G.~V.}\
  \bibnamefont {Buklanov}}, \bibinfo {author} {\bibfnamefont {S.~V.}\
  \bibnamefont {Shishkin}}, \bibinfo {author} {\bibfnamefont {V.~I.}\
  \bibnamefont {Chepygin}}, \bibinfo {author} {\bibfnamefont {G.~K.}\
  \bibnamefont {Vostokin}}, \bibinfo {author} {\bibfnamefont {N.~V.}\
  \bibnamefont {Aksenov}}, \bibinfo {author} {\bibfnamefont {M.}~\bibnamefont
  {Hussonnois}}, \bibinfo {author} {\bibfnamefont {K.}~\bibnamefont {Subotic}},
  \bibinfo {author} {\bibfnamefont {V.~I.}\ \bibnamefont {Zagrebaev}}, \bibinfo
  {author} {\bibfnamefont {K.~J.}\ \bibnamefont {Moody}}, \bibinfo {author}
  {\bibfnamefont {J.~B.}\ \bibnamefont {Patin}}, \bibinfo {author}
  {\bibfnamefont {J.~F.}\ \bibnamefont {Wild}}, \bibinfo {author}
  {\bibfnamefont {M.~A.}\ \bibnamefont {Stoyer}}, \bibinfo {author}
  {\bibfnamefont {N.~J.}\ \bibnamefont {Stoyer}}, \bibinfo {author}
  {\bibfnamefont {D.~A.}\ \bibnamefont {Shaughnessy}}, \bibinfo {author}
  {\bibfnamefont {J.~M.}\ \bibnamefont {Kenneally}}, \bibinfo {author}
  {\bibfnamefont {P.~A.}\ \bibnamefont {Wilk}}, \bibinfo {author}
  {\bibfnamefont {R.~W.}\ \bibnamefont {Lougheed}}, \bibinfo {author}
  {\bibfnamefont {H.~W.}\ \bibnamefont {Gäggeler}}, \bibinfo {author}
  {\bibfnamefont {D.}~\bibnamefont {Schumann}}, \bibinfo {author}
  {\bibfnamefont {H.}~\bibnamefont {Bruchertseifer}},\ and\ \bibinfo {author}
  {\bibfnamefont {R.}~\bibnamefont {Eichler}},\ }\href
  {https://doi.org/10.1103/PhysRevC.72.034611} {\bibfield  {journal} {\bibinfo
  {journal} {Physical Review C}\ }\textbf {\bibinfo {volume} {72}},\ \bibinfo
  {pages} {034611} (\bibinfo {year} {2005})}\BibitemShut {NoStop}%
\bibitem [{\citenamefont {Oganessian}\ \emph {et~al.}(2006)\citenamefont
  {Oganessian}, \citenamefont {Utyonkov}, \citenamefont {Lobanov},
  \citenamefont {Abdullin}, \citenamefont {Polyakov}, \citenamefont {Sagaidak},
  \citenamefont {Shirokovsky}, \citenamefont {Tsyganov}, \citenamefont
  {Voinov}, \citenamefont {Gulbekian}, \citenamefont {Bogomolov}, \citenamefont
  {Gikal}, \citenamefont {Mezentsev}, \citenamefont {Iliev}, \citenamefont
  {Subbotin}, \citenamefont {Sukhov}, \citenamefont {Subotic}, \citenamefont
  {Zagrebaev}, \citenamefont {Vostokin}, \citenamefont {Itkis}, \citenamefont
  {Moody}, \citenamefont {Patin}, \citenamefont {Shaughnessy}, \citenamefont
  {Stoyer}, \citenamefont {Stoyer}, \citenamefont {Wilk}, \citenamefont
  {Kenneally}, \citenamefont {Landrum}, \citenamefont {Wild},\ and\
  \citenamefont {Lougheed}}]{RN313}%
  \BibitemOpen
  \bibfield  {author} {\bibinfo {author} {\bibfnamefont {Y.~T.}\ \bibnamefont
  {Oganessian}}, \bibinfo {author} {\bibfnamefont {V.~K.}\ \bibnamefont
  {Utyonkov}}, \bibinfo {author} {\bibfnamefont {Y.~V.}\ \bibnamefont
  {Lobanov}}, \bibinfo {author} {\bibfnamefont {F.~S.}\ \bibnamefont
  {Abdullin}}, \bibinfo {author} {\bibfnamefont {A.~N.}\ \bibnamefont
  {Polyakov}}, \bibinfo {author} {\bibfnamefont {R.~N.}\ \bibnamefont
  {Sagaidak}}, \bibinfo {author} {\bibfnamefont {I.~V.}\ \bibnamefont
  {Shirokovsky}}, \bibinfo {author} {\bibfnamefont {Y.~S.}\ \bibnamefont
  {Tsyganov}}, \bibinfo {author} {\bibfnamefont {A.~A.}\ \bibnamefont
  {Voinov}}, \bibinfo {author} {\bibfnamefont {G.~G.}\ \bibnamefont
  {Gulbekian}}, \bibinfo {author} {\bibfnamefont {S.~L.}\ \bibnamefont
  {Bogomolov}}, \bibinfo {author} {\bibfnamefont {B.~N.}\ \bibnamefont
  {Gikal}}, \bibinfo {author} {\bibfnamefont {A.~N.}\ \bibnamefont
  {Mezentsev}}, \bibinfo {author} {\bibfnamefont {S.}~\bibnamefont {Iliev}},
  \bibinfo {author} {\bibfnamefont {V.~G.}\ \bibnamefont {Subbotin}}, \bibinfo
  {author} {\bibfnamefont {A.~M.}\ \bibnamefont {Sukhov}}, \bibinfo {author}
  {\bibfnamefont {K.}~\bibnamefont {Subotic}}, \bibinfo {author} {\bibfnamefont
  {V.~I.}\ \bibnamefont {Zagrebaev}}, \bibinfo {author} {\bibfnamefont {G.~K.}\
  \bibnamefont {Vostokin}}, \bibinfo {author} {\bibfnamefont {M.~G.}\
  \bibnamefont {Itkis}}, \bibinfo {author} {\bibfnamefont {K.~J.}\ \bibnamefont
  {Moody}}, \bibinfo {author} {\bibfnamefont {J.~B.}\ \bibnamefont {Patin}},
  \bibinfo {author} {\bibfnamefont {D.~A.}\ \bibnamefont {Shaughnessy}},
  \bibinfo {author} {\bibfnamefont {M.~A.}\ \bibnamefont {Stoyer}}, \bibinfo
  {author} {\bibfnamefont {N.~J.}\ \bibnamefont {Stoyer}}, \bibinfo {author}
  {\bibfnamefont {P.~A.}\ \bibnamefont {Wilk}}, \bibinfo {author}
  {\bibfnamefont {J.~M.}\ \bibnamefont {Kenneally}}, \bibinfo {author}
  {\bibfnamefont {J.~H.}\ \bibnamefont {Landrum}}, \bibinfo {author}
  {\bibfnamefont {J.~F.}\ \bibnamefont {Wild}},\ and\ \bibinfo {author}
  {\bibfnamefont {R.~W.}\ \bibnamefont {Lougheed}},\ }\href
  {https://doi.org/10.1103/PhysRevC.74.044602} {\bibfield  {journal} {\bibinfo
  {journal} {Physical Review C}\ }\textbf {\bibinfo {volume} {74}},\ \bibinfo
  {pages} {044602} (\bibinfo {year} {2006})}\BibitemShut {NoStop}%
\bibitem [{\citenamefont {Oganessian}\ \emph {et~al.}(2011)\citenamefont
  {Oganessian}, \citenamefont {Abdullin}, \citenamefont {Bailey}, \citenamefont
  {Benker}, \citenamefont {Bennett}, \citenamefont {Dmitriev}, \citenamefont
  {Ezold}, \citenamefont {Hamilton}, \citenamefont {Henderson}, \citenamefont
  {Itkis}, \citenamefont {Lobanov}, \citenamefont {Mezentsev}, \citenamefont
  {Moody}, \citenamefont {Nelson}, \citenamefont {Polyakov}, \citenamefont
  {Porter}, \citenamefont {Ramayya}, \citenamefont {Riley}, \citenamefont
  {Roberto}, \citenamefont {Ryabinin}, \citenamefont {Rykaczewski},
  \citenamefont {Sagaidak}, \citenamefont {Shaughnessy}, \citenamefont
  {Shirokovsky}, \citenamefont {Stoyer}, \citenamefont {Subbotin},
  \citenamefont {Sudowe}, \citenamefont {Sukhov}, \citenamefont {Taylor},
  \citenamefont {Tsyganov}, \citenamefont {Utyonkov}, \citenamefont {Voinov},
  \citenamefont {Vostokin},\ and\ \citenamefont {Wilk}}]{RN314}%
  \BibitemOpen
  \bibfield  {author} {\bibinfo {author} {\bibfnamefont {Y.~T.}\ \bibnamefont
  {Oganessian}}, \bibinfo {author} {\bibfnamefont {F.~S.}\ \bibnamefont
  {Abdullin}}, \bibinfo {author} {\bibfnamefont {P.~D.}\ \bibnamefont
  {Bailey}}, \bibinfo {author} {\bibfnamefont {D.~E.}\ \bibnamefont {Benker}},
  \bibinfo {author} {\bibfnamefont {M.~E.}\ \bibnamefont {Bennett}}, \bibinfo
  {author} {\bibfnamefont {S.~N.}\ \bibnamefont {Dmitriev}}, \bibinfo {author}
  {\bibfnamefont {J.~G.}\ \bibnamefont {Ezold}}, \bibinfo {author}
  {\bibfnamefont {J.~H.}\ \bibnamefont {Hamilton}}, \bibinfo {author}
  {\bibfnamefont {R.~A.}\ \bibnamefont {Henderson}}, \bibinfo {author}
  {\bibfnamefont {M.~G.}\ \bibnamefont {Itkis}}, \bibinfo {author}
  {\bibfnamefont {Y.~V.}\ \bibnamefont {Lobanov}}, \bibinfo {author}
  {\bibfnamefont {A.~N.}\ \bibnamefont {Mezentsev}}, \bibinfo {author}
  {\bibfnamefont {K.~J.}\ \bibnamefont {Moody}}, \bibinfo {author}
  {\bibfnamefont {S.~L.}\ \bibnamefont {Nelson}}, \bibinfo {author}
  {\bibfnamefont {A.~N.}\ \bibnamefont {Polyakov}}, \bibinfo {author}
  {\bibfnamefont {C.~E.}\ \bibnamefont {Porter}}, \bibinfo {author}
  {\bibfnamefont {A.~V.}\ \bibnamefont {Ramayya}}, \bibinfo {author}
  {\bibfnamefont {F.~D.}\ \bibnamefont {Riley}}, \bibinfo {author}
  {\bibfnamefont {J.~B.}\ \bibnamefont {Roberto}}, \bibinfo {author}
  {\bibfnamefont {M.~A.}\ \bibnamefont {Ryabinin}}, \bibinfo {author}
  {\bibfnamefont {K.~P.}\ \bibnamefont {Rykaczewski}}, \bibinfo {author}
  {\bibfnamefont {R.~N.}\ \bibnamefont {Sagaidak}}, \bibinfo {author}
  {\bibfnamefont {D.~A.}\ \bibnamefont {Shaughnessy}}, \bibinfo {author}
  {\bibfnamefont {I.~V.}\ \bibnamefont {Shirokovsky}}, \bibinfo {author}
  {\bibfnamefont {M.~A.}\ \bibnamefont {Stoyer}}, \bibinfo {author}
  {\bibfnamefont {V.~G.}\ \bibnamefont {Subbotin}}, \bibinfo {author}
  {\bibfnamefont {R.}~\bibnamefont {Sudowe}}, \bibinfo {author} {\bibfnamefont
  {A.~M.}\ \bibnamefont {Sukhov}}, \bibinfo {author} {\bibfnamefont
  {R.}~\bibnamefont {Taylor}}, \bibinfo {author} {\bibfnamefont {Y.~S.}\
  \bibnamefont {Tsyganov}}, \bibinfo {author} {\bibfnamefont {V.~K.}\
  \bibnamefont {Utyonkov}}, \bibinfo {author} {\bibfnamefont {A.~A.}\
  \bibnamefont {Voinov}}, \bibinfo {author} {\bibfnamefont {G.~K.}\
  \bibnamefont {Vostokin}},\ and\ \bibinfo {author} {\bibfnamefont {P.~A.}\
  \bibnamefont {Wilk}},\ }\href {https://doi.org/10.1103/PhysRevC.83.054315}
  {\bibfield  {journal} {\bibinfo  {journal} {Physical Review C}\ }\textbf
  {\bibinfo {volume} {83}},\ \bibinfo {pages} {054315} (\bibinfo {year}
  {2011})}\BibitemShut {NoStop}%
\bibitem [{\citenamefont {Oganessian}\ \emph {et~al.}(2013)\citenamefont
  {Oganessian}, \citenamefont {Abdullin}, \citenamefont {Alexander},
  \citenamefont {Binder}, \citenamefont {Boll}, \citenamefont {Dmitriev},
  \citenamefont {Ezold}, \citenamefont {Felker}, \citenamefont {Gostic},
  \citenamefont {Grzywacz}, \citenamefont {Hamilton}, \citenamefont
  {Henderson}, \citenamefont {Itkis}, \citenamefont {Miernik}, \citenamefont
  {Miller}, \citenamefont {Moody}, \citenamefont {Polyakov}, \citenamefont
  {Ramayya}, \citenamefont {Roberto}, \citenamefont {Ryabinin}, \citenamefont
  {Rykaczewski}, \citenamefont {Sagaidak}, \citenamefont {Shaughnessy},
  \citenamefont {Shirokovsky}, \citenamefont {Shumeiko}, \citenamefont
  {Stoyer}, \citenamefont {Stoyer}, \citenamefont {Subbotin}, \citenamefont
  {Sukhov}, \citenamefont {Tsyganov}, \citenamefont {Utyonkov}, \citenamefont
  {Voinov},\ and\ \citenamefont {Vostokin}}]{RN315}%
  \BibitemOpen
  \bibfield  {author} {\bibinfo {author} {\bibfnamefont {Y.~T.}\ \bibnamefont
  {Oganessian}}, \bibinfo {author} {\bibfnamefont {F.~S.}\ \bibnamefont
  {Abdullin}}, \bibinfo {author} {\bibfnamefont {C.}~\bibnamefont {Alexander}},
  \bibinfo {author} {\bibfnamefont {J.}~\bibnamefont {Binder}}, \bibinfo
  {author} {\bibfnamefont {R.~A.}\ \bibnamefont {Boll}}, \bibinfo {author}
  {\bibfnamefont {S.~N.}\ \bibnamefont {Dmitriev}}, \bibinfo {author}
  {\bibfnamefont {J.}~\bibnamefont {Ezold}}, \bibinfo {author} {\bibfnamefont
  {K.}~\bibnamefont {Felker}}, \bibinfo {author} {\bibfnamefont {J.~M.}\
  \bibnamefont {Gostic}}, \bibinfo {author} {\bibfnamefont {R.~K.}\
  \bibnamefont {Grzywacz}}, \bibinfo {author} {\bibfnamefont {J.~H.}\
  \bibnamefont {Hamilton}}, \bibinfo {author} {\bibfnamefont {R.~A.}\
  \bibnamefont {Henderson}}, \bibinfo {author} {\bibfnamefont {M.~G.}\
  \bibnamefont {Itkis}}, \bibinfo {author} {\bibfnamefont {K.}~\bibnamefont
  {Miernik}}, \bibinfo {author} {\bibfnamefont {D.}~\bibnamefont {Miller}},
  \bibinfo {author} {\bibfnamefont {K.~J.}\ \bibnamefont {Moody}}, \bibinfo
  {author} {\bibfnamefont {A.~N.}\ \bibnamefont {Polyakov}}, \bibinfo {author}
  {\bibfnamefont {A.~V.}\ \bibnamefont {Ramayya}}, \bibinfo {author}
  {\bibfnamefont {J.~B.}\ \bibnamefont {Roberto}}, \bibinfo {author}
  {\bibfnamefont {M.~A.}\ \bibnamefont {Ryabinin}}, \bibinfo {author}
  {\bibfnamefont {K.~P.}\ \bibnamefont {Rykaczewski}}, \bibinfo {author}
  {\bibfnamefont {R.~N.}\ \bibnamefont {Sagaidak}}, \bibinfo {author}
  {\bibfnamefont {D.~A.}\ \bibnamefont {Shaughnessy}}, \bibinfo {author}
  {\bibfnamefont {I.~V.}\ \bibnamefont {Shirokovsky}}, \bibinfo {author}
  {\bibfnamefont {M.~V.}\ \bibnamefont {Shumeiko}}, \bibinfo {author}
  {\bibfnamefont {M.~A.}\ \bibnamefont {Stoyer}}, \bibinfo {author}
  {\bibfnamefont {N.~J.}\ \bibnamefont {Stoyer}}, \bibinfo {author}
  {\bibfnamefont {V.~G.}\ \bibnamefont {Subbotin}}, \bibinfo {author}
  {\bibfnamefont {A.~M.}\ \bibnamefont {Sukhov}}, \bibinfo {author}
  {\bibfnamefont {Y.~S.}\ \bibnamefont {Tsyganov}}, \bibinfo {author}
  {\bibfnamefont {V.~K.}\ \bibnamefont {Utyonkov}}, \bibinfo {author}
  {\bibfnamefont {A.~A.}\ \bibnamefont {Voinov}},\ and\ \bibinfo {author}
  {\bibfnamefont {G.~K.}\ \bibnamefont {Vostokin}},\ }\href
  {https://doi.org/10.1103/PhysRevC.87.054621} {\bibfield  {journal} {\bibinfo
  {journal} {Physical Review C}\ }\textbf {\bibinfo {volume} {87}},\ \bibinfo
  {pages} {054621} (\bibinfo {year} {2013})}\BibitemShut {NoStop}%
\bibitem [{\citenamefont {Gates}\ \emph {et~al.}(2011)\citenamefont {Gates},
  \citenamefont {Düllmann}, \citenamefont {Schädel}, \citenamefont
  {Yakushev}, \citenamefont {Türler}, \citenamefont {Eberhardt}, \citenamefont
  {Kratz}, \citenamefont {Ackermann}, \citenamefont {Andersson}, \citenamefont
  {Block}, \citenamefont {Brüchle}, \citenamefont {Dvorak}, \citenamefont
  {Essel}, \citenamefont {Ellison}, \citenamefont {Even}, \citenamefont
  {Forsberg}, \citenamefont {Gellanki}, \citenamefont {Gorshkov}, \citenamefont
  {Graeger}, \citenamefont {Gregorich}, \citenamefont {Hartmann}, \citenamefont
  {Herzberg}, \citenamefont {Heßberger}, \citenamefont {Hild}, \citenamefont
  {Hübner}, \citenamefont {Jäger}, \citenamefont {Khuyagbaatar},
  \citenamefont {Kindler}, \citenamefont {Krier}, \citenamefont {Kurz},
  \citenamefont {Lahiri}, \citenamefont {Liebe}, \citenamefont {Lommel},
  \citenamefont {Maiti}, \citenamefont {Nitsche}, \citenamefont {Omtvedt},
  \citenamefont {Parr}, \citenamefont {Rudolph}, \citenamefont {Runke},
  \citenamefont {Schaffner}, \citenamefont {Schausten}, \citenamefont
  {Schimpf}, \citenamefont {Semchenkov}, \citenamefont {Steiner}, \citenamefont
  {Thörle-Pospiech}, \citenamefont {Uusitalo}, \citenamefont {Wegrzecki},\
  and\ \citenamefont {Wiehl}}]{RN316}%
  \BibitemOpen
  \bibfield  {author} {\bibinfo {author} {\bibfnamefont {J.~M.}\ \bibnamefont
  {Gates}}, \bibinfo {author} {\bibfnamefont {C.~E.}\ \bibnamefont
  {Düllmann}}, \bibinfo {author} {\bibfnamefont {M.}~\bibnamefont {Schädel}},
  \bibinfo {author} {\bibfnamefont {A.}~\bibnamefont {Yakushev}}, \bibinfo
  {author} {\bibfnamefont {A.}~\bibnamefont {Türler}}, \bibinfo {author}
  {\bibfnamefont {K.}~\bibnamefont {Eberhardt}}, \bibinfo {author}
  {\bibfnamefont {J.~V.}\ \bibnamefont {Kratz}}, \bibinfo {author}
  {\bibfnamefont {D.}~\bibnamefont {Ackermann}}, \bibinfo {author}
  {\bibfnamefont {L.~L.}\ \bibnamefont {Andersson}}, \bibinfo {author}
  {\bibfnamefont {M.}~\bibnamefont {Block}}, \bibinfo {author} {\bibfnamefont
  {W.}~\bibnamefont {Brüchle}}, \bibinfo {author} {\bibfnamefont
  {J.}~\bibnamefont {Dvorak}}, \bibinfo {author} {\bibfnamefont {H.~G.}\
  \bibnamefont {Essel}}, \bibinfo {author} {\bibfnamefont {P.~A.}\ \bibnamefont
  {Ellison}}, \bibinfo {author} {\bibfnamefont {J.}~\bibnamefont {Even}},
  \bibinfo {author} {\bibfnamefont {U.}~\bibnamefont {Forsberg}}, \bibinfo
  {author} {\bibfnamefont {J.}~\bibnamefont {Gellanki}}, \bibinfo {author}
  {\bibfnamefont {A.}~\bibnamefont {Gorshkov}}, \bibinfo {author}
  {\bibfnamefont {R.}~\bibnamefont {Graeger}}, \bibinfo {author} {\bibfnamefont
  {K.~E.}\ \bibnamefont {Gregorich}}, \bibinfo {author} {\bibfnamefont
  {W.}~\bibnamefont {Hartmann}}, \bibinfo {author} {\bibfnamefont {R.~D.}\
  \bibnamefont {Herzberg}}, \bibinfo {author} {\bibfnamefont {F.~P.}\
  \bibnamefont {Heßberger}}, \bibinfo {author} {\bibfnamefont
  {D.}~\bibnamefont {Hild}}, \bibinfo {author} {\bibfnamefont {A.}~\bibnamefont
  {Hübner}}, \bibinfo {author} {\bibfnamefont {E.}~\bibnamefont {Jäger}},
  \bibinfo {author} {\bibfnamefont {J.}~\bibnamefont {Khuyagbaatar}}, \bibinfo
  {author} {\bibfnamefont {B.}~\bibnamefont {Kindler}}, \bibinfo {author}
  {\bibfnamefont {J.}~\bibnamefont {Krier}}, \bibinfo {author} {\bibfnamefont
  {N.}~\bibnamefont {Kurz}}, \bibinfo {author} {\bibfnamefont {S.}~\bibnamefont
  {Lahiri}}, \bibinfo {author} {\bibfnamefont {D.}~\bibnamefont {Liebe}},
  \bibinfo {author} {\bibfnamefont {B.}~\bibnamefont {Lommel}}, \bibinfo
  {author} {\bibfnamefont {M.}~\bibnamefont {Maiti}}, \bibinfo {author}
  {\bibfnamefont {H.}~\bibnamefont {Nitsche}}, \bibinfo {author} {\bibfnamefont
  {J.~P.}\ \bibnamefont {Omtvedt}}, \bibinfo {author} {\bibfnamefont
  {E.}~\bibnamefont {Parr}}, \bibinfo {author} {\bibfnamefont {D.}~\bibnamefont
  {Rudolph}}, \bibinfo {author} {\bibfnamefont {J.}~\bibnamefont {Runke}},
  \bibinfo {author} {\bibfnamefont {H.}~\bibnamefont {Schaffner}}, \bibinfo
  {author} {\bibfnamefont {B.}~\bibnamefont {Schausten}}, \bibinfo {author}
  {\bibfnamefont {E.}~\bibnamefont {Schimpf}}, \bibinfo {author} {\bibfnamefont
  {A.}~\bibnamefont {Semchenkov}}, \bibinfo {author} {\bibfnamefont
  {J.}~\bibnamefont {Steiner}}, \bibinfo {author} {\bibfnamefont
  {P.}~\bibnamefont {Thörle-Pospiech}}, \bibinfo {author} {\bibfnamefont
  {J.}~\bibnamefont {Uusitalo}}, \bibinfo {author} {\bibfnamefont
  {M.}~\bibnamefont {Wegrzecki}},\ and\ \bibinfo {author} {\bibfnamefont
  {N.}~\bibnamefont {Wiehl}},\ }\href
  {https://doi.org/10.1103/PhysRevC.83.054618} {\bibfield  {journal} {\bibinfo
  {journal} {Physical Review C}\ }\textbf {\bibinfo {volume} {83}},\ \bibinfo
  {pages} {054618} (\bibinfo {year} {2011})}\BibitemShut {NoStop}%
\bibitem [{\citenamefont {Oganessian}\ \emph
  {et~al.}(2012{\natexlab{a}})\citenamefont {Oganessian}, \citenamefont
  {Abdullin}, \citenamefont {Alexander}, \citenamefont {Binder}, \citenamefont
  {Boll}, \citenamefont {Dmitriev}, \citenamefont {Ezold}, \citenamefont
  {Felker}, \citenamefont {Gostic}, \citenamefont {Grzywacz}, \citenamefont
  {Hamilton}, \citenamefont {Henderson}, \citenamefont {Itkis}, \citenamefont
  {Miernik}, \citenamefont {Miller}, \citenamefont {Moody}, \citenamefont
  {Polyakov}, \citenamefont {Ramayya}, \citenamefont {Roberto}, \citenamefont
  {Ryabinin}, \citenamefont {Rykaczewski}, \citenamefont {Sagaidak},
  \citenamefont {Shaughnessy}, \citenamefont {Shirokovsky}, \citenamefont
  {Shumeiko}, \citenamefont {Stoyer}, \citenamefont {Stoyer}, \citenamefont
  {Subbotin}, \citenamefont {Sukhov}, \citenamefont {Tsyganov}, \citenamefont
  {Utyonkov}, \citenamefont {Voinov},\ and\ \citenamefont {Vostokin}}]{RN317}%
  \BibitemOpen
  \bibfield  {author} {\bibinfo {author} {\bibfnamefont {Y.~T.}\ \bibnamefont
  {Oganessian}}, \bibinfo {author} {\bibfnamefont {F.~S.}\ \bibnamefont
  {Abdullin}}, \bibinfo {author} {\bibfnamefont {C.}~\bibnamefont {Alexander}},
  \bibinfo {author} {\bibfnamefont {J.}~\bibnamefont {Binder}}, \bibinfo
  {author} {\bibfnamefont {R.~A.}\ \bibnamefont {Boll}}, \bibinfo {author}
  {\bibfnamefont {S.~N.}\ \bibnamefont {Dmitriev}}, \bibinfo {author}
  {\bibfnamefont {J.}~\bibnamefont {Ezold}}, \bibinfo {author} {\bibfnamefont
  {K.}~\bibnamefont {Felker}}, \bibinfo {author} {\bibfnamefont {J.~M.}\
  \bibnamefont {Gostic}}, \bibinfo {author} {\bibfnamefont {R.~K.}\
  \bibnamefont {Grzywacz}}, \bibinfo {author} {\bibfnamefont {J.~H.}\
  \bibnamefont {Hamilton}}, \bibinfo {author} {\bibfnamefont {R.~A.}\
  \bibnamefont {Henderson}}, \bibinfo {author} {\bibfnamefont {M.~G.}\
  \bibnamefont {Itkis}}, \bibinfo {author} {\bibfnamefont {K.}~\bibnamefont
  {Miernik}}, \bibinfo {author} {\bibfnamefont {D.}~\bibnamefont {Miller}},
  \bibinfo {author} {\bibfnamefont {K.~J.}\ \bibnamefont {Moody}}, \bibinfo
  {author} {\bibfnamefont {A.~N.}\ \bibnamefont {Polyakov}}, \bibinfo {author}
  {\bibfnamefont {A.~V.}\ \bibnamefont {Ramayya}}, \bibinfo {author}
  {\bibfnamefont {J.~B.}\ \bibnamefont {Roberto}}, \bibinfo {author}
  {\bibfnamefont {M.~A.}\ \bibnamefont {Ryabinin}}, \bibinfo {author}
  {\bibfnamefont {K.~P.}\ \bibnamefont {Rykaczewski}}, \bibinfo {author}
  {\bibfnamefont {R.~N.}\ \bibnamefont {Sagaidak}}, \bibinfo {author}
  {\bibfnamefont {D.~A.}\ \bibnamefont {Shaughnessy}}, \bibinfo {author}
  {\bibfnamefont {I.~V.}\ \bibnamefont {Shirokovsky}}, \bibinfo {author}
  {\bibfnamefont {M.~V.}\ \bibnamefont {Shumeiko}}, \bibinfo {author}
  {\bibfnamefont {M.~A.}\ \bibnamefont {Stoyer}}, \bibinfo {author}
  {\bibfnamefont {N.~J.}\ \bibnamefont {Stoyer}}, \bibinfo {author}
  {\bibfnamefont {V.~G.}\ \bibnamefont {Subbotin}}, \bibinfo {author}
  {\bibfnamefont {A.~M.}\ \bibnamefont {Sukhov}}, \bibinfo {author}
  {\bibfnamefont {Y.~S.}\ \bibnamefont {Tsyganov}}, \bibinfo {author}
  {\bibfnamefont {V.~K.}\ \bibnamefont {Utyonkov}}, \bibinfo {author}
  {\bibfnamefont {A.~A.}\ \bibnamefont {Voinov}},\ and\ \bibinfo {author}
  {\bibfnamefont {G.~K.}\ \bibnamefont {Vostokin}},\ }\href
  {https://doi.org/10.1103/PhysRevLett.109.162501} {\bibfield  {journal}
  {\bibinfo  {journal} {Physical Review Letters}\ }\textbf {\bibinfo {volume}
  {109}},\ \bibinfo {pages} {162501} (\bibinfo {year}
  {2012}{\natexlab{a}})}\BibitemShut {NoStop}%
\bibitem [{\citenamefont {Ellison}\ \emph {et~al.}(2010)\citenamefont
  {Ellison}, \citenamefont {Gregorich}, \citenamefont {Berryman}, \citenamefont
  {Bleuel}, \citenamefont {Clark}, \citenamefont {Dragojević}, \citenamefont
  {Dvorak}, \citenamefont {Fallon}, \citenamefont {Fineman-Sotomayor},
  \citenamefont {Gates}, \citenamefont {Gothe}, \citenamefont {Lee},
  \citenamefont {Loveland}, \citenamefont {McLaughlin}, \citenamefont
  {Paschalis}, \citenamefont {Petri}, \citenamefont {Qian}, \citenamefont
  {Stavsetra}, \citenamefont {Wiedeking},\ and\ \citenamefont
  {Nitsche}}]{RN318}%
  \BibitemOpen
  \bibfield  {author} {\bibinfo {author} {\bibfnamefont {P.~A.}\ \bibnamefont
  {Ellison}}, \bibinfo {author} {\bibfnamefont {K.~E.}\ \bibnamefont
  {Gregorich}}, \bibinfo {author} {\bibfnamefont {J.~S.}\ \bibnamefont
  {Berryman}}, \bibinfo {author} {\bibfnamefont {D.~L.}\ \bibnamefont
  {Bleuel}}, \bibinfo {author} {\bibfnamefont {R.~M.}\ \bibnamefont {Clark}},
  \bibinfo {author} {\bibfnamefont {I.}~\bibnamefont {Dragojević}}, \bibinfo
  {author} {\bibfnamefont {J.}~\bibnamefont {Dvorak}}, \bibinfo {author}
  {\bibfnamefont {P.}~\bibnamefont {Fallon}}, \bibinfo {author} {\bibfnamefont
  {C.}~\bibnamefont {Fineman-Sotomayor}}, \bibinfo {author} {\bibfnamefont
  {J.~M.}\ \bibnamefont {Gates}}, \bibinfo {author} {\bibfnamefont {O.~R.}\
  \bibnamefont {Gothe}}, \bibinfo {author} {\bibfnamefont {I.~Y.}\ \bibnamefont
  {Lee}}, \bibinfo {author} {\bibfnamefont {W.~D.}\ \bibnamefont {Loveland}},
  \bibinfo {author} {\bibfnamefont {J.~P.}\ \bibnamefont {McLaughlin}},
  \bibinfo {author} {\bibfnamefont {S.}~\bibnamefont {Paschalis}}, \bibinfo
  {author} {\bibfnamefont {M.}~\bibnamefont {Petri}}, \bibinfo {author}
  {\bibfnamefont {J.}~\bibnamefont {Qian}}, \bibinfo {author} {\bibfnamefont
  {L.}~\bibnamefont {Stavsetra}}, \bibinfo {author} {\bibfnamefont
  {M.}~\bibnamefont {Wiedeking}},\ and\ \bibinfo {author} {\bibfnamefont
  {H.}~\bibnamefont {Nitsche}},\ }\href
  {https://doi.org/10.1103/PhysRevLett.105.182701} {\bibfield  {journal}
  {\bibinfo  {journal} {Physical Review Letters}\ }\textbf {\bibinfo {volume}
  {105}},\ \bibinfo {pages} {182701} (\bibinfo {year} {2010})}\BibitemShut
  {NoStop}%
\bibitem [{\citenamefont {Stavsetra}\ \emph {et~al.}(2009)\citenamefont
  {Stavsetra}, \citenamefont {Gregorich}, \citenamefont {Dvorak}, \citenamefont
  {Ellison}, \citenamefont {Dragojević}, \citenamefont {Garcia},\ and\
  \citenamefont {Nitsche}}]{RN319}%
  \BibitemOpen
  \bibfield  {author} {\bibinfo {author} {\bibfnamefont {L.}~\bibnamefont
  {Stavsetra}}, \bibinfo {author} {\bibfnamefont {K.~E.}\ \bibnamefont
  {Gregorich}}, \bibinfo {author} {\bibfnamefont {J.}~\bibnamefont {Dvorak}},
  \bibinfo {author} {\bibfnamefont {P.~A.}\ \bibnamefont {Ellison}}, \bibinfo
  {author} {\bibfnamefont {I.}~\bibnamefont {Dragojević}}, \bibinfo {author}
  {\bibfnamefont {M.~A.}\ \bibnamefont {Garcia}},\ and\ \bibinfo {author}
  {\bibfnamefont {H.}~\bibnamefont {Nitsche}},\ }\href
  {https://doi.org/10.1103/PhysRevLett.103.132502} {\bibfield  {journal}
  {\bibinfo  {journal} {Physical Review Letters}\ }\textbf {\bibinfo {volume}
  {103}},\ \bibinfo {pages} {132502} (\bibinfo {year} {2009})}\BibitemShut
  {NoStop}%
\bibitem [{\citenamefont {Düllmann}\ \emph {et~al.}(2010)\citenamefont
  {Düllmann}, \citenamefont {Schädel}, \citenamefont {Yakushev},
  \citenamefont {Türler}, \citenamefont {Eberhardt}, \citenamefont {Kratz},
  \citenamefont {Ackermann}, \citenamefont {Andersson}, \citenamefont {Block},
  \citenamefont {Brüchle}, \citenamefont {Dvorak}, \citenamefont {Essel},
  \citenamefont {Ellison}, \citenamefont {Even}, \citenamefont {Gates},
  \citenamefont {Gorshkov}, \citenamefont {Graeger}, \citenamefont {Gregorich},
  \citenamefont {Hartmann}, \citenamefont {Herzberg}, \citenamefont
  {Heßberger}, \citenamefont {Hild}, \citenamefont {Hübner}, \citenamefont
  {Jäger}, \citenamefont {Khuyagbaatar}, \citenamefont {Kindler},
  \citenamefont {Krier}, \citenamefont {Kurz}, \citenamefont {Lahiri},
  \citenamefont {Liebe}, \citenamefont {Lommel}, \citenamefont {Maiti},
  \citenamefont {Nitsche}, \citenamefont {Omtvedt}, \citenamefont {Parr},
  \citenamefont {Rudolph}, \citenamefont {Runke}, \citenamefont {Schausten},
  \citenamefont {Schimpf}, \citenamefont {Semchenkov}, \citenamefont {Steiner},
  \citenamefont {Thörle-Pospiech}, \citenamefont {Uusitalo}, \citenamefont
  {Wegrzecki},\ and\ \citenamefont {Wiehl}}]{RN320}%
  \BibitemOpen
  \bibfield  {author} {\bibinfo {author} {\bibfnamefont {C.~E.}\ \bibnamefont
  {Düllmann}}, \bibinfo {author} {\bibfnamefont {M.}~\bibnamefont {Schädel}},
  \bibinfo {author} {\bibfnamefont {A.}~\bibnamefont {Yakushev}}, \bibinfo
  {author} {\bibfnamefont {A.}~\bibnamefont {Türler}}, \bibinfo {author}
  {\bibfnamefont {K.}~\bibnamefont {Eberhardt}}, \bibinfo {author}
  {\bibfnamefont {J.~V.}\ \bibnamefont {Kratz}}, \bibinfo {author}
  {\bibfnamefont {D.}~\bibnamefont {Ackermann}}, \bibinfo {author}
  {\bibfnamefont {L.~L.}\ \bibnamefont {Andersson}}, \bibinfo {author}
  {\bibfnamefont {M.}~\bibnamefont {Block}}, \bibinfo {author} {\bibfnamefont
  {W.}~\bibnamefont {Brüchle}}, \bibinfo {author} {\bibfnamefont
  {J.}~\bibnamefont {Dvorak}}, \bibinfo {author} {\bibfnamefont {H.~G.}\
  \bibnamefont {Essel}}, \bibinfo {author} {\bibfnamefont {P.~A.}\ \bibnamefont
  {Ellison}}, \bibinfo {author} {\bibfnamefont {J.}~\bibnamefont {Even}},
  \bibinfo {author} {\bibfnamefont {J.~M.}\ \bibnamefont {Gates}}, \bibinfo
  {author} {\bibfnamefont {A.}~\bibnamefont {Gorshkov}}, \bibinfo {author}
  {\bibfnamefont {R.}~\bibnamefont {Graeger}}, \bibinfo {author} {\bibfnamefont
  {K.~E.}\ \bibnamefont {Gregorich}}, \bibinfo {author} {\bibfnamefont
  {W.}~\bibnamefont {Hartmann}}, \bibinfo {author} {\bibfnamefont {R.~D.}\
  \bibnamefont {Herzberg}}, \bibinfo {author} {\bibfnamefont {F.~P.}\
  \bibnamefont {Heßberger}}, \bibinfo {author} {\bibfnamefont
  {D.}~\bibnamefont {Hild}}, \bibinfo {author} {\bibfnamefont {A.}~\bibnamefont
  {Hübner}}, \bibinfo {author} {\bibfnamefont {E.}~\bibnamefont {Jäger}},
  \bibinfo {author} {\bibfnamefont {J.}~\bibnamefont {Khuyagbaatar}}, \bibinfo
  {author} {\bibfnamefont {B.}~\bibnamefont {Kindler}}, \bibinfo {author}
  {\bibfnamefont {J.}~\bibnamefont {Krier}}, \bibinfo {author} {\bibfnamefont
  {N.}~\bibnamefont {Kurz}}, \bibinfo {author} {\bibfnamefont {S.}~\bibnamefont
  {Lahiri}}, \bibinfo {author} {\bibfnamefont {D.}~\bibnamefont {Liebe}},
  \bibinfo {author} {\bibfnamefont {B.}~\bibnamefont {Lommel}}, \bibinfo
  {author} {\bibfnamefont {M.}~\bibnamefont {Maiti}}, \bibinfo {author}
  {\bibfnamefont {H.}~\bibnamefont {Nitsche}}, \bibinfo {author} {\bibfnamefont
  {J.~P.}\ \bibnamefont {Omtvedt}}, \bibinfo {author} {\bibfnamefont
  {E.}~\bibnamefont {Parr}}, \bibinfo {author} {\bibfnamefont {D.}~\bibnamefont
  {Rudolph}}, \bibinfo {author} {\bibfnamefont {J.}~\bibnamefont {Runke}},
  \bibinfo {author} {\bibfnamefont {B.}~\bibnamefont {Schausten}}, \bibinfo
  {author} {\bibfnamefont {E.}~\bibnamefont {Schimpf}}, \bibinfo {author}
  {\bibfnamefont {A.}~\bibnamefont {Semchenkov}}, \bibinfo {author}
  {\bibfnamefont {J.}~\bibnamefont {Steiner}}, \bibinfo {author} {\bibfnamefont
  {P.}~\bibnamefont {Thörle-Pospiech}}, \bibinfo {author} {\bibfnamefont
  {J.}~\bibnamefont {Uusitalo}}, \bibinfo {author} {\bibfnamefont
  {M.}~\bibnamefont {Wegrzecki}},\ and\ \bibinfo {author} {\bibfnamefont
  {N.}~\bibnamefont {Wiehl}},\ }\href
  {https://doi.org/10.1103/PhysRevLett.104.252701} {\bibfield  {journal}
  {\bibinfo  {journal} {Physical Review Letters}\ }\textbf {\bibinfo {volume}
  {104}},\ \bibinfo {pages} {252701} (\bibinfo {year} {2010})}\BibitemShut
  {NoStop}%
\bibitem [{\citenamefont {Hofmann}\ \emph {et~al.}(2007)\citenamefont
  {Hofmann}, \citenamefont {Ackermann}, \citenamefont {Antalic}, \citenamefont
  {Burkhard}, \citenamefont {Comas}, \citenamefont {Dressler}, \citenamefont
  {Gan}, \citenamefont {Heinz}, \citenamefont {Heredia}, \citenamefont
  {Heßberger}, \citenamefont {Khuyagbaatar}, \citenamefont {Kindler},
  \citenamefont {Kojouharov}, \citenamefont {Kuusiniemi}, \citenamefont
  {Leino}, \citenamefont {Lommel}, \citenamefont {Mann}, \citenamefont
  {Münzenberg}, \citenamefont {Nishio}, \citenamefont {Popeko}, \citenamefont
  {Saro}, \citenamefont {Schött}, \citenamefont {Streicher}, \citenamefont
  {Sulignano}, \citenamefont {Uusitalo}, \citenamefont {Venhart},\ and\
  \citenamefont {Yeremin}}]{RN323}%
  \BibitemOpen
  \bibfield  {author} {\bibinfo {author} {\bibfnamefont {S.}~\bibnamefont
  {Hofmann}}, \bibinfo {author} {\bibfnamefont {D.}~\bibnamefont {Ackermann}},
  \bibinfo {author} {\bibfnamefont {S.}~\bibnamefont {Antalic}}, \bibinfo
  {author} {\bibfnamefont {H.~G.}\ \bibnamefont {Burkhard}}, \bibinfo {author}
  {\bibfnamefont {V.~F.}\ \bibnamefont {Comas}}, \bibinfo {author}
  {\bibfnamefont {R.}~\bibnamefont {Dressler}}, \bibinfo {author}
  {\bibfnamefont {Z.}~\bibnamefont {Gan}}, \bibinfo {author} {\bibfnamefont
  {S.}~\bibnamefont {Heinz}}, \bibinfo {author} {\bibfnamefont {J.~A.}\
  \bibnamefont {Heredia}}, \bibinfo {author} {\bibfnamefont {F.~P.}\
  \bibnamefont {Heßberger}}, \bibinfo {author} {\bibfnamefont
  {J.}~\bibnamefont {Khuyagbaatar}}, \bibinfo {author} {\bibfnamefont
  {B.}~\bibnamefont {Kindler}}, \bibinfo {author} {\bibfnamefont
  {I.}~\bibnamefont {Kojouharov}}, \bibinfo {author} {\bibfnamefont
  {P.}~\bibnamefont {Kuusiniemi}}, \bibinfo {author} {\bibfnamefont
  {M.}~\bibnamefont {Leino}}, \bibinfo {author} {\bibfnamefont
  {B.}~\bibnamefont {Lommel}}, \bibinfo {author} {\bibfnamefont
  {R.}~\bibnamefont {Mann}}, \bibinfo {author} {\bibfnamefont {G.}~\bibnamefont
  {Münzenberg}}, \bibinfo {author} {\bibfnamefont {K.}~\bibnamefont {Nishio}},
  \bibinfo {author} {\bibfnamefont {A.~G.}\ \bibnamefont {Popeko}}, \bibinfo
  {author} {\bibfnamefont {S.}~\bibnamefont {Saro}}, \bibinfo {author}
  {\bibfnamefont {H.~J.}\ \bibnamefont {Schött}}, \bibinfo {author}
  {\bibfnamefont {B.}~\bibnamefont {Streicher}}, \bibinfo {author}
  {\bibfnamefont {B.}~\bibnamefont {Sulignano}}, \bibinfo {author}
  {\bibfnamefont {J.}~\bibnamefont {Uusitalo}}, \bibinfo {author}
  {\bibfnamefont {M.}~\bibnamefont {Venhart}},\ and\ \bibinfo {author}
  {\bibfnamefont {A.~V.}\ \bibnamefont {Yeremin}},\ }\href
  {https://doi.org/10.1140/epja/i2007-10373-x} {\bibfield  {journal} {\bibinfo
  {journal} {The European Physical Journal A}\ }\textbf {\bibinfo {volume}
  {32}},\ \bibinfo {pages} {251} (\bibinfo {year} {2007})}\BibitemShut
  {NoStop}%
\bibitem [{\citenamefont {Utyonkov}\ \emph {et~al.}(2015)\citenamefont
  {Utyonkov}, \citenamefont {Brewer}, \citenamefont {Oganessian}, \citenamefont
  {Rykaczewski}, \citenamefont {Abdullin}, \citenamefont {Dmitriev},
  \citenamefont {Grzywacz}, \citenamefont {Itkis}, \citenamefont {Miernik},
  \citenamefont {Polyakov}, \citenamefont {Roberto}, \citenamefont {Sagaidak},
  \citenamefont {Shirokovsky}, \citenamefont {Shumeiko}, \citenamefont
  {Tsyganov}, \citenamefont {Voinov}, \citenamefont {Subbotin}, \citenamefont
  {Sukhov}, \citenamefont {Sabel'nikov}, \citenamefont {Vostokin},
  \citenamefont {Hamilton}, \citenamefont {Stoyer},\ and\ \citenamefont
  {Strauss}}]{RN324}%
  \BibitemOpen
  \bibfield  {author} {\bibinfo {author} {\bibfnamefont {V.~K.}\ \bibnamefont
  {Utyonkov}}, \bibinfo {author} {\bibfnamefont {N.~T.}\ \bibnamefont
  {Brewer}}, \bibinfo {author} {\bibfnamefont {Y.~T.}\ \bibnamefont
  {Oganessian}}, \bibinfo {author} {\bibfnamefont {K.~P.}\ \bibnamefont
  {Rykaczewski}}, \bibinfo {author} {\bibfnamefont {F.~S.}\ \bibnamefont
  {Abdullin}}, \bibinfo {author} {\bibfnamefont {S.~N.}\ \bibnamefont
  {Dmitriev}}, \bibinfo {author} {\bibfnamefont {R.~K.}\ \bibnamefont
  {Grzywacz}}, \bibinfo {author} {\bibfnamefont {M.~G.}\ \bibnamefont {Itkis}},
  \bibinfo {author} {\bibfnamefont {K.}~\bibnamefont {Miernik}}, \bibinfo
  {author} {\bibfnamefont {A.~N.}\ \bibnamefont {Polyakov}}, \bibinfo {author}
  {\bibfnamefont {J.~B.}\ \bibnamefont {Roberto}}, \bibinfo {author}
  {\bibfnamefont {R.~N.}\ \bibnamefont {Sagaidak}}, \bibinfo {author}
  {\bibfnamefont {I.~V.}\ \bibnamefont {Shirokovsky}}, \bibinfo {author}
  {\bibfnamefont {M.~V.}\ \bibnamefont {Shumeiko}}, \bibinfo {author}
  {\bibfnamefont {Y.~S.}\ \bibnamefont {Tsyganov}}, \bibinfo {author}
  {\bibfnamefont {A.~A.}\ \bibnamefont {Voinov}}, \bibinfo {author}
  {\bibfnamefont {V.~G.}\ \bibnamefont {Subbotin}}, \bibinfo {author}
  {\bibfnamefont {A.~M.}\ \bibnamefont {Sukhov}}, \bibinfo {author}
  {\bibfnamefont {A.~V.}\ \bibnamefont {Sabel'nikov}}, \bibinfo {author}
  {\bibfnamefont {G.~K.}\ \bibnamefont {Vostokin}}, \bibinfo {author}
  {\bibfnamefont {J.~H.}\ \bibnamefont {Hamilton}}, \bibinfo {author}
  {\bibfnamefont {M.~A.}\ \bibnamefont {Stoyer}},\ and\ \bibinfo {author}
  {\bibfnamefont {S.~Y.}\ \bibnamefont {Strauss}},\ }\href
  {https://doi.org/10.1103/PhysRevC.92.034609} {\bibfield  {journal} {\bibinfo
  {journal} {Physical Review C}\ }\textbf {\bibinfo {volume} {92}},\ \bibinfo
  {pages} {034609} (\bibinfo {year} {2015})}\BibitemShut {NoStop}%
\bibitem [{\citenamefont {Hofmann}\ \emph {et~al.}(2012)\citenamefont
  {Hofmann}, \citenamefont {Heinz}, \citenamefont {Mann}, \citenamefont
  {Maurer}, \citenamefont {Khuyagbaatar}, \citenamefont {Ackermann},
  \citenamefont {Antalic}, \citenamefont {Barth}, \citenamefont {Block},
  \citenamefont {Burkhard}, \citenamefont {Comas}, \citenamefont {Dahl},
  \citenamefont {Eberhardt}, \citenamefont {Gostic}, \citenamefont {Henderson},
  \citenamefont {Heredia}, \citenamefont {Heßberger}, \citenamefont
  {Kenneally}, \citenamefont {Kindler}, \citenamefont {Kojouharov},
  \citenamefont {Kratz}, \citenamefont {Lang}, \citenamefont {Leino},
  \citenamefont {Lommel}, \citenamefont {Moody}, \citenamefont {Münzenberg},
  \citenamefont {Nelson}, \citenamefont {Nishio}, \citenamefont {Popeko},
  \citenamefont {Runke}, \citenamefont {Saro}, \citenamefont {Shaughnessy},
  \citenamefont {Stoyer}, \citenamefont {Thörle-Pospiech}, \citenamefont
  {Tinschert}, \citenamefont {Trautmann}, \citenamefont {Uusitalo},
  \citenamefont {Wilk},\ and\ \citenamefont {Yeremin}}]{RN325}%
  \BibitemOpen
  \bibfield  {author} {\bibinfo {author} {\bibfnamefont {S.}~\bibnamefont
  {Hofmann}}, \bibinfo {author} {\bibfnamefont {S.}~\bibnamefont {Heinz}},
  \bibinfo {author} {\bibfnamefont {R.}~\bibnamefont {Mann}}, \bibinfo {author}
  {\bibfnamefont {J.}~\bibnamefont {Maurer}}, \bibinfo {author} {\bibfnamefont
  {J.}~\bibnamefont {Khuyagbaatar}}, \bibinfo {author} {\bibfnamefont
  {D.}~\bibnamefont {Ackermann}}, \bibinfo {author} {\bibfnamefont
  {S.}~\bibnamefont {Antalic}}, \bibinfo {author} {\bibfnamefont
  {W.}~\bibnamefont {Barth}}, \bibinfo {author} {\bibfnamefont
  {M.}~\bibnamefont {Block}}, \bibinfo {author} {\bibfnamefont {H.~G.}\
  \bibnamefont {Burkhard}}, \bibinfo {author} {\bibfnamefont {V.~F.}\
  \bibnamefont {Comas}}, \bibinfo {author} {\bibfnamefont {L.}~\bibnamefont
  {Dahl}}, \bibinfo {author} {\bibfnamefont {K.}~\bibnamefont {Eberhardt}},
  \bibinfo {author} {\bibfnamefont {J.}~\bibnamefont {Gostic}}, \bibinfo
  {author} {\bibfnamefont {R.~A.}\ \bibnamefont {Henderson}}, \bibinfo {author}
  {\bibfnamefont {J.~A.}\ \bibnamefont {Heredia}}, \bibinfo {author}
  {\bibfnamefont {F.~P.}\ \bibnamefont {Heßberger}}, \bibinfo {author}
  {\bibfnamefont {J.~M.}\ \bibnamefont {Kenneally}}, \bibinfo {author}
  {\bibfnamefont {B.}~\bibnamefont {Kindler}}, \bibinfo {author} {\bibfnamefont
  {I.}~\bibnamefont {Kojouharov}}, \bibinfo {author} {\bibfnamefont {J.~V.}\
  \bibnamefont {Kratz}}, \bibinfo {author} {\bibfnamefont {R.}~\bibnamefont
  {Lang}}, \bibinfo {author} {\bibfnamefont {M.}~\bibnamefont {Leino}},
  \bibinfo {author} {\bibfnamefont {B.}~\bibnamefont {Lommel}}, \bibinfo
  {author} {\bibfnamefont {K.~J.}\ \bibnamefont {Moody}}, \bibinfo {author}
  {\bibfnamefont {G.}~\bibnamefont {Münzenberg}}, \bibinfo {author}
  {\bibfnamefont {S.~L.}\ \bibnamefont {Nelson}}, \bibinfo {author}
  {\bibfnamefont {K.}~\bibnamefont {Nishio}}, \bibinfo {author} {\bibfnamefont
  {A.~G.}\ \bibnamefont {Popeko}}, \bibinfo {author} {\bibfnamefont
  {J.}~\bibnamefont {Runke}}, \bibinfo {author} {\bibfnamefont
  {S.}~\bibnamefont {Saro}}, \bibinfo {author} {\bibfnamefont {D.~A.}\
  \bibnamefont {Shaughnessy}}, \bibinfo {author} {\bibfnamefont {M.~A.}\
  \bibnamefont {Stoyer}}, \bibinfo {author} {\bibfnamefont {P.}~\bibnamefont
  {Thörle-Pospiech}}, \bibinfo {author} {\bibfnamefont {K.}~\bibnamefont
  {Tinschert}}, \bibinfo {author} {\bibfnamefont {N.}~\bibnamefont
  {Trautmann}}, \bibinfo {author} {\bibfnamefont {J.}~\bibnamefont {Uusitalo}},
  \bibinfo {author} {\bibfnamefont {P.~A.}\ \bibnamefont {Wilk}},\ and\
  \bibinfo {author} {\bibfnamefont {A.~V.}\ \bibnamefont {Yeremin}},\ }\href
  {https://doi.org/10.1140/epja/i2012-12062-1} {\bibfield  {journal} {\bibinfo
  {journal} {The European Physical Journal A}\ }\textbf {\bibinfo {volume}
  {48}},\ \bibinfo {pages} {62} (\bibinfo {year} {2012})}\BibitemShut {NoStop}%
\bibitem [{\citenamefont {Oganessian}\ \emph
  {et~al.}(2004{\natexlab{b}})\citenamefont {Oganessian}, \citenamefont
  {Utyonkov}, \citenamefont {Lobanov}, \citenamefont {Abdullin}, \citenamefont
  {Polyakov}, \citenamefont {Shirokovsky}, \citenamefont {Tsyganov},
  \citenamefont {Gulbekian}, \citenamefont {Bogomolov}, \citenamefont {Gikal},
  \citenamefont {Mezentsev}, \citenamefont {Iliev}, \citenamefont {Subbotin},
  \citenamefont {Sukhov}, \citenamefont {Voinov}, \citenamefont {Buklanov},
  \citenamefont {Subotic}, \citenamefont {Zagrebaev}, \citenamefont {Itkis},
  \citenamefont {Patin}, \citenamefont {Moody}, \citenamefont {Wild},
  \citenamefont {Stoyer}, \citenamefont {Stoyer}, \citenamefont {Shaughnessy},
  \citenamefont {Kenneally},\ and\ \citenamefont {Lougheed}}]{RN326}%
  \BibitemOpen
  \bibfield  {author} {\bibinfo {author} {\bibfnamefont {Y.~T.}\ \bibnamefont
  {Oganessian}}, \bibinfo {author} {\bibfnamefont {V.~K.}\ \bibnamefont
  {Utyonkov}}, \bibinfo {author} {\bibfnamefont {Y.~V.}\ \bibnamefont
  {Lobanov}}, \bibinfo {author} {\bibfnamefont {F.~S.}\ \bibnamefont
  {Abdullin}}, \bibinfo {author} {\bibfnamefont {A.~N.}\ \bibnamefont
  {Polyakov}}, \bibinfo {author} {\bibfnamefont {I.~V.}\ \bibnamefont
  {Shirokovsky}}, \bibinfo {author} {\bibfnamefont {Y.~S.}\ \bibnamefont
  {Tsyganov}}, \bibinfo {author} {\bibfnamefont {G.~G.}\ \bibnamefont
  {Gulbekian}}, \bibinfo {author} {\bibfnamefont {S.~L.}\ \bibnamefont
  {Bogomolov}}, \bibinfo {author} {\bibfnamefont {B.~N.}\ \bibnamefont
  {Gikal}}, \bibinfo {author} {\bibfnamefont {A.~N.}\ \bibnamefont
  {Mezentsev}}, \bibinfo {author} {\bibfnamefont {S.}~\bibnamefont {Iliev}},
  \bibinfo {author} {\bibfnamefont {V.~G.}\ \bibnamefont {Subbotin}}, \bibinfo
  {author} {\bibfnamefont {A.~M.}\ \bibnamefont {Sukhov}}, \bibinfo {author}
  {\bibfnamefont {A.~A.}\ \bibnamefont {Voinov}}, \bibinfo {author}
  {\bibfnamefont {G.~V.}\ \bibnamefont {Buklanov}}, \bibinfo {author}
  {\bibfnamefont {K.}~\bibnamefont {Subotic}}, \bibinfo {author} {\bibfnamefont
  {V.~I.}\ \bibnamefont {Zagrebaev}}, \bibinfo {author} {\bibfnamefont {M.~G.}\
  \bibnamefont {Itkis}}, \bibinfo {author} {\bibfnamefont {J.~B.}\ \bibnamefont
  {Patin}}, \bibinfo {author} {\bibfnamefont {K.~J.}\ \bibnamefont {Moody}},
  \bibinfo {author} {\bibfnamefont {J.~F.}\ \bibnamefont {Wild}}, \bibinfo
  {author} {\bibfnamefont {M.~A.}\ \bibnamefont {Stoyer}}, \bibinfo {author}
  {\bibfnamefont {N.~J.}\ \bibnamefont {Stoyer}}, \bibinfo {author}
  {\bibfnamefont {D.~A.}\ \bibnamefont {Shaughnessy}}, \bibinfo {author}
  {\bibfnamefont {J.~M.}\ \bibnamefont {Kenneally}},\ and\ \bibinfo {author}
  {\bibfnamefont {R.~W.}\ \bibnamefont {Lougheed}},\ }\href
  {https://doi.org/10.1103/PhysRevC.69.054607} {\bibfield  {journal} {\bibinfo
  {journal} {Physical Review C}\ }\textbf {\bibinfo {volume} {69}},\ \bibinfo
  {pages} {054607} (\bibinfo {year} {2004}{\natexlab{b}})}\BibitemShut
  {NoStop}%
\bibitem [{\citenamefont {Oganessian}\ \emph
  {et~al.}(2012{\natexlab{b}})\citenamefont {Oganessian}, \citenamefont
  {Abdullin}, \citenamefont {Dmitriev}, \citenamefont {Gostic}, \citenamefont
  {Hamilton}, \citenamefont {Henderson}, \citenamefont {Itkis}, \citenamefont
  {Moody}, \citenamefont {Polyakov}, \citenamefont {Ramayya}, \citenamefont
  {Roberto}, \citenamefont {Rykaczewski}, \citenamefont {Sagaidak},
  \citenamefont {Shaughnessy}, \citenamefont {Shirokovsky}, \citenamefont
  {Stoyer}, \citenamefont {Subbotin}, \citenamefont {Sukhov}, \citenamefont
  {Tsyganov}, \citenamefont {Utyonkov}, \citenamefont {Voinov},\ and\
  \citenamefont {Vostokin}}]{RN327}%
  \BibitemOpen
  \bibfield  {author} {\bibinfo {author} {\bibfnamefont {Y.~T.}\ \bibnamefont
  {Oganessian}}, \bibinfo {author} {\bibfnamefont {F.~S.}\ \bibnamefont
  {Abdullin}}, \bibinfo {author} {\bibfnamefont {S.~N.}\ \bibnamefont
  {Dmitriev}}, \bibinfo {author} {\bibfnamefont {J.~M.}\ \bibnamefont
  {Gostic}}, \bibinfo {author} {\bibfnamefont {J.~H.}\ \bibnamefont
  {Hamilton}}, \bibinfo {author} {\bibfnamefont {R.~A.}\ \bibnamefont
  {Henderson}}, \bibinfo {author} {\bibfnamefont {M.~G.}\ \bibnamefont
  {Itkis}}, \bibinfo {author} {\bibfnamefont {K.~J.}\ \bibnamefont {Moody}},
  \bibinfo {author} {\bibfnamefont {A.~N.}\ \bibnamefont {Polyakov}}, \bibinfo
  {author} {\bibfnamefont {A.~V.}\ \bibnamefont {Ramayya}}, \bibinfo {author}
  {\bibfnamefont {J.~B.}\ \bibnamefont {Roberto}}, \bibinfo {author}
  {\bibfnamefont {K.~P.}\ \bibnamefont {Rykaczewski}}, \bibinfo {author}
  {\bibfnamefont {R.~N.}\ \bibnamefont {Sagaidak}}, \bibinfo {author}
  {\bibfnamefont {D.~A.}\ \bibnamefont {Shaughnessy}}, \bibinfo {author}
  {\bibfnamefont {I.~V.}\ \bibnamefont {Shirokovsky}}, \bibinfo {author}
  {\bibfnamefont {M.~A.}\ \bibnamefont {Stoyer}}, \bibinfo {author}
  {\bibfnamefont {V.~G.}\ \bibnamefont {Subbotin}}, \bibinfo {author}
  {\bibfnamefont {A.~M.}\ \bibnamefont {Sukhov}}, \bibinfo {author}
  {\bibfnamefont {Y.~S.}\ \bibnamefont {Tsyganov}}, \bibinfo {author}
  {\bibfnamefont {V.~K.}\ \bibnamefont {Utyonkov}}, \bibinfo {author}
  {\bibfnamefont {A.~A.}\ \bibnamefont {Voinov}},\ and\ \bibinfo {author}
  {\bibfnamefont {G.~K.}\ \bibnamefont {Vostokin}},\ }\href
  {https://doi.org/10.1103/PhysRevLett.108.022502} {\bibfield  {journal}
  {\bibinfo  {journal} {Physical Review Letters}\ }\textbf {\bibinfo {volume}
  {108}},\ \bibinfo {pages} {022502} (\bibinfo {year}
  {2012}{\natexlab{b}})}\BibitemShut {NoStop}%
\bibitem [{\citenamefont {Adamian}\ \emph {et~al.}(1998)\citenamefont
  {Adamian}, \citenamefont {Antonenko}, \citenamefont {Scheid},\ and\
  \citenamefont {Volkov}}]{RN328}%
  \BibitemOpen
  \bibfield  {author} {\bibinfo {author} {\bibfnamefont {G.~G.}\ \bibnamefont
  {Adamian}}, \bibinfo {author} {\bibfnamefont {N.~V.}\ \bibnamefont
  {Antonenko}}, \bibinfo {author} {\bibfnamefont {W.}~\bibnamefont {Scheid}},\
  and\ \bibinfo {author} {\bibfnamefont {V.~V.}\ \bibnamefont {Volkov}},\
  }\href {https://doi.org/https://doi.org/10.1016/S0375-9474(98)00124-9}
  {\bibfield  {journal} {\bibinfo  {journal} {Nuclear Physics A}\ }\textbf
  {\bibinfo {volume} {633}},\ \bibinfo {pages} {409} (\bibinfo {year}
  {1998})}\BibitemShut {NoStop}%
\bibitem [{\citenamefont {Adamian}\ \emph {et~al.}(2000)\citenamefont
  {Adamian}, \citenamefont {Antonenko},\ and\ \citenamefont {Scheid}}]{RN329}%
  \BibitemOpen
  \bibfield  {author} {\bibinfo {author} {\bibfnamefont {G.~G.}\ \bibnamefont
  {Adamian}}, \bibinfo {author} {\bibfnamefont {N.~V.}\ \bibnamefont
  {Antonenko}},\ and\ \bibinfo {author} {\bibfnamefont {W.}~\bibnamefont
  {Scheid}},\ }\href
  {https://doi.org/https://doi.org/10.1016/S0375-9474(00)00317-1} {\bibfield
  {journal} {\bibinfo  {journal} {Nuclear Physics A}\ }\textbf {\bibinfo
  {volume} {678}},\ \bibinfo {pages} {24} (\bibinfo {year} {2000})}\BibitemShut
  {NoStop}%
\bibitem [{\citenamefont {Adamian}\ \emph {et~al.}(2009)\citenamefont
  {Adamian}, \citenamefont {Antonenko},\ and\ \citenamefont {Scheid}}]{RN330}%
  \BibitemOpen
  \bibfield  {author} {\bibinfo {author} {\bibfnamefont {G.~G.}\ \bibnamefont
  {Adamian}}, \bibinfo {author} {\bibfnamefont {N.~V.}\ \bibnamefont
  {Antonenko}},\ and\ \bibinfo {author} {\bibfnamefont {W.}~\bibnamefont
  {Scheid}},\ }\href {https://doi.org/10.1140/epja/i2009-10795-4} {\bibfield
  {journal} {\bibinfo  {journal} {The European Physical Journal A}\ }\textbf
  {\bibinfo {volume} {41}},\ \bibinfo {pages} {235} (\bibinfo {year}
  {2009})}\BibitemShut {NoStop}%
\bibitem [{\citenamefont {Feng}\ \emph {et~al.}(2006)\citenamefont {Feng},
  \citenamefont {Jin}, \citenamefont {Fu},\ and\ \citenamefont {Li}}]{RN331}%
  \BibitemOpen
  \bibfield  {author} {\bibinfo {author} {\bibfnamefont {Z.-Q.}\ \bibnamefont
  {Feng}}, \bibinfo {author} {\bibfnamefont {G.-M.}\ \bibnamefont {Jin}},
  \bibinfo {author} {\bibfnamefont {F.}~\bibnamefont {Fu}},\ and\ \bibinfo
  {author} {\bibfnamefont {J.-Q.}\ \bibnamefont {Li}},\ }\href
  {https://doi.org/https://doi.org/10.1016/j.nuclphysa.2006.03.002} {\bibfield
  {journal} {\bibinfo  {journal} {Nuclear Physics A}\ }\textbf {\bibinfo
  {volume} {771}},\ \bibinfo {pages} {50} (\bibinfo {year} {2006})}\BibitemShut
  {NoStop}%
\bibitem [{\citenamefont {Gupta}\ \emph {et~al.}(2004)\citenamefont {Gupta},
  \citenamefont {Singh},\ and\ \citenamefont {Manhas}}]{RN332}%
  \BibitemOpen
  \bibfield  {author} {\bibinfo {author} {\bibfnamefont {R.~K.}\ \bibnamefont
  {Gupta}}, \bibinfo {author} {\bibfnamefont {N.}~\bibnamefont {Singh}},\ and\
  \bibinfo {author} {\bibfnamefont {M.}~\bibnamefont {Manhas}},\ }\href
  {https://doi.org/10.1103/PhysRevC.70.034608} {\bibfield  {journal} {\bibinfo
  {journal} {Physical Review C}\ }\textbf {\bibinfo {volume} {70}},\ \bibinfo
  {pages} {034608} (\bibinfo {year} {2004})}\BibitemShut {NoStop}%
\bibitem [{\citenamefont {Loveland}(2007)}]{RN333}%
  \BibitemOpen
  \bibfield  {author} {\bibinfo {author} {\bibfnamefont {W.}~\bibnamefont
  {Loveland}},\ }\href {https://doi.org/10.1103/PhysRevC.76.014612} {\bibfield
  {journal} {\bibinfo  {journal} {Physical Review C}\ }\textbf {\bibinfo
  {volume} {76}},\ \bibinfo {pages} {014612} (\bibinfo {year}
  {2007})}\BibitemShut {NoStop}%
\bibitem [{\citenamefont {Liu}\ and\ \citenamefont
  {Bao}(2009{\natexlab{a}})}]{RN334}%
  \BibitemOpen
  \bibfield  {author} {\bibinfo {author} {\bibfnamefont {Z.~H.}\ \bibnamefont
  {Liu}}\ and\ \bibinfo {author} {\bibfnamefont {J.-D.}\ \bibnamefont {Bao}},\
  }\href {https://doi.org/10.1103/PhysRevC.80.034601} {\bibfield  {journal}
  {\bibinfo  {journal} {Physical Review C}\ }\textbf {\bibinfo {volume} {80}},\
  \bibinfo {pages} {034601} (\bibinfo {year} {2009}{\natexlab{a}})}\BibitemShut
  {NoStop}%
\bibitem [{\citenamefont {Dutt}\ and\ \citenamefont {Puri}(2010)}]{RN335}%
  \BibitemOpen
  \bibfield  {author} {\bibinfo {author} {\bibfnamefont {I.}~\bibnamefont
  {Dutt}}\ and\ \bibinfo {author} {\bibfnamefont {R.~K.}\ \bibnamefont
  {Puri}},\ }\href {https://doi.org/10.1103/PhysRevC.81.064609} {\bibfield
  {journal} {\bibinfo  {journal} {Physical Review C}\ }\textbf {\bibinfo
  {volume} {81}},\ \bibinfo {pages} {064609} (\bibinfo {year}
  {2010})}\BibitemShut {NoStop}%
\bibitem [{\citenamefont {Ghodsi}\ and\ \citenamefont {Lari}(2014)}]{RN336}%
  \BibitemOpen
  \bibfield  {author} {\bibinfo {author} {\bibfnamefont {O.}~\bibnamefont
  {Ghodsi}}\ and\ \bibinfo {author} {\bibfnamefont {F.}~\bibnamefont {Lari}},\
  }\href {https://doi.org/10.1103/PhysRevC.89.054607} {\bibfield  {journal}
  {\bibinfo  {journal} {Physical Review C}\ }\textbf {\bibinfo {volume} {89}},\
  \bibinfo {pages} {054607} (\bibinfo {year} {2014})}\BibitemShut {NoStop}%
\bibitem [{\citenamefont {Bao}\ \emph {et~al.}(2015)\citenamefont {Bao},
  \citenamefont {Gao}, \citenamefont {Li},\ and\ \citenamefont
  {Zhang}}]{RN338}%
  \BibitemOpen
  \bibfield  {author} {\bibinfo {author} {\bibfnamefont {X.~J.}\ \bibnamefont
  {Bao}}, \bibinfo {author} {\bibfnamefont {Y.}~\bibnamefont {Gao}}, \bibinfo
  {author} {\bibfnamefont {J.~Q.}\ \bibnamefont {Li}},\ and\ \bibinfo {author}
  {\bibfnamefont {H.~F.}\ \bibnamefont {Zhang}},\ }\href
  {https://doi.org/10.1103/PhysRevC.91.011603} {\bibfield  {journal} {\bibinfo
  {journal} {Physical Review C}\ }\textbf {\bibinfo {volume} {91}},\ \bibinfo
  {pages} {011603} (\bibinfo {year} {2015})}\BibitemShut {NoStop}%
\bibitem [{\citenamefont {Ghahramany}\ and\ \citenamefont
  {Ansari}(2016)}]{RN339}%
  \BibitemOpen
  \bibfield  {author} {\bibinfo {author} {\bibfnamefont {N.}~\bibnamefont
  {Ghahramany}}\ and\ \bibinfo {author} {\bibfnamefont {A.}~\bibnamefont
  {Ansari}},\ }\href {https://doi.org/10.1140/epja/i2016-16287-6} {\bibfield
  {journal} {\bibinfo  {journal} {The European Physical Journal A}\ }\textbf
  {\bibinfo {volume} {52}},\ \bibinfo {pages} {287} (\bibinfo {year}
  {2016})}\BibitemShut {NoStop}%
\bibitem [{\citenamefont {Hagino}(2018)}]{RN341}%
  \BibitemOpen
  \bibfield  {author} {\bibinfo {author} {\bibfnamefont {K.}~\bibnamefont
  {Hagino}},\ }\href {https://doi.org/10.1103/PhysRevC.98.014607} {\bibfield
  {journal} {\bibinfo  {journal} {Physical Review C}\ }\textbf {\bibinfo
  {volume} {98}},\ \bibinfo {pages} {014607} (\bibinfo {year}
  {2018})}\BibitemShut {NoStop}%
\bibitem [{\citenamefont {Liu}\ \emph {et~al.}(2016)\citenamefont {Liu},
  \citenamefont {Shen}, \citenamefont {Li}, \citenamefont {Tu}, \citenamefont
  {Wang},\ and\ \citenamefont {Wang}}]{RN342}%
  \BibitemOpen
  \bibfield  {author} {\bibinfo {author} {\bibfnamefont {L.}~\bibnamefont
  {Liu}}, \bibinfo {author} {\bibfnamefont {C.}~\bibnamefont {Shen}}, \bibinfo
  {author} {\bibfnamefont {Q.}~\bibnamefont {Li}}, \bibinfo {author}
  {\bibfnamefont {Y.}~\bibnamefont {Tu}}, \bibinfo {author} {\bibfnamefont
  {X.}~\bibnamefont {Wang}},\ and\ \bibinfo {author} {\bibfnamefont
  {Y.}~\bibnamefont {Wang}},\ }\href
  {https://doi.org/10.1140/epja/i2016-16035-0} {\bibfield  {journal} {\bibinfo
  {journal} {The European Physical Journal A}\ }\textbf {\bibinfo {volume}
  {52}},\ \bibinfo {pages} {35} (\bibinfo {year} {2016})}\BibitemShut {NoStop}%
\bibitem [{\citenamefont {Liu}\ and\ \citenamefont {Bao}(2013)}]{RN343}%
  \BibitemOpen
  \bibfield  {author} {\bibinfo {author} {\bibfnamefont {Z.-H.}\ \bibnamefont
  {Liu}}\ and\ \bibinfo {author} {\bibfnamefont {J.-D.}\ \bibnamefont {Bao}},\
  }\href {https://doi.org/10.1103/PhysRevC.87.034616} {\bibfield  {journal}
  {\bibinfo  {journal} {Physical Review C}\ }\textbf {\bibinfo {volume} {87}},\
  \bibinfo {pages} {034616} (\bibinfo {year} {2013})}\BibitemShut {NoStop}%
\bibitem [{\citenamefont {Nasirov}\ \emph {et~al.}(2011)\citenamefont
  {Nasirov}, \citenamefont {Mandaglio}, \citenamefont {Giardina}, \citenamefont
  {Sobiczewski},\ and\ \citenamefont {Muminov}}]{RN344}%
  \BibitemOpen
  \bibfield  {author} {\bibinfo {author} {\bibfnamefont {A.~K.}\ \bibnamefont
  {Nasirov}}, \bibinfo {author} {\bibfnamefont {G.}~\bibnamefont {Mandaglio}},
  \bibinfo {author} {\bibfnamefont {G.}~\bibnamefont {Giardina}}, \bibinfo
  {author} {\bibfnamefont {A.}~\bibnamefont {Sobiczewski}},\ and\ \bibinfo
  {author} {\bibfnamefont {A.~I.}\ \bibnamefont {Muminov}},\ }\href
  {https://doi.org/10.1103/PhysRevC.84.044612} {\bibfield  {journal} {\bibinfo
  {journal} {Physical Review C}\ }\textbf {\bibinfo {volume} {84}},\ \bibinfo
  {pages} {044612} (\bibinfo {year} {2011})}\BibitemShut {NoStop}%
\bibitem [{\citenamefont {Santhosh}\ and\ \citenamefont
  {Safoora}(2016)}]{RN345}%
  \BibitemOpen
  \bibfield  {author} {\bibinfo {author} {\bibfnamefont {K.}~\bibnamefont
  {Santhosh}}\ and\ \bibinfo {author} {\bibfnamefont {V.}~\bibnamefont
  {Safoora}},\ }\href {https://doi.org/10.1103/PhysRevC.94.024623} {\bibfield
  {journal} {\bibinfo  {journal} {Physical Review C}\ }\textbf {\bibinfo
  {volume} {94}},\ \bibinfo {pages} {024623} (\bibinfo {year}
  {2016})}\BibitemShut {NoStop}%
\bibitem [{\citenamefont {Santhosh}\ and\ \citenamefont
  {Safoora}(2017{\natexlab{a}})}]{RN347}%
  \BibitemOpen
  \bibfield  {author} {\bibinfo {author} {\bibfnamefont {K.}~\bibnamefont
  {Santhosh}}\ and\ \bibinfo {author} {\bibfnamefont {V.}~\bibnamefont
  {Safoora}},\ }\href {https://doi.org/10.1103/PhysRevC.95.064611} {\bibfield
  {journal} {\bibinfo  {journal} {Physical Review C}\ }\textbf {\bibinfo
  {volume} {95}},\ \bibinfo {pages} {064611} (\bibinfo {year}
  {2017}{\natexlab{a}})}\BibitemShut {NoStop}%
\bibitem [{\citenamefont {Santhosh}\ and\ \citenamefont
  {Safoora}(2017{\natexlab{b}})}]{RN348}%
  \BibitemOpen
  \bibfield  {author} {\bibinfo {author} {\bibfnamefont {K.}~\bibnamefont
  {Santhosh}}\ and\ \bibinfo {author} {\bibfnamefont {V.}~\bibnamefont
  {Safoora}},\ }\href {https://doi.org/10.1103/PhysRevC.96.034610} {\bibfield
  {journal} {\bibinfo  {journal} {Physical Review C}\ }\textbf {\bibinfo
  {volume} {96}},\ \bibinfo {pages} {034610} (\bibinfo {year}
  {2017}{\natexlab{b}})}\BibitemShut {NoStop}%
\bibitem [{\citenamefont {Santhosh}\ and\ \citenamefont
  {Safoora}(2018)}]{RN349}%
  \BibitemOpen
  \bibfield  {author} {\bibinfo {author} {\bibfnamefont {K.}~\bibnamefont
  {Santhosh}}\ and\ \bibinfo {author} {\bibfnamefont {V.}~\bibnamefont
  {Safoora}},\ }\href {https://doi.org/10.1140/epja/i2018-12512-8} {\bibfield
  {journal} {\bibinfo  {journal} {The European Physical Journal A}\ }\textbf
  {\bibinfo {volume} {54}},\ \bibinfo {pages} {80} (\bibinfo {year}
  {2018})}\BibitemShut {NoStop}%
\bibitem [{\citenamefont {Santhosh}\ and\ \citenamefont
  {Safoora}(2021)}]{RN350}%
  \BibitemOpen
  \bibfield  {author} {\bibinfo {author} {\bibfnamefont {K.}~\bibnamefont
  {Santhosh}}\ and\ \bibinfo {author} {\bibfnamefont {V.}~\bibnamefont
  {Safoora}},\ }\href {https://doi.org/10.1007/s13538-020-00830-2} {\bibfield
  {journal} {\bibinfo  {journal} {Brazilian Journal of Physics}\ }\textbf
  {\bibinfo {volume} {51}},\ \bibinfo {pages} {90} (\bibinfo {year}
  {2021})}\BibitemShut {NoStop}%
\bibitem [{\citenamefont {Zagrebaev}(2001)}]{RN352}%
  \BibitemOpen
  \bibfield  {author} {\bibinfo {author} {\bibfnamefont {V.}~\bibnamefont
  {Zagrebaev}},\ }\href {https://doi.org/10.1103/PhysRevC.64.034606} {\bibfield
   {journal} {\bibinfo  {journal} {Physical Review C}\ }\textbf {\bibinfo
  {volume} {64}},\ \bibinfo {pages} {034606} (\bibinfo {year}
  {2001})}\BibitemShut {NoStop}%
\bibitem [{\citenamefont {Lv}\ \emph {et~al.}(2021)\citenamefont {Lv},
  \citenamefont {Yue}, \citenamefont {Zhao},\ and\ \citenamefont
  {Wang}}]{RN354}%
  \BibitemOpen
  \bibfield  {author} {\bibinfo {author} {\bibfnamefont {X.-J.}\ \bibnamefont
  {Lv}}, \bibinfo {author} {\bibfnamefont {Z.-Y.}\ \bibnamefont {Yue}},
  \bibinfo {author} {\bibfnamefont {W.-J.}\ \bibnamefont {Zhao}},\ and\
  \bibinfo {author} {\bibfnamefont {B.}~\bibnamefont {Wang}},\ }\href
  {https://doi.org/10.1103/PhysRevC.103.064616} {\bibfield  {journal} {\bibinfo
   {journal} {Physical Review C}\ }\textbf {\bibinfo {volume} {103}},\ \bibinfo
  {pages} {064616} (\bibinfo {year} {2021})}\BibitemShut {NoStop}%
\bibitem [{\citenamefont {Siwek-Wilczyńska}\ \emph {et~al.}(2012)\citenamefont
  {Siwek-Wilczyńska}, \citenamefont {Cap}, \citenamefont {Kowal},
  \citenamefont {Sobiczewski},\ and\ \citenamefont {Wilczyński}}]{RN355}%
  \BibitemOpen
  \bibfield  {author} {\bibinfo {author} {\bibfnamefont {K.}~\bibnamefont
  {Siwek-Wilczyńska}}, \bibinfo {author} {\bibfnamefont {T.}~\bibnamefont
  {Cap}}, \bibinfo {author} {\bibfnamefont {M.}~\bibnamefont {Kowal}}, \bibinfo
  {author} {\bibfnamefont {A.}~\bibnamefont {Sobiczewski}},\ and\ \bibinfo
  {author} {\bibfnamefont {J.}~\bibnamefont {Wilczyński}},\ }\href
  {https://doi.org/10.1103/PhysRevC.86.014611} {\bibfield  {journal} {\bibinfo
  {journal} {Physical Review C}\ }\textbf {\bibinfo {volume} {86}},\ \bibinfo
  {pages} {014611} (\bibinfo {year} {2012})}\BibitemShut {NoStop}%
\bibitem [{\citenamefont {Zagrebaev}\ and\ \citenamefont
  {Greiner}(2008)}]{RN357}%
  \BibitemOpen
  \bibfield  {author} {\bibinfo {author} {\bibfnamefont {V.}~\bibnamefont
  {Zagrebaev}}\ and\ \bibinfo {author} {\bibfnamefont {W.}~\bibnamefont
  {Greiner}},\ }\href {https://doi.org/10.1103/PhysRevC.78.034610} {\bibfield
  {journal} {\bibinfo  {journal} {Physical Review C}\ }\textbf {\bibinfo
  {volume} {78}},\ \bibinfo {pages} {034610} (\bibinfo {year}
  {2008})}\BibitemShut {NoStop}%
\bibitem [{\citenamefont {Zhu}\ \emph {et~al.}(2014)\citenamefont {Zhu},
  \citenamefont {Xie},\ and\ \citenamefont {Zhang}}]{RN359}%
  \BibitemOpen
  \bibfield  {author} {\bibinfo {author} {\bibfnamefont {L.}~\bibnamefont
  {Zhu}}, \bibinfo {author} {\bibfnamefont {W.-J.}\ \bibnamefont {Xie}},\ and\
  \bibinfo {author} {\bibfnamefont {F.-S.}\ \bibnamefont {Zhang}},\ }\href
  {https://doi.org/10.1103/PhysRevC.89.024615} {\bibfield  {journal} {\bibinfo
  {journal} {Physical Review C}\ }\textbf {\bibinfo {volume} {89}},\ \bibinfo
  {pages} {024615} (\bibinfo {year} {2014})}\BibitemShut {NoStop}%
\bibitem [{\citenamefont {Zubov}\ \emph {et~al.}(2002)\citenamefont {Zubov},
  \citenamefont {Adamian}, \citenamefont {Antonenko}, \citenamefont {Ivanova},\
  and\ \citenamefont {Scheid}}]{RN361}%
  \BibitemOpen
  \bibfield  {author} {\bibinfo {author} {\bibfnamefont {A.~S.}\ \bibnamefont
  {Zubov}}, \bibinfo {author} {\bibfnamefont {G.~G.}\ \bibnamefont {Adamian}},
  \bibinfo {author} {\bibfnamefont {N.~V.}\ \bibnamefont {Antonenko}}, \bibinfo
  {author} {\bibfnamefont {S.~P.}\ \bibnamefont {Ivanova}},\ and\ \bibinfo
  {author} {\bibfnamefont {W.}~\bibnamefont {Scheid}},\ }\href
  {https://doi.org/10.1103/PhysRevC.65.024308} {\bibfield  {journal} {\bibinfo
  {journal} {Physical Review C}\ }\textbf {\bibinfo {volume} {65}},\ \bibinfo
  {pages} {024308} (\bibinfo {year} {2002})}\BibitemShut {NoStop}%
\bibitem [{\citenamefont {Amritanshu~Shukla}(2021)}]{RN362}%
  \BibitemOpen
  \bibfield  {author} {\bibinfo {author} {\bibfnamefont {S.~K.~P.}\
  \bibnamefont {Amritanshu~Shukla}},\ }\href
  {https://doi.org/https://doi.org/10.1201/9780429288647} {\emph {\bibinfo
  {title} {Nuclear Structure Physics}}}\ (\bibinfo  {publisher} {CRC Press},\
  \bibinfo {year} {2021})\BibitemShut {NoStop}%
\bibitem [{\citenamefont {SHEN}\ \emph {et~al.}(2008)\citenamefont {SHEN},
  \citenamefont {ABE}, \citenamefont {BOILLEY}, \citenamefont {KOSENKO},\ and\
  \citenamefont {ZHAO}}]{RN363}%
  \BibitemOpen
  \bibfield  {author} {\bibinfo {author} {\bibfnamefont {C.}~\bibnamefont
  {SHEN}}, \bibinfo {author} {\bibfnamefont {Y.}~\bibnamefont {ABE}}, \bibinfo
  {author} {\bibfnamefont {D.}~\bibnamefont {BOILLEY}}, \bibinfo {author}
  {\bibfnamefont {G.}~\bibnamefont {KOSENKO}},\ and\ \bibinfo {author}
  {\bibfnamefont {E.}~\bibnamefont {ZHAO}},\ }\href
  {https://doi.org/10.1142/s0218301308011768} {\bibfield  {journal} {\bibinfo
  {journal} {International Journal of Modern Physics E}\ }\textbf {\bibinfo
  {volume} {17}},\ \bibinfo {pages} {66} (\bibinfo {year} {2008})}\BibitemShut
  {NoStop}%
\bibitem [{\citenamefont {Liu}\ and\ \citenamefont
  {Bao}(2009{\natexlab{b}})}]{RN364}%
  \BibitemOpen
  \bibfield  {author} {\bibinfo {author} {\bibfnamefont {Z.~H.}\ \bibnamefont
  {Liu}}\ and\ \bibinfo {author} {\bibfnamefont {J.-D.}\ \bibnamefont {Bao}},\
  }\href {https://doi.org/10.1103/PhysRevC.80.054608} {\bibfield  {journal}
  {\bibinfo  {journal} {Physical Review C}\ }\textbf {\bibinfo {volume} {80}},\
  \bibinfo {pages} {054608} (\bibinfo {year} {2009}{\natexlab{b}})}\BibitemShut
  {NoStop}%
\bibitem [{\citenamefont {SIWEK-WILCZYŃSKA}\ \emph {et~al.}(2010)\citenamefont
  {SIWEK-WILCZYŃSKA}, \citenamefont {CAP},\ and\ \citenamefont
  {WILCZYŃSKI}}]{RN365}%
  \BibitemOpen
  \bibfield  {author} {\bibinfo {author} {\bibfnamefont {K.}~\bibnamefont
  {SIWEK-WILCZYŃSKA}}, \bibinfo {author} {\bibfnamefont {T.}~\bibnamefont
  {CAP}},\ and\ \bibinfo {author} {\bibfnamefont {J.}~\bibnamefont
  {WILCZYŃSKI}},\ }\href {https://doi.org/10.1142/s021830131001490x}
  {\bibfield  {journal} {\bibinfo  {journal} {International Journal of Modern
  Physics E}\ }\textbf {\bibinfo {volume} {19}},\ \bibinfo {pages} {500}
  (\bibinfo {year} {2010})}\BibitemShut {NoStop}%
\bibitem [{\citenamefont {Blocki}\ and\ \citenamefont
  {Świątecki}(1981)}]{RN366}%
  \BibitemOpen
  \bibfield  {author} {\bibinfo {author} {\bibfnamefont {J.}~\bibnamefont
  {Blocki}}\ and\ \bibinfo {author} {\bibfnamefont {W.~J.}\ \bibnamefont
  {Świątecki}},\ }\href
  {https://doi.org/https://doi.org/10.1016/0003-4916(81)90268-2} {\bibfield
  {journal} {\bibinfo  {journal} {Annals of Physics}\ }\textbf {\bibinfo
  {volume} {132}},\ \bibinfo {pages} {53} (\bibinfo {year} {1981})}\BibitemShut
  {NoStop}%
\bibitem [{\citenamefont {Myers}\ and\ \citenamefont
  {Swiatecki}(1966)}]{RN368}%
  \BibitemOpen
  \bibfield  {author} {\bibinfo {author} {\bibfnamefont {W.~D.}\ \bibnamefont
  {Myers}}\ and\ \bibinfo {author} {\bibfnamefont {W.~J.}\ \bibnamefont
  {Swiatecki}},\ }\href
  {https://doi.org/https://doi.org/10.1016/S0029-5582(66)80001-9} {\bibfield
  {journal} {\bibinfo  {journal} {Nuclear Physics}\ }\textbf {\bibinfo {volume}
  {81}},\ \bibinfo {pages} {1} (\bibinfo {year} {1966})}\BibitemShut {NoStop}%
\bibitem [{\citenamefont {Myers}\ and\ \citenamefont
  {Swiatecki}(1967)}]{RN369}%
  \BibitemOpen
  \bibfield  {author} {\bibinfo {author} {\bibfnamefont {W.~D.}\ \bibnamefont
  {Myers}}\ and\ \bibinfo {author} {\bibfnamefont {W.}~\bibnamefont
  {Swiatecki}},\ }\href {https://escholarship.org/uc/item/3g69c9s2} {\bibfield
  {journal} {\bibinfo  {journal} {Ark. Fys}\ }\textbf {\bibinfo {volume}
  {36}},\ \bibinfo {pages} {343} (\bibinfo {year} {1967})}\BibitemShut
  {NoStop}%
\bibitem [{\citenamefont {Möller}\ and\ \citenamefont {Nix}(1976)}]{RN370}%
  \BibitemOpen
  \bibfield  {author} {\bibinfo {author} {\bibfnamefont {P.}~\bibnamefont
  {Möller}}\ and\ \bibinfo {author} {\bibfnamefont {J.~R.}\ \bibnamefont
  {Nix}},\ }\href
  {https://doi.org/https://doi.org/10.1016/0375-9474(76)90345-6} {\bibfield
  {journal} {\bibinfo  {journal} {Nuclear Physics A}\ }\textbf {\bibinfo
  {volume} {272}},\ \bibinfo {pages} {502} (\bibinfo {year}
  {1976})}\BibitemShut {NoStop}%
\bibitem [{\citenamefont {Krappe}\ \emph {et~al.}(1979)\citenamefont {Krappe},
  \citenamefont {Nix},\ and\ \citenamefont {Sierk}}]{RN371}%
  \BibitemOpen
  \bibfield  {author} {\bibinfo {author} {\bibfnamefont {H.~J.}\ \bibnamefont
  {Krappe}}, \bibinfo {author} {\bibfnamefont {J.~R.}\ \bibnamefont {Nix}},\
  and\ \bibinfo {author} {\bibfnamefont {A.~J.}\ \bibnamefont {Sierk}},\ }\href
  {https://doi.org/10.1103/PhysRevC.20.992} {\bibfield  {journal} {\bibinfo
  {journal} {Physical Review C}\ }\textbf {\bibinfo {volume} {20}},\ \bibinfo
  {pages} {992} (\bibinfo {year} {1979})}\BibitemShut {NoStop}%
\bibitem [{\citenamefont {Möller}\ and\ \citenamefont {Nix}(1981)}]{RN372}%
  \BibitemOpen
  \bibfield  {author} {\bibinfo {author} {\bibfnamefont {P.}~\bibnamefont
  {Möller}}\ and\ \bibinfo {author} {\bibfnamefont {J.~R.}\ \bibnamefont
  {Nix}},\ }\href
  {https://doi.org/https://doi.org/10.1016/0375-9474(81)90473-5} {\bibfield
  {journal} {\bibinfo  {journal} {Nuclear Physics A}\ }\textbf {\bibinfo
  {volume} {361}},\ \bibinfo {pages} {117} (\bibinfo {year}
  {1981})}\BibitemShut {NoStop}%
\bibitem [{\citenamefont {Royer}\ and\ \citenamefont {Remaud}(1984)}]{RN373}%
  \BibitemOpen
  \bibfield  {author} {\bibinfo {author} {\bibfnamefont {G.}~\bibnamefont
  {Royer}}\ and\ \bibinfo {author} {\bibfnamefont {B.}~\bibnamefont {Remaud}},\
  }\href {https://doi.org/10.1088/0305-4616/10/8/011} {\bibfield  {journal}
  {\bibinfo  {journal} {Journal of Physics G: Nuclear Physics}\ }\textbf
  {\bibinfo {volume} {10}},\ \bibinfo {pages} {1057} (\bibinfo {year}
  {1984})}\BibitemShut {NoStop}%
\bibitem [{\citenamefont {Möller}\ and\ \citenamefont {Nix}(1995)}]{RN374}%
  \BibitemOpen
  \bibfield  {author} {\bibinfo {author} {\bibfnamefont {P.}~\bibnamefont
  {Möller}}\ and\ \bibinfo {author} {\bibfnamefont {J.}~\bibnamefont {Nix}},\
  }\href {https://doi.org/10.1006/adnd.1995.1002} {\bibfield  {journal}
  {\bibinfo  {journal} {Atom. Data and Nucl. Data Tables}\ }\textbf {\bibinfo
  {volume} {59}},\ \bibinfo {pages} {185} (\bibinfo {year} {1995})}\BibitemShut
  {NoStop}%
\bibitem [{\citenamefont {Pomorski}\ and\ \citenamefont {Dudek}(2003)}]{RN375}%
  \BibitemOpen
  \bibfield  {author} {\bibinfo {author} {\bibfnamefont {K.}~\bibnamefont
  {Pomorski}}\ and\ \bibinfo {author} {\bibfnamefont {J.}~\bibnamefont
  {Dudek}},\ }\href {https://doi.org/10.1103/PhysRevC.67.044316} {\bibfield
  {journal} {\bibinfo  {journal} {Physical Review C}\ }\textbf {\bibinfo
  {volume} {67}},\ \bibinfo {pages} {044316} (\bibinfo {year}
  {2003})}\BibitemShut {NoStop}%
\bibitem [{\citenamefont {Sobiczewski}\ \emph {et~al.}(1966)\citenamefont
  {Sobiczewski}, \citenamefont {Gareev},\ and\ \citenamefont
  {Kalinkin}}]{RN376}%
  \BibitemOpen
  \bibfield  {author} {\bibinfo {author} {\bibfnamefont {A.}~\bibnamefont
  {Sobiczewski}}, \bibinfo {author} {\bibfnamefont {F.~A.}\ \bibnamefont
  {Gareev}},\ and\ \bibinfo {author} {\bibfnamefont {B.~N.}\ \bibnamefont
  {Kalinkin}},\ }\href
  {https://doi.org/https://doi.org/10.1016/0031-9163(66)91243-1} {\bibfield
  {journal} {\bibinfo  {journal} {Physics Letters}\ }\textbf {\bibinfo {volume}
  {22}},\ \bibinfo {pages} {500} (\bibinfo {year} {1966})}\BibitemShut
  {NoStop}%
\bibitem [{\citenamefont {Meldner}(1966)}]{RN377}%
  \BibitemOpen
  \bibfield  {author} {\bibinfo {author} {\bibfnamefont {H.}~\bibnamefont
  {Meldner}},\ }\href {https://escholarship.org/uc/item/28q2j00w} {\bibfield
  {journal} {\bibinfo  {journal} {Ark. Fys}\ } (\bibinfo {year}
  {1966})}\BibitemShut {NoStop}%
\bibitem [{\citenamefont {Stoyer}(2006)}]{RN378}%
  \BibitemOpen
  \bibfield  {author} {\bibinfo {author} {\bibfnamefont {M.~A.}\ \bibnamefont
  {Stoyer}},\ }\href {https://doi.org/10.1038/442876a} {\bibfield  {journal}
  {\bibinfo  {journal} {Nature}\ }\textbf {\bibinfo {volume} {442}},\ \bibinfo
  {pages} {876} (\bibinfo {year} {2006})}\BibitemShut {NoStop}%
\bibitem [{\citenamefont {Roberto}\ \emph {et~al.}(2015)\citenamefont
  {Roberto}, \citenamefont {Alexander}, \citenamefont {Boll}, \citenamefont
  {Burns}, \citenamefont {Ezold}, \citenamefont {Felker}, \citenamefont
  {Hogle},\ and\ \citenamefont {Rykaczewski}}]{RN379}%
  \BibitemOpen
  \bibfield  {author} {\bibinfo {author} {\bibfnamefont {J.~B.}\ \bibnamefont
  {Roberto}}, \bibinfo {author} {\bibfnamefont {C.~W.}\ \bibnamefont
  {Alexander}}, \bibinfo {author} {\bibfnamefont {R.~A.}\ \bibnamefont {Boll}},
  \bibinfo {author} {\bibfnamefont {J.~D.}\ \bibnamefont {Burns}}, \bibinfo
  {author} {\bibfnamefont {J.~G.}\ \bibnamefont {Ezold}}, \bibinfo {author}
  {\bibfnamefont {L.~K.}\ \bibnamefont {Felker}}, \bibinfo {author}
  {\bibfnamefont {S.~L.}\ \bibnamefont {Hogle}},\ and\ \bibinfo {author}
  {\bibfnamefont {K.~P.}\ \bibnamefont {Rykaczewski}},\ }\href
  {https://doi.org/https://doi.org/10.1016/j.nuclphysa.2015.06.009} {\bibfield
  {journal} {\bibinfo  {journal} {Nuclear Physics A}\ }\textbf {\bibinfo
  {volume} {944}},\ \bibinfo {pages} {99} (\bibinfo {year} {2015})}\BibitemShut
  {NoStop}%
\bibitem [{\citenamefont {Zagrebaev}\ and\ \citenamefont
  {Greiner}(2015)}]{RN381}%
  \BibitemOpen
  \bibfield  {author} {\bibinfo {author} {\bibfnamefont {V.~I.}\ \bibnamefont
  {Zagrebaev}}\ and\ \bibinfo {author} {\bibfnamefont {W.}~\bibnamefont
  {Greiner}},\ }\href
  {https://doi.org/https://doi.org/10.1016/j.nuclphysa.2015.02.010} {\bibfield
  {journal} {\bibinfo  {journal} {Nuclear Physics A}\ }\textbf {\bibinfo
  {volume} {944}},\ \bibinfo {pages} {257} (\bibinfo {year}
  {2015})}\BibitemShut {NoStop}%
\bibitem [{\citenamefont {Oganessian}\ \emph {et~al.}(2009)\citenamefont
  {Oganessian}, \citenamefont {Utyonkov}, \citenamefont {Lobanov},
  \citenamefont {Abdullin}, \citenamefont {Polyakov}, \citenamefont {Sagaidak},
  \citenamefont {Shirokovsky}, \citenamefont {Tsyganov}, \citenamefont
  {Voinov}, \citenamefont {Mezentsev}, \citenamefont {Subbotin}, \citenamefont
  {Sukhov}, \citenamefont {Subotic}, \citenamefont {Zagrebaev}, \citenamefont
  {Dmitriev}, \citenamefont {Henderson}, \citenamefont {Moody}, \citenamefont
  {Kenneally}, \citenamefont {Landrum}, \citenamefont {Shaughnessy},
  \citenamefont {Stoyer}, \citenamefont {Stoyer},\ and\ \citenamefont
  {Wilk}}]{RN380}%
  \BibitemOpen
  \bibfield  {author} {\bibinfo {author} {\bibfnamefont {Y.~T.}\ \bibnamefont
  {Oganessian}}, \bibinfo {author} {\bibfnamefont {V.~K.}\ \bibnamefont
  {Utyonkov}}, \bibinfo {author} {\bibfnamefont {Y.~V.}\ \bibnamefont
  {Lobanov}}, \bibinfo {author} {\bibfnamefont {F.~S.}\ \bibnamefont
  {Abdullin}}, \bibinfo {author} {\bibfnamefont {A.~N.}\ \bibnamefont
  {Polyakov}}, \bibinfo {author} {\bibfnamefont {R.~N.}\ \bibnamefont
  {Sagaidak}}, \bibinfo {author} {\bibfnamefont {I.~V.}\ \bibnamefont
  {Shirokovsky}}, \bibinfo {author} {\bibfnamefont {Y.~S.}\ \bibnamefont
  {Tsyganov}}, \bibinfo {author} {\bibfnamefont {A.~A.}\ \bibnamefont
  {Voinov}}, \bibinfo {author} {\bibfnamefont {A.~N.}\ \bibnamefont
  {Mezentsev}}, \bibinfo {author} {\bibfnamefont {V.~G.}\ \bibnamefont
  {Subbotin}}, \bibinfo {author} {\bibfnamefont {A.~M.}\ \bibnamefont
  {Sukhov}}, \bibinfo {author} {\bibfnamefont {K.}~\bibnamefont {Subotic}},
  \bibinfo {author} {\bibfnamefont {V.~I.}\ \bibnamefont {Zagrebaev}}, \bibinfo
  {author} {\bibfnamefont {S.~N.}\ \bibnamefont {Dmitriev}}, \bibinfo {author}
  {\bibfnamefont {R.~A.}\ \bibnamefont {Henderson}}, \bibinfo {author}
  {\bibfnamefont {K.~J.}\ \bibnamefont {Moody}}, \bibinfo {author}
  {\bibfnamefont {J.~M.}\ \bibnamefont {Kenneally}}, \bibinfo {author}
  {\bibfnamefont {J.~H.}\ \bibnamefont {Landrum}}, \bibinfo {author}
  {\bibfnamefont {D.~A.}\ \bibnamefont {Shaughnessy}}, \bibinfo {author}
  {\bibfnamefont {M.~A.}\ \bibnamefont {Stoyer}}, \bibinfo {author}
  {\bibfnamefont {N.~J.}\ \bibnamefont {Stoyer}},\ and\ \bibinfo {author}
  {\bibfnamefont {P.~A.}\ \bibnamefont {Wilk}},\ }\href
  {https://doi.org/10.1103/PhysRevC.79.024603} {\bibfield  {journal} {\bibinfo
  {journal} {Physical Review C}\ }\textbf {\bibinfo {volume} {79}},\ \bibinfo
  {pages} {024603} (\bibinfo {year} {2009})}\BibitemShut {NoStop}%
\bibitem [{\citenamefont {Hofmann}\ \emph {et~al.}(2016)\citenamefont
  {Hofmann}, \citenamefont {Heinz}, \citenamefont {Mann}, \citenamefont
  {Maurer}, \citenamefont {Münzenberg}, \citenamefont {Antalic}, \citenamefont
  {Barth}, \citenamefont {Burkhard}, \citenamefont {Dahl}, \citenamefont
  {Eberhardt}, \citenamefont {Grzywacz}, \citenamefont {Hamilton},
  \citenamefont {Henderson}, \citenamefont {Kenneally}, \citenamefont
  {Kindler}, \citenamefont {Kojouharov}, \citenamefont {Lang}, \citenamefont
  {Lommel}, \citenamefont {Miernik}, \citenamefont {Miller}, \citenamefont
  {Moody}, \citenamefont {Morita}, \citenamefont {Nishio}, \citenamefont
  {Popeko}, \citenamefont {Roberto}, \citenamefont {Runke}, \citenamefont
  {Rykaczewski}, \citenamefont {Saro}, \citenamefont {Scheidenberger},
  \citenamefont {Schött}, \citenamefont {Shaughnessy}, \citenamefont {Stoyer},
  \citenamefont {Thörle-Pospiech}, \citenamefont {Tinschert}, \citenamefont
  {Trautmann}, \citenamefont {Uusitalo},\ and\ \citenamefont
  {Yeremin}}]{RN395}%
  \BibitemOpen
  \bibfield  {author} {\bibinfo {author} {\bibfnamefont {S.}~\bibnamefont
  {Hofmann}}, \bibinfo {author} {\bibfnamefont {S.}~\bibnamefont {Heinz}},
  \bibinfo {author} {\bibfnamefont {R.}~\bibnamefont {Mann}}, \bibinfo {author}
  {\bibfnamefont {J.}~\bibnamefont {Maurer}}, \bibinfo {author} {\bibfnamefont
  {G.}~\bibnamefont {Münzenberg}}, \bibinfo {author} {\bibfnamefont
  {S.}~\bibnamefont {Antalic}}, \bibinfo {author} {\bibfnamefont
  {W.}~\bibnamefont {Barth}}, \bibinfo {author} {\bibfnamefont {H.~G.}\
  \bibnamefont {Burkhard}}, \bibinfo {author} {\bibfnamefont {L.}~\bibnamefont
  {Dahl}}, \bibinfo {author} {\bibfnamefont {K.}~\bibnamefont {Eberhardt}},
  \bibinfo {author} {\bibfnamefont {R.}~\bibnamefont {Grzywacz}}, \bibinfo
  {author} {\bibfnamefont {J.~H.}\ \bibnamefont {Hamilton}}, \bibinfo {author}
  {\bibfnamefont {R.~A.}\ \bibnamefont {Henderson}}, \bibinfo {author}
  {\bibfnamefont {J.~M.}\ \bibnamefont {Kenneally}}, \bibinfo {author}
  {\bibfnamefont {B.}~\bibnamefont {Kindler}}, \bibinfo {author} {\bibfnamefont
  {I.}~\bibnamefont {Kojouharov}}, \bibinfo {author} {\bibfnamefont
  {R.}~\bibnamefont {Lang}}, \bibinfo {author} {\bibfnamefont {B.}~\bibnamefont
  {Lommel}}, \bibinfo {author} {\bibfnamefont {K.}~\bibnamefont {Miernik}},
  \bibinfo {author} {\bibfnamefont {D.}~\bibnamefont {Miller}}, \bibinfo
  {author} {\bibfnamefont {K.~J.}\ \bibnamefont {Moody}}, \bibinfo {author}
  {\bibfnamefont {K.}~\bibnamefont {Morita}}, \bibinfo {author} {\bibfnamefont
  {K.}~\bibnamefont {Nishio}}, \bibinfo {author} {\bibfnamefont {A.~G.}\
  \bibnamefont {Popeko}}, \bibinfo {author} {\bibfnamefont {J.~B.}\
  \bibnamefont {Roberto}}, \bibinfo {author} {\bibfnamefont {J.}~\bibnamefont
  {Runke}}, \bibinfo {author} {\bibfnamefont {K.~P.}\ \bibnamefont
  {Rykaczewski}}, \bibinfo {author} {\bibfnamefont {S.}~\bibnamefont {Saro}},
  \bibinfo {author} {\bibfnamefont {C.}~\bibnamefont {Scheidenberger}},
  \bibinfo {author} {\bibfnamefont {H.~J.}\ \bibnamefont {Schött}}, \bibinfo
  {author} {\bibfnamefont {D.~A.}\ \bibnamefont {Shaughnessy}}, \bibinfo
  {author} {\bibfnamefont {M.~A.}\ \bibnamefont {Stoyer}}, \bibinfo {author}
  {\bibfnamefont {P.}~\bibnamefont {Thörle-Pospiech}}, \bibinfo {author}
  {\bibfnamefont {K.}~\bibnamefont {Tinschert}}, \bibinfo {author}
  {\bibfnamefont {N.}~\bibnamefont {Trautmann}}, \bibinfo {author}
  {\bibfnamefont {J.}~\bibnamefont {Uusitalo}},\ and\ \bibinfo {author}
  {\bibfnamefont {A.~V.}\ \bibnamefont {Yeremin}},\ }\href
  {https://doi.org/10.1140/epja/i2016-16180-4} {\bibfield  {journal} {\bibinfo
  {journal} {The European Physical Journal A}\ }\textbf {\bibinfo {volume}
  {52}},\ \bibinfo {pages} {180} (\bibinfo {year} {2016})}\BibitemShut
  {NoStop}%
\bibitem [{\citenamefont {Khuyagbaatar}\ \emph {et~al.}(2020)\citenamefont
  {Khuyagbaatar}, \citenamefont {Yakushev}, \citenamefont {Düllmann},
  \citenamefont {Ackermann}, \citenamefont {Andersson}, \citenamefont {Asai},
  \citenamefont {Block}, \citenamefont {Boll}, \citenamefont {Brand},
  \citenamefont {Cox}, \citenamefont {Dasgupta}, \citenamefont {Derkx},
  \citenamefont {Di~Nitto}, \citenamefont {Eberhardt}, \citenamefont {Even},
  \citenamefont {Evers}, \citenamefont {Fahlander}, \citenamefont {Forsberg},
  \citenamefont {Gates}, \citenamefont {Gharibyan}, \citenamefont {Golubev},
  \citenamefont {Gregorich}, \citenamefont {Hamilton}, \citenamefont
  {Hartmann}, \citenamefont {Herzberg}, \citenamefont {Heßberger},
  \citenamefont {Hinde}, \citenamefont {Hoffmann}, \citenamefont {Hollinger},
  \citenamefont {Hübner}, \citenamefont {Jäger}, \citenamefont {Kindler},
  \citenamefont {Kratz}, \citenamefont {Krier}, \citenamefont {Kurz},
  \citenamefont {Laatiaoui}, \citenamefont {Lahiri}, \citenamefont {Lang},
  \citenamefont {Lommel}, \citenamefont {Maiti}, \citenamefont {Miernik},
  \citenamefont {Minami}, \citenamefont {Mistry}, \citenamefont {Mokry},
  \citenamefont {Nitsche}, \citenamefont {Omtvedt}, \citenamefont {Pang},
  \citenamefont {Papadakis}, \citenamefont {Renisch}, \citenamefont {Roberto},
  \citenamefont {Rudolph}, \citenamefont {Runke}, \citenamefont {Rykaczewski},
  \citenamefont {Sarmiento}, \citenamefont {Schädel}, \citenamefont
  {Schausten}, \citenamefont {Semchenkov}, \citenamefont {Shaughnessy},
  \citenamefont {Steinegger}, \citenamefont {Steiner}, \citenamefont
  {Tereshatov}, \citenamefont {Thörle-Pospiech}, \citenamefont {Tinschert},
  \citenamefont {Torres De~Heidenreich}, \citenamefont {Trautmann},
  \citenamefont {Türler}, \citenamefont {Uusitalo}, \citenamefont {Wegrzecki},
  \citenamefont {Wiehl}, \citenamefont {Van~Cleve},\ and\ \citenamefont
  {Yakusheva}}]{RN396}%
  \BibitemOpen
  \bibfield  {author} {\bibinfo {author} {\bibfnamefont {J.}~\bibnamefont
  {Khuyagbaatar}}, \bibinfo {author} {\bibfnamefont {A.}~\bibnamefont
  {Yakushev}}, \bibinfo {author} {\bibfnamefont {C.~E.}\ \bibnamefont
  {Düllmann}}, \bibinfo {author} {\bibfnamefont {D.}~\bibnamefont
  {Ackermann}}, \bibinfo {author} {\bibfnamefont {L.~L.}\ \bibnamefont
  {Andersson}}, \bibinfo {author} {\bibfnamefont {M.}~\bibnamefont {Asai}},
  \bibinfo {author} {\bibfnamefont {M.}~\bibnamefont {Block}}, \bibinfo
  {author} {\bibfnamefont {R.~A.}\ \bibnamefont {Boll}}, \bibinfo {author}
  {\bibfnamefont {H.}~\bibnamefont {Brand}}, \bibinfo {author} {\bibfnamefont
  {D.~M.}\ \bibnamefont {Cox}}, \bibinfo {author} {\bibfnamefont
  {M.}~\bibnamefont {Dasgupta}}, \bibinfo {author} {\bibfnamefont
  {X.}~\bibnamefont {Derkx}}, \bibinfo {author} {\bibfnamefont
  {A.}~\bibnamefont {Di~Nitto}}, \bibinfo {author} {\bibfnamefont
  {K.}~\bibnamefont {Eberhardt}}, \bibinfo {author} {\bibfnamefont
  {J.}~\bibnamefont {Even}}, \bibinfo {author} {\bibfnamefont {M.}~\bibnamefont
  {Evers}}, \bibinfo {author} {\bibfnamefont {C.}~\bibnamefont {Fahlander}},
  \bibinfo {author} {\bibfnamefont {U.}~\bibnamefont {Forsberg}}, \bibinfo
  {author} {\bibfnamefont {J.~M.}\ \bibnamefont {Gates}}, \bibinfo {author}
  {\bibfnamefont {N.}~\bibnamefont {Gharibyan}}, \bibinfo {author}
  {\bibfnamefont {P.}~\bibnamefont {Golubev}}, \bibinfo {author} {\bibfnamefont
  {K.~E.}\ \bibnamefont {Gregorich}}, \bibinfo {author} {\bibfnamefont {J.~H.}\
  \bibnamefont {Hamilton}}, \bibinfo {author} {\bibfnamefont {W.}~\bibnamefont
  {Hartmann}}, \bibinfo {author} {\bibfnamefont {R.~D.}\ \bibnamefont
  {Herzberg}}, \bibinfo {author} {\bibfnamefont {F.~P.}\ \bibnamefont
  {Heßberger}}, \bibinfo {author} {\bibfnamefont {D.~J.}\ \bibnamefont
  {Hinde}}, \bibinfo {author} {\bibfnamefont {J.}~\bibnamefont {Hoffmann}},
  \bibinfo {author} {\bibfnamefont {R.}~\bibnamefont {Hollinger}}, \bibinfo
  {author} {\bibfnamefont {A.}~\bibnamefont {Hübner}}, \bibinfo {author}
  {\bibfnamefont {E.}~\bibnamefont {Jäger}}, \bibinfo {author} {\bibfnamefont
  {B.}~\bibnamefont {Kindler}}, \bibinfo {author} {\bibfnamefont {J.~V.}\
  \bibnamefont {Kratz}}, \bibinfo {author} {\bibfnamefont {J.}~\bibnamefont
  {Krier}}, \bibinfo {author} {\bibfnamefont {N.}~\bibnamefont {Kurz}},
  \bibinfo {author} {\bibfnamefont {M.}~\bibnamefont {Laatiaoui}}, \bibinfo
  {author} {\bibfnamefont {S.}~\bibnamefont {Lahiri}}, \bibinfo {author}
  {\bibfnamefont {R.}~\bibnamefont {Lang}}, \bibinfo {author} {\bibfnamefont
  {B.}~\bibnamefont {Lommel}}, \bibinfo {author} {\bibfnamefont
  {M.}~\bibnamefont {Maiti}}, \bibinfo {author} {\bibfnamefont
  {K.}~\bibnamefont {Miernik}}, \bibinfo {author} {\bibfnamefont
  {S.}~\bibnamefont {Minami}}, \bibinfo {author} {\bibfnamefont {A.~K.}\
  \bibnamefont {Mistry}}, \bibinfo {author} {\bibfnamefont {C.}~\bibnamefont
  {Mokry}}, \bibinfo {author} {\bibfnamefont {H.}~\bibnamefont {Nitsche}},
  \bibinfo {author} {\bibfnamefont {J.~P.}\ \bibnamefont {Omtvedt}}, \bibinfo
  {author} {\bibfnamefont {G.~K.}\ \bibnamefont {Pang}}, \bibinfo {author}
  {\bibfnamefont {P.}~\bibnamefont {Papadakis}}, \bibinfo {author}
  {\bibfnamefont {D.}~\bibnamefont {Renisch}}, \bibinfo {author} {\bibfnamefont
  {J.~B.}\ \bibnamefont {Roberto}}, \bibinfo {author} {\bibfnamefont
  {D.}~\bibnamefont {Rudolph}}, \bibinfo {author} {\bibfnamefont
  {J.}~\bibnamefont {Runke}}, \bibinfo {author} {\bibfnamefont {K.~P.}\
  \bibnamefont {Rykaczewski}}, \bibinfo {author} {\bibfnamefont {L.~G.}\
  \bibnamefont {Sarmiento}}, \bibinfo {author} {\bibfnamefont {M.}~\bibnamefont
  {Schädel}}, \bibinfo {author} {\bibfnamefont {B.}~\bibnamefont {Schausten}},
  \bibinfo {author} {\bibfnamefont {A.}~\bibnamefont {Semchenkov}}, \bibinfo
  {author} {\bibfnamefont {D.~A.}\ \bibnamefont {Shaughnessy}}, \bibinfo
  {author} {\bibfnamefont {P.}~\bibnamefont {Steinegger}}, \bibinfo {author}
  {\bibfnamefont {J.}~\bibnamefont {Steiner}}, \bibinfo {author} {\bibfnamefont
  {E.~E.}\ \bibnamefont {Tereshatov}}, \bibinfo {author} {\bibfnamefont
  {P.}~\bibnamefont {Thörle-Pospiech}}, \bibinfo {author} {\bibfnamefont
  {K.}~\bibnamefont {Tinschert}}, \bibinfo {author} {\bibfnamefont
  {T.}~\bibnamefont {Torres De~Heidenreich}}, \bibinfo {author} {\bibfnamefont
  {N.}~\bibnamefont {Trautmann}}, \bibinfo {author} {\bibfnamefont
  {A.}~\bibnamefont {Türler}}, \bibinfo {author} {\bibfnamefont
  {J.}~\bibnamefont {Uusitalo}}, \bibinfo {author} {\bibfnamefont
  {M.}~\bibnamefont {Wegrzecki}}, \bibinfo {author} {\bibfnamefont
  {N.}~\bibnamefont {Wiehl}}, \bibinfo {author} {\bibfnamefont {S.~M.}\
  \bibnamefont {Van~Cleve}},\ and\ \bibinfo {author} {\bibfnamefont
  {V.}~\bibnamefont {Yakusheva}},\ }\href
  {https://doi.org/10.1103/PhysRevC.102.064602} {\bibfield  {journal} {\bibinfo
   {journal} {Physical Review C}\ }\textbf {\bibinfo {volume} {102}},\ \bibinfo
  {pages} {064602} (\bibinfo {year} {2020})}\BibitemShut {NoStop}%
\bibitem [{\citenamefont {GIARDINA}\ \emph {et~al.}(2010)\citenamefont
  {GIARDINA}, \citenamefont {FAZIO}, \citenamefont {MANDAGLIO}, \citenamefont
  {MANGANARO}, \citenamefont {NASIROV}, \citenamefont {ROMANIUK},\ and\
  \citenamefont {SACCÀ}}]{RN382}%
  \BibitemOpen
  \bibfield  {author} {\bibinfo {author} {\bibfnamefont {G.}~\bibnamefont
  {GIARDINA}}, \bibinfo {author} {\bibfnamefont {G.}~\bibnamefont {FAZIO}},
  \bibinfo {author} {\bibfnamefont {G.}~\bibnamefont {MANDAGLIO}}, \bibinfo
  {author} {\bibfnamefont {M.}~\bibnamefont {MANGANARO}}, \bibinfo {author}
  {\bibfnamefont {A.~K.}\ \bibnamefont {NASIROV}}, \bibinfo {author}
  {\bibfnamefont {M.~V.}\ \bibnamefont {ROMANIUK}},\ and\ \bibinfo {author}
  {\bibfnamefont {C.}~\bibnamefont {SACCÀ}},\ }\href
  {https://doi.org/10.1142/s0218301310015333} {\bibfield  {journal} {\bibinfo
  {journal} {International Journal of Modern Physics E}\ }\textbf {\bibinfo
  {volume} {19}},\ \bibinfo {pages} {882} (\bibinfo {year} {2010})}\BibitemShut
  {NoStop}%
\bibitem [{\citenamefont {Wang}\ \emph {et~al.}(2011)\citenamefont {Wang},
  \citenamefont {Tian},\ and\ \citenamefont {Scheid}}]{RN383}%
  \BibitemOpen
  \bibfield  {author} {\bibinfo {author} {\bibfnamefont {N.}~\bibnamefont
  {Wang}}, \bibinfo {author} {\bibfnamefont {J.}~\bibnamefont {Tian}},\ and\
  \bibinfo {author} {\bibfnamefont {W.}~\bibnamefont {Scheid}},\ }\href
  {https://doi.org/10.1103/PhysRevC.84.061601} {\bibfield  {journal} {\bibinfo
  {journal} {Physical Review C}\ }\textbf {\bibinfo {volume} {84}},\ \bibinfo
  {pages} {061601} (\bibinfo {year} {2011})}\BibitemShut {NoStop}%
\bibitem [{\citenamefont {Wang}\ \emph {et~al.}(2012)\citenamefont {Wang},
  \citenamefont {Zhao}, \citenamefont {Scheid},\ and\ \citenamefont
  {Zhou}}]{RN384}%
  \BibitemOpen
  \bibfield  {author} {\bibinfo {author} {\bibfnamefont {N.}~\bibnamefont
  {Wang}}, \bibinfo {author} {\bibfnamefont {E.-G.}\ \bibnamefont {Zhao}},
  \bibinfo {author} {\bibfnamefont {W.}~\bibnamefont {Scheid}},\ and\ \bibinfo
  {author} {\bibfnamefont {S.-G.}\ \bibnamefont {Zhou}},\ }\href
  {https://doi.org/10.1103/PhysRevC.85.041601} {\bibfield  {journal} {\bibinfo
  {journal} {Physical Review C}\ }\textbf {\bibinfo {volume} {85}},\ \bibinfo
  {pages} {041601} (\bibinfo {year} {2012})}\BibitemShut {NoStop}%
\bibitem [{\citenamefont {Zhang}\ \emph {et~al.}(2013)\citenamefont {Zhang},
  \citenamefont {Wang},\ and\ \citenamefont {Ren}}]{RN385}%
  \BibitemOpen
  \bibfield  {author} {\bibinfo {author} {\bibfnamefont {J.}~\bibnamefont
  {Zhang}}, \bibinfo {author} {\bibfnamefont {C.}~\bibnamefont {Wang}},\ and\
  \bibinfo {author} {\bibfnamefont {Z.}~\bibnamefont {Ren}},\ }\href
  {https://doi.org/https://doi.org/10.1016/j.nuclphysa.2013.04.010} {\bibfield
  {journal} {\bibinfo  {journal} {Nuclear Physics A}\ }\textbf {\bibinfo
  {volume} {909}},\ \bibinfo {pages} {36} (\bibinfo {year} {2013})}\BibitemShut
  {NoStop}%
\bibitem [{\citenamefont {Gan}\ \emph {et~al.}(2011)\citenamefont {Gan},
  \citenamefont {Zhou}, \citenamefont {Huang}, \citenamefont {Feng},\ and\
  \citenamefont {Li}}]{RN389}%
  \BibitemOpen
  \bibfield  {author} {\bibinfo {author} {\bibfnamefont {Z.}~\bibnamefont
  {Gan}}, \bibinfo {author} {\bibfnamefont {X.}~\bibnamefont {Zhou}}, \bibinfo
  {author} {\bibfnamefont {M.}~\bibnamefont {Huang}}, \bibinfo {author}
  {\bibfnamefont {Z.}~\bibnamefont {Feng}},\ and\ \bibinfo {author}
  {\bibfnamefont {J.}~\bibnamefont {Li}},\ }\href
  {https://doi.org/10.1007/s11433-011-4436-4} {\bibfield  {journal} {\bibinfo
  {journal} {Science China Physics, Mechanics and Astronomy}\ }\textbf
  {\bibinfo {volume} {54}},\ \bibinfo {pages} {61} (\bibinfo {year}
  {2011})}\BibitemShut {NoStop}%
\bibitem [{\citenamefont {Wong}(1973)}]{RN390}%
  \BibitemOpen
  \bibfield  {author} {\bibinfo {author} {\bibfnamefont {C.~Y.}\ \bibnamefont
  {Wong}},\ }\href {https://doi.org/10.1103/PhysRevLett.31.766} {\bibfield
  {journal} {\bibinfo  {journal} {Physical Review Letters}\ }\textbf {\bibinfo
  {volume} {31}},\ \bibinfo {pages} {766} (\bibinfo {year} {1973})}\BibitemShut
  {NoStop}%
\bibitem [{\citenamefont {Dutt}(2011)}]{RN367}%
  \BibitemOpen
  \bibfield  {author} {\bibinfo {author} {\bibfnamefont {I.}~\bibnamefont
  {Dutt}},\ }\href {https://doi.org/10.1007/s12043-011-0094-3} {\bibfield
  {journal} {\bibinfo  {journal} {Pramana}\ }\textbf {\bibinfo {volume} {76}},\
  \bibinfo {pages} {921} (\bibinfo {year} {2011})}\BibitemShut {NoStop}%
\bibitem [{\citenamefont {Hill}\ and\ \citenamefont {Wheeler}(1953)}]{RN391}%
  \BibitemOpen
  \bibfield  {author} {\bibinfo {author} {\bibfnamefont {D.~L.}\ \bibnamefont
  {Hill}}\ and\ \bibinfo {author} {\bibfnamefont {J.~A.}\ \bibnamefont
  {Wheeler}},\ }\href {https://doi.org/10.1103/PhysRev.89.1102} {\bibfield
  {journal} {\bibinfo  {journal} {Physical Review}\ }\textbf {\bibinfo {volume}
  {89}},\ \bibinfo {pages} {1102} (\bibinfo {year} {1953})}\BibitemShut
  {NoStop}%
\bibitem [{\citenamefont {Möller}\ \emph {et~al.}(2016)\citenamefont
  {Möller}, \citenamefont {Sierk}, \citenamefont {Ichikawa},\ and\
  \citenamefont {Sagawa}}]{RN392}%
  \BibitemOpen
  \bibfield  {author} {\bibinfo {author} {\bibfnamefont {P.}~\bibnamefont
  {Möller}}, \bibinfo {author} {\bibfnamefont {A.~J.}\ \bibnamefont {Sierk}},
  \bibinfo {author} {\bibfnamefont {T.}~\bibnamefont {Ichikawa}},\ and\
  \bibinfo {author} {\bibfnamefont {H.}~\bibnamefont {Sagawa}},\ }\href
  {https://doi.org/https://doi.org/10.1016/j.adt.2015.10.002} {\bibfield
  {journal} {\bibinfo  {journal} {Atomic Data and Nuclear Data Tables}\
  }\textbf {\bibinfo {volume} {109-110}},\ \bibinfo {pages} {1} (\bibinfo
  {year} {2016})}\BibitemShut {NoStop}%
\bibitem [{\citenamefont {Vandenbosch}\ and\ \citenamefont
  {Huizenga}(1974)}]{RN393}%
  \BibitemOpen
  \bibfield  {author} {\bibinfo {author} {\bibfnamefont {R.}~\bibnamefont
  {Vandenbosch}}\ and\ \bibinfo {author} {\bibfnamefont {R.}~\bibnamefont
  {Huizenga}, \bibfnamefont {John}},\ }\href@noop {} {\emph {\bibinfo {title}
  {Nuclear Fission Academic Press}}},\ Vol.~\bibinfo {volume} {29}\ (\bibinfo
  {year} {1974})\ pp.\ \bibinfo {pages} {864--865}\BibitemShut {NoStop}%
\bibitem [{\citenamefont {Jackson}(1956)}]{RN394}%
  \BibitemOpen
  \bibfield  {author} {\bibinfo {author} {\bibfnamefont {J.~D.}\ \bibnamefont
  {Jackson}},\ }\href {https://doi.org/10.1139/p56-087} {\bibfield  {journal}
  {\bibinfo  {journal} {Canadian Journal of Physics}\ }\textbf {\bibinfo
  {volume} {34}},\ \bibinfo {pages} {767} (\bibinfo {year} {1956})}\BibitemShut
  {NoStop}%
\end{thebibliography}%

\end{document}